\documentclass[12pt]{article}
\usepackage{amsmath}
\usepackage{latexsym}
\usepackage{bbm}
\usepackage{amssymb}
\usepackage{lipsum}

\usepackage{tikz}
\usetikzlibrary{decorations.pathmorphing}
\usetikzlibrary{decorations.markings}

\usepackage{hyperref}
\usepackage[all]{hypcap}

\def\a{\alpha}

\def\b{\beta}

\def\d{\delta}

\def\e{\varepsilon}

\def\z{\zeta}
\def\et{\eta}
\def\th{\theta}

\def\l{\lambda}
\def\m{\mu}
\def\n{\nu}

\def\r{\rho}

\def\s{\sigma}
\def\t{\tau}

\def\D{\Delta}

\def\O{\Omega}

\def\pa{\partial}

\def\half{\frac{1}{2}}

\def\tr{{\rm tr}}

\def\and{{\rm and}}

\def\ie{{\it i.e.,} }

\def\IC{\mathbbm C}

\newcommand{\be}{\begin{equation}}
\newcommand{\bea}{\begin{eqnarray}}

\newcommand{\ee}{\end{equation}}
\newcommand{\eea}{\end{eqnarray}}
\newcommand{\blfootnote}[1]{{\renewcommand{\thefootnote}{\roman{footnote}}\footnotetext[0]{#1}}}

\begin{document}
\vspace*{-1.0in}
\thispagestyle{empty}
\begin{flushright}
CALT-TH-2017-049
\end{flushright}

%\large
\normalsize
\baselineskip = 18pt
\parskip = 6pt

\vspace{1.0in}

{\Large \begin{center}
{\bf M5-Brane and D-Brane Scattering Amplitudes}
\end{center}}

\vspace{.25in}

\begin{center}
Matthew Heydeman\blfootnote{mheydema@caltech.edu},$^a$
John H. Schwarz\blfootnote{jhs@theory.caltech.edu},$^a$
and Congkao Wen\blfootnote{cwen@caltech.edu}$^{a,b}$
\\
\emph{${}^a$ Walter Burke Institute for Theoretical Physics\\
California Institute of Technology 452-48\\ Pasadena, CA  91125, USA \\
${}^b$Mani L. Bhaumik Institute for Theoretical Physics, \\[-1mm]
			 Department of Physics and Astronomy, UCLA, Los Angeles, CA 90095, USA}
\end{center}
\vspace{.25in}

\begin{center}
\textbf{Abstract}
\end{center}
\begin{quotation}

We present tree-level $n$-particle on-shell scattering amplitudes of
various brane theories with $16$ conserved supercharges. These include the
world-volume theory of a probe D3-brane or D5-brane in 10D Minkowski spacetime
as well as a probe M5-brane in 11D
Minkowski spacetime, which describes self interactions of an abelian tensor
supermultiplet with 6D $(2,0)$ supersymmetry.
Twistor-string-like formulas are proposed for tree-level scattering amplitudes
of all multiplicities for each of these theories. The R symmetry
of the D3-brane theory is shown to be $SU(4) \times U(1)$, and the $U(1)$ factor
implies that its amplitudes are helicity conserving.
Each of 6D theories (D5-brane and M5-brane) reduces to the
D3-brane theory by dimensional reduction.
As special cases of the general
M5-brane amplitudes, we present compact formulas for examples involving
only the self-dual $B$ field with $n=4,6,8$.

\end{quotation}

\newpage

\pagenumbering{arabic}

\tableofcontents

\newpage

\section{Introduction}

This paper proposes explicit formulas for on-shell $n$-particle scattering amplitudes
in the tree approximation
%\footnote{As far as we know, these theories only exist as classical field
%theories, so it would not make sense to construct loop amplitudes.}
for three massless field theories, each of which is a maximally supersymmetric matter
theory with $16$ unbroken supersymmetries and
16 additional spontaneously broken supersymmetries. The fermions in these theories are
Goldstone particles (or Goldstinos) of the type first considered by Volkov and
Akulov \cite{Volkov:1973ix}\cite{Kallosh:1997aw}. These three theories arise
naturally in string theory as the world-volume theories of branes. The first theory is the
world-volume theory of a probe D3-brane (of type IIB superstring theory)
in a 10D Minkowski-space background. The second theory is the
world-volume theory of a probe D5-brane (of type IIB superstring theory)
in a 10D Minkowski-space background.
The third theory is the world-volume theory of a probe M5-brane
(of M theory) in an 11D Minkowski-space background.
We will refer to these theories as the D3 theory, the D5 theory, and the
M5 theory. These three theories are closely related. Specifically, both of the
6D theories (D5 and M5) can be truncated (by a procedure called dimensional reduction)
to give rise to the 4D theory (D3). These relationships, which are predicted by string
theory, will provide powerful checks of the results, as well as a role in their derivation.
Another important feature that all three of these theories have in common is that nonvanishing on-shell
scattering amplitudes require an even number of particles, \ie $n$ must be even.

The D3 theory is a 4D Dirac--Born--Infeld (DBI) theory, with ${\cal N}=4$ supersymmetry,
which some authors call sDBI theory. It it a self-interacting theory of a massless abelian
${\cal N} =4$ vector supermultiplet, which consists of a vector, four spinors, and six
scalars. Its R-symmetry group is $SU(4) \times U(1)$. Although the helicity-conservation property of scattering amplitudes of the D3 theory has also been understood previously \cite{Rosly:2002jt} using  the electric magnetic duality of D3-brane action \cite{Gibbons:1995cv}, the additional $U(1)$ factor in  the R-symmetry group has not been noted previously.\footnote{We will explain later why the D3 theory has a larger
R-symmetry group than ${\cal N}=4$ super Yang--Mills theory. Of course, there
are many other differences. For example, the D3 theory is not conformal.}
The action for the D3 theory was derived in \cite{Aganagic:1996nn}
by dimensional reduction of the action for the D9-brane, which was constructed using
string-theoretic techniques. (See
\cite{Cederwall:1996pv}\cite{Aganagic:1996pe}\cite{Cederwall:1996ri}\cite{Bergshoeff:1996tu}
for related work.) The D3 theory has been examined in some detail
recently in \cite{Bergshoeff:2013pia}. There has been a recent proposal
for the tree amplitudes of this theory in \cite{He:2016vfi}\cite{Cachazo:2016njl}. Our formulas will
take a different form, for reasons that will be explained.

The action for the D5 theory also can be obtained by dimensional reduction of the
D9-brane action. This theory is a self-interacting theory of a single vector
supermultiplet with $(1,1)$ supersymmetry in 6D.
The vector supermultiplet consists of a vector, four spinors,
and four scalars. The R-symmetry group of the D5 theory is $SU(2) \times SU(2)$.

The M5 theory is a self-interacting theory of a single tensor supermultiplet
with $(2,0)$ supersymmetry in 6D. This multiplet contains a two-form
field $B$ with a self-dual field strength ($H=dB=\star H$) as well as four spinors and
five scalars. There is an analog of the Born--Infeld action that
describes self interactions of the $B$ field, which was constructed in \cite{Perry:1996mk}.
This theory has 6D Lorentz invariance, though the action only has manifest
5D Lorentz invariance.  The five additional Lorentz transformations that involve a
particular (arbitrarily chosen) direction are not obvious symmetries. These transformations
of the Lagrangian give a total derivative. Dimensional reduction of this theory to five
dimensions gives pure Born--Infeld theory. The action for the supersymmetric extension
of the 6D theory that incorporates the complete $(2,0)$ supermultiplet, \ie the M5 theory,
was constructed in \cite{Aganagic:1997zq}.
(See \cite{Howe:1996yn}\cite{Pasti:1997gx}\cite{Bandos:1997ui}\cite{Howe:1997fb}
for related work.) The R-symmetry group of this theory is $USp(4)$. Certain lower-point
amplitudes for the M5 theory have been discussed previously, for example in
\cite{Huang:2010rn}\cite{Czech:2011dk}\cite{Elvang:2012st}\cite{Chen:2015hpa}\cite{Bianchi:2016viy}.
The requirement that they give D3 amplitudes
after dimensional reduction to 4D will play an important role in our analysis.

Another feature that these three theories have in common
is that they inherit their symmetries from those of the parent
theories, \ie M-theory in flat space and Type IIB
superstring theory in flat space. By positioning
the probe branes in the ambient space, some of the symmetries
of the parent theory are spontaneously broken. Broken symmetries include
translations perpendicular to the branes and half
of the supersymmetries. These symmetries are
realized nonlinearly in the brane theories. Thus,
the scalars and spinors in these theories are Goldstone particles.
As a result, the amplitudes of these theories satisfy various soft theorems.
The vector and tensor gauge symmetries are inherited from the background NS-NS 2-form
of Type IIB and the M-theory 3-form, respectively \cite{Strominger:1995ac}.

One of the challenges in formulating on-shell scattering amplitudes for these
theories is to make their various required symmetries manifest. As has become
conventional for massless particles, we use twistor-like spinor-helicity coordinates
to represent momenta and supercharges. These introduce a little-group symmetry
for each of the scattered particles. As we will explain, this group is
$SU(2)\times SU(2)$ for the D5 theory, $SU(2)$ for the M5 theory, and $U(1)$ for the D3 theory.
The use of spinor-helicity variables allows us to construct on-shell amplitudes with
manifest Lorentz invariance even for chiral theories, such as the M5 theory,
which has well-known obstructions to constructing a useful Lorentz-invariant action.
In addition to super-Poincar\'e symmetry, each of these theories has an
R-symmetry group: $SU(2)\times SU(2)$ for the D5 theory,
$USp(4)$ for the M5 theory, and $SU(4) \times U(1)$ for the D3 theory.

Our formulas for scattering
amplitudes in each of the three theories take forms that are similar to the
twistor-string formulation of 4D $\mathcal{N}=4$ super Yang--Mills amplitudes (SYM)
in Witten's classic twistor-string paper \cite{Witten:2003nn}. 
The twistor-string formulation of 4D $\mathcal{N}=4$ SYM is given by the Witten-RSV formula, 
which was studied in detail in \cite{Roiban:2004vt, Roiban:2004ka, Roiban:2004yf}. In particular,
we associate a coordinate $\s_i$
on the Riemann sphere to the $i$th particle in an $n$-particle scattering amplitude.
The formula for the amplitude is required to be invariant under a simultaneous
$SL(2,\IC)$ transformation of these coordinates. Following Cachazo
et al. \cite{Cachazo:2013iaa}, in the twistor-string-like
formalism that we use, certain rational functions of $\s_i$ are associated
to the $i$th particle. These functions are restricted by delta-function
constraints in such a way that the number of bosonic delta functions is equal to
the number of bosonic integrations. Thus, the formulas are actually algebraic, as they
should be for tree amplitudes. Furthermore, the delta-function constraints
imply the scattering equations \cite{Cachazo:2013gna},
which are $\sum_j p_i \cdot p_j / \s_{ij} = 0$, where $\s_{ij} = \s_i - \s_j$.
This approach allows us to formulate all of the amplitudes for the three theories
in a uniform way. It also is convenient for verifying some of their essential
properties.

Our main results are general formulas for the $n$-particle on-shell tree
amplitudes for each of the three theories. These formulas make most of the
required symmetries manifest, or at least easy to understand. The exception
is the R symmetry, where only a subgroup is manifest.
The supermultiplets are incorporated by associating four Grassmann coordinates, with
specified transformation properties, to each external particle. The key to making
the full R-symmetry group manifest is to carry out a Fourier transformation for half of the
Grassmann coordinates -- two per particle. The price that one pays for making R symmetry
manifest is that
%the little-group symmetry is no longer manifest, and
the formulas become
somewhat more complicated for the 6D theories.

The paper is organized as follows: We begin in \hyperref[sec:symmetry]{section 2} with a
discussion of general properties, such as symmetries, conserved charges, and on-shell
states, for each of the three theories considered in this paper.
We utilize the 4D spinor-helicity
formalism for the D3 theory and the 6D one for the M5 theory and the D5 theory.
To illustrate the structures and ideas, \hyperref[sec:4pts]{section 3} examines
the four-particle amplitudes for these theories. \hyperref[sec:nptD3]{Section 4}
presents a general formula for the $n$-particle amplitudes
of the D3 theory. As mentioned previously, our formulas for scattering amplitudes in each of the
three theories take forms that are similar to the twistor-string formulation
of 4D $\mathcal{N}=4$ super Yang--Mills amplitudes \cite{Witten:2003nn}.
This formulation of the D3 theory is somewhat different from those in the
literature. It is more suitable for the generalization to 6D,
which is required for the M5 and D5 theories.

In \hyperref[sec:nptM5]{section 5} we propose a new formula, given in Eq.~(\ref{eq:(2,0)}),
which gives all of the tree amplitudes
of the M5 theory and generalizes the D3 formula in a way that is consistent with
dimensional reduction of $\mathcal{N} = (2,0)$ in 6D to $\mathcal{N} = 4$ in 4D.
This is our most novel result, providing a mathematical formula for the complete
tree-level S-matrix for a theory whose Lagrangian description has well-known issues
mentioned earlier. This section also describes
various checks of the formula, including symmetries, soft theorems, and factorization.
Using knowledge of the lower-point amplitudes and factorization, we obtain compact analytic
expressions for certain amplitudes of the self-dual $B$ fields for $n=6$ and $n=8$. These
agree perfectly with the general integral formula and give explicit consistency checks.
Despite the apparent differences between the M5 and D5 theories, in \hyperref[sec:nptD5]{section 6}
we present a similar integral formula for the D5-brane amplitudes, which reproduces what one
obtains from the D5-brane action. Finally, our conclusions and remarks concerning future
directions are presented in \hyperref[sec:conclusion]{section 7}. Further technical details
and an analysis of the R symmetries are presented in the \hyperref[appendix:Jacobi]{Appendices}.

\section{Symmetries, conserved charges, and supermultiplets} \label{sec:symmetry}

The three theories that we are considering have three types of conserved charges, which form a nice
superalgebra in each case. These charges, are the momenta $p_i$, supersymmetry charges $q_i$,
and R-symmetry charges $R_i$, where the index $i=1, 2, \ldots, n$ labels the $n$
particles participating in an $n$-particle on-shell scattering amplitude $A_n$.
By treating all of the
particles symmetrically as ingoing, conservation of these charges is simply the statement that
\begin{equation}
\sum_{i=1}^n p_i = 0, \quad \sum_{i=1}^n q_i = 0, \quad \sum_{i=1}^n R_i =0.
\end{equation}
In practice, some of these conservation laws are implemented by including appropriate delta functions
in the formula for $A_n$. The other charges are represented by differential operators and
their conservation is achieved by requiring that $A_n$ is annihilated by the
appropriate sums of these differential operators. Lorentz invariance will be manifest in all formulas.

\subsection{M5 theory} \label{section:introM5theory}

The world-volume theory of a probe M5 in an 11D
Minkowski space background has $(2,0)$ 6D
supersymmetry. This theory describes a single massless
self-interacting tensor supermultiplet. This supermultiplet contains a
two-form field $B_{\m\n}$, with a three-form field strength
$H = dB$, which is self-dual in the free-theory limit. Such a field
gives rise to three on-shell degrees of freedom.
The tensor supermultiplet also contains four fermions and five scalars.
Altogether, there are eight bosonic and eight fermionic
on-shell degrees of freedom. The three multiplicities (1, 4, 5)
correspond to representations of the
$USp(4) = Spin(5)$ R-symmetry group, which is an unbroken
global symmetry of the M5 theory. This symmetry can be thought
of as arising from rotations in the five spatial dimensions
that are orthogonal to a flat M5 in 11D Minkowski spacetime.
The little group for massless particles in $d$ dimensions is $Spin(d-2)$. Thus,
in 6D it is $SU(2) \times SU(2)$.
However, in the special case of the tensor multiplet all of the
on-shell particles are singlets of one of the two
$SU(2)$ factors. Specifically, the self-dual tensor transforms as $(3,1)$,
the spinors, which are also chiral, transform as $(2,1)$ and the scalars
transform as $(1,1)$. Therefore we shall ignore the trivial $SU(2)$
and refer to the nontrivial $SU(2)$ as the little group of this theory.
In the case of the D5 theory, considered in the next subsection,
both $SU(2)$ factors will be required.

It is convenient to introduce four Grassmann coordinates,
such that the entire on-shell supermultiplet can be
described by a single scalar expression. There are various ways
to do this. One obvious choice is to introduce four Grassmann coordinates $\eta^I$,
which transform as the fundamental four-dimensional representation of
the $USp(4)$ R-symmetry group. In this way, one can make the R symmetry manifest,
and we first discuss this formulation.
However, because amplitudes for massless particles are labeled by incoming momenta
and little-group indices,
in most formulas we will make use of a second description of the supermultiplet
that makes little group symmetry manifest.

For theories involving massless particles, it is also convenient to introduce
eight bosonic spinor-helicity coordinates
$\l^A_a$, where $A = 1,2,3,4$ labels a
spinor representation of the 6D Lorentz group $Spin(5,1)$ and $a=\pm$ labels
a doublet of the chiral little group discussed above. These coordinates
belong to a real representation of the product group, because the spinor representation
of the Lorentz group and the doublet little-group representation
are both pseudoreal. In terms of these coordinates
the momentum of an on-shell massless particle is written \cite{Cheung:2009dc},
\begin{equation}
p^{AB} = \e^{ab} \l^A_a \l^B_b = \l^A_a \l^{Ba}= \l_+^A \l_-^B - \l_-^A \l_+^B.
\end{equation}
This formula is invariant under the $SU(2)$ little group,
and therefore three of the eight
$\l$ coordinates are redundant, leaving five nontrivial
degrees of freedom, as appropriate for
the momentum of a massless particle in 6D.
Note also that $p^{AB} = - p^{BA}$ is a six-vector
of the Lorentz group. $p^2$, which gives the square of the mass,
is proportional to the Pfaffian of
$p^{AB}$. This vanishes because the $4\times 4$ matrix $p^{AB}$ has
rank two. When we describe $n$-particle
scattering amplitudes we attach labels
$i,j, \ldots$, which take the values $1,2,\ldots, n$,
to the coordinates. Thus, the $i$th particle is associated to
$\l^A_{i+}$, $\l^A_{i-}$, and $\eta^I_i$.

The 16 supersymmetry charges of the M5 theory can be
represented by\footnote{6D $\mathcal{N}=(2,0)$ on-shell superspace was first discussed in \cite{Huang:2010rn}.}
\begin{equation} \label{qAI}
q^{AI} =  \l^A_+ \eta^I  - \O^{IJ}\l_-^A \frac{\pa}{\pa \eta^J},
\end{equation}
where the antisymmetric matrix $\O^{IJ}$ is the symplectic metric. We will find
it convenient later to choose $\O^{13} = \O^{24} =1$. This formula
can be recast as
\bea
q^{AI} =\e^{ab}\l_a^A \eta_b^I = \l_a^A \eta^{Ia},
\eea
where $\eta_-^I = \eta^I$ and $\eta_+^I = \O^{IJ} \pa/\pa \eta^J$. Then
\begin{equation}
\{\eta_a^I, \eta_b^J\} = \e_{ab} \O^{IJ}.
\end{equation}
This makes the little-group invariance of the supercharges manifest.
Note that the supercharges belong to a chiral representation of the Lorentz group,
and the opposite chirality representation does not appear. This is what is meant by
saying that the theory has $(2,0)$ supersymmetry. As usual, the supercharges anticommute to
give the momenta
\begin{equation}
\{q^{AI}, q^{BJ}\} = \O^{IJ} p^{AB}.
\end{equation}
The ten R charges, $R^{IJ} =R^{JI}$,  are represented by
\begin{equation}
R^{IJ} =  \e^{ab}\eta_a^I \eta_b^J = \eta_a^I \eta^{Ja}
= \eta^I \O^{JK} \frac{\pa}{\pa \eta^K}
+  \eta^J \O^{IK} \frac{\pa}{\pa \eta^K}.
\end{equation}
These charges generate $USp(4)$ and they transform the supercharges appropriately
\begin{equation}
[R^{IJ}, q^{AK}] =  \O^{IK} q^{AJ} + \O^{JK} q^{AI}.
\end{equation}

The on-shell supermultiplet consists of three kinds of particles: a helicity triplet $B^{ab} = B^{ba}$,
which is an R-symmetry singlet, a helicity doublet $\psi^a_I$, which an R-symmetry quartet, and a
helicity singlet $\phi_{IJ} = - \phi_{JI}$, $\O^{IJ} \phi_{IJ} =0$, which is an R-symmetry quintet.
These can be combined into a single R-symmetry invariant expression:
\begin{equation} \label{Bpp}
\Phi (\eta) = B^{++} + \eta^I \psi^+_I + \half\eta^I \eta^J \phi_{IJ}
+ \half(\eta \cdot\eta) B^{+-}
+ (\eta\cdot\eta) \eta^I \psi^-_I + \half(\eta\cdot\eta)^2 B^{--},
\end{equation}
where we have defined
\begin{equation}
\eta \cdot \eta = \half \O_{IJ} \eta^I \eta^J = \et^1 \eta^3 + \eta^2 \eta^4.
\end{equation}
Note that each $+$ or $-$ superscript correspond to half a unit of $H_3$, the third
component of the
little group $SU(2)$ algebra. Each term in $\Phi$, and hence $\Phi$ itself,
carries a total of one unit of $H_3$ if we assign
a half unit of $H_3$ to each factor of $\eta$. This was to be expected because $\eta^I$ was
introduced as a renaming of $\eta^I_- = \eta^{I+}$.

This description of the supermultiplet has two deficiencies: first,
it is not invariant under the little group; second, little-group multiplets
are split up among different terms in the expansion.
As noted already, both of these deficiencies can be overcome by
using a different formulation of the supermultiplet. The price to be paid will be that only an $SU(2)$
subgroup of the $USp(4)$ R-symmetry group will be manifest.

The $SU(2)$ little group is not a global symmetry of the M5 theory. Rather, it is a redundancy in
the formalism, analogous to a local symmetry,
which is not manifest in the preceding equations. It can be made manifest by Fourier
transforming half of the $\eta$ coordinates. A Fourier transform replaces a Grassmann coordinate
by a Grassmann derivative and vice versa. As before, we choose
$\O^{13} = - \O^{31} = \O^{24} = -\O^{42} =1$, while  all other components of $\O^{IJ}$ vanish.
Then we replace $\eta^3$ and $\eta^4$ by derivatives with respect to $\tilde\eta^1$
and $\tilde\eta^2$ and rename $(\eta^I, \tilde\eta^I)$
as $(\eta^I_-, \eta^I_+)$. Altogether, the four coordinates $\eta^I$ are replaced by
four coordinates $\eta_a^I$, which now transform as a doublet of the little group and as a doublet
of an $SU(2)$ subgroup of the R symmetry group. The formulas for the 16 supercharges become
\begin{equation}
q^{AI} =  \l^A_a \eta^{Ia} \quad {\rm and} \quad \tilde q^A_I
= \l_a^A \frac{\pa}{\pa \eta^I_a} \quad I=1,2.
\end{equation}
As promised, we have traded manifest $USp(4)$ R symmetry for little group $SU(2)$ symmetry.
This is also the case for the on-shell supermultiplet formula, which is a Grassmann Fourier transform
of the one in Eq.~(\ref{Bpp}). It now takes the form
\begin{equation} \label{tildephi}
\tilde\Phi (\eta) =  \phi + \eta^I_a \psi_I^{a} + \e_{IJ} \eta^I_a \eta^J_b B^{ab} +
\eta^I_a \eta^{Ja} \phi_{IJ} + (\eta^3)^I_a \tilde\psi^a_I + (\eta^4) \phi'  ,
\end{equation}
where $ (\eta^3)^I_a = \e_{JK}\eta^I_b \eta^{Jb} \eta_a^K$ and
$ (\eta^4) = \e_{IJ} \e_{KL} \eta_a^I\eta_b^J \eta^{Ka} \eta^{Lb}$.
Recall that in $\Phi (\eta)$ the index $I$ takes four values, whereas in $\tilde\Phi (\eta)$
it takes two values. (We prefer not to introduce another symbol.)
The five scalars are split $1+3+1$ and the four spinors are split $2+2$ even though they form
irreducible R-symmetry multiplets. To summarize,
the $\Phi$ representation has manifest R symmetry,
whereas the $\tilde\Phi$ representation has manifest little-group symmetry.
The latter representation will turn out to be the easier one to deal with,
and our main formulas for scattering amplitudes
will use this superfield description.

\subsection{D5 theory}

The world-volume theory of a probe D5 in a 10D
Minkowski space background has $(1,1)$ 6D supersymmetry.
On-shell superspace with $(1,1)$ 6D supersymmetry has been used for
studying 6D super Yang--Mills theory,
see e.g. \cite{Dennen:2009vk, Bern:2010qa, Brandhuber:2010mm}. This theory,
which is nonchiral, \ie parity invariant, describes a single massless
self-interacting vector supermultiplet. This supermultiplet contains a
one-form field $A_{\m}$, with a two-form field strength
$F = dA$. Such a field gives rise to four on-shell degrees of freedom.
The vector supermultiplet also contains four fermions and four scalars.
Altogether, there are eight bosonic and eight fermionic
on-shell degrees of freedom. The three multiplicities (1, 4, 4)
correspond to representations of the
$SU(2) \times SU(2) = Spin(4)$ R-symmetry group, which is an unbroken
global symmetry of the D5 theory. The representations are $(1,1)$ for the vector,
$(2,2)$ for the scalars and $(1,2) + (2,1)$ for the fermions.
This symmetry can be thought of as arising from rotations in the four spatial dimensions
that are orthogonal to a flat D5 in 10D Minkowski spacetime.

As discussed earlier, the little group in 6D is also
$SU(2) \times SU(2)$. Altogether, in terms of four $SU(2)$ factors,
with the first two referring to the little group
and the second two to the R-symmetry group,
the vector supermultiplet contains the following representations:
\begin{equation}
(2,2;1,1) + (1,1;2,2) +(2,1;1,2) + (1,2; 2,1).
\end{equation}
Note that, unlike the M5 theory, the D5 theory involves nontrivial
representations of both $SU(2)$ factors of the little group. In terms of on-shell
fields, these representations correspond to $A^{a \hat a}$, $\phi^{I \hat I}$,
$\psi^{a \hat I}$, and $\psi^{\hat a I}$, in a notation that should be
self-explanatory.

As before, we can introduce eight bosonic expressions
$\l^A_a$, where $A = 1,2,3,4$ labels a
spinor representation of the 6D Lorentz group
$Spin(5,1)$ and $a=\pm$ labels a doublet of the first $SU(2)$ factor in the
little group. In terms of these coordinates
the momentum of an on-shell massless particle can be written in the form
\begin{equation}
p^{AB} = \e^{ab} \l^A_a \l^B_b = \l^A_a \l^{Ba} .
\end{equation}
Three of the eight $\l$ coordinates are redundant, leaving five nontrivial
degrees of freedom, as appropriate for a massless particle in 6D.
Unlike the case of the M5 theory, this is not sufficient. The Lorentz group
has a second four-dimensional spinor representation, corresponding to the
opposite chirality, and the little group has a second
$SU(2)$ factor, both of which are utilized (on an equal footing)
in the D5 theory. Therefore, it is natural to introduce
an alternative formula for the momentum utilizing them
\begin{equation}
\hat p_{ A  B} =
\e^{\hat a \hat b} \hat\l_{A \hat a} \hat\l_{B \hat b}
= \hat\l_{A \hat a} \hat\l_B^{\hat a}.
\end{equation}
The two sets of spinor-helicity variables are orthogonal in the sense that
\bea \label{lalahat}
\l^A_a \hat\l_{A \hat b} =0.
\eea
This combination must vanish in order that $p^2 \sim  p^{AB} \hat p_{A B} =0$.

Since the momentum six-vector $p_i^{\m}$ is given by
$\half\s^\m_{AB} p_i^{AB} = \half\hat\s^{\m A B} \hat p_{i A B}$,
where $\s$ and $\hat\s$ are the appropriate
Lorentz-invariant tensors, the information encoded in $\l$ and $\hat \l$,
modulo little-group transformations, is the same. In fact, if one of them
is given, the other is determined up to a little-group transformation.
The two four-dimensional representations of the 6D
Lorentz group, labeled by the upper and lower indices $A$ and $B$, are inequivalent.
If the group were $SU(4)$ they would be complex conjugates of another,
but for Lorentzian signature the group is $Spin(5,1)$ and each of these
representations is pseudoreal. Nonetheless, for either signature it is a fact
that the Kronecker product of these two representation gives the adjoint plus a singlet.
It will be important in the analysis of the M5 theory that $\l$ determines
$\hat\l$ up to a little-group transformation and that the Lorentz invariant 
combination of $\l$ and $\hat \l$ in Eq.~(\ref{lalahat}) vanishes. 

In the notation introduced above the 16 supercharges are given by $q^{AI}$ and
$\hat q^{\hat I}_A$. Then the $(1,1)$ supersymmetry algebra is
\begin{equation}
\{q^{AI}, q^{BJ}\} = p^{AB} \e^{IJ}, \quad
\{\hat q^{\hat I}_A, \hat q^{ \hat J}_B\} =
\hat p_{AB} \e^{\hat I \hat J},
\quad \{q^{AI}, \hat q^{\hat J}_B\} = 0.
\end{equation}
These are conveniently represented by
\begin{equation}
q^{AI} = \e^{ab} \l_a^A \eta_b^I = \l_a^A \eta^{Ia}, \quad
\hat q^{\hat I}_A = \e^{\hat a \hat b} \hat\l_{A\hat a} \hat\eta^{\hat I}_{\hat b}
= \hat\l_{A\hat a} \hat\eta^{\hat I\hat a},
\end{equation}
where the Grassmann coordinates satisfy
\begin{equation}
\{\eta^I_a, \eta^J_b\} = \e_{ab} \e^{IJ}, \quad
\{\hat\eta^{\hat I}_{\hat a}, \hat\eta^{\hat J}_{\hat b}\}
= \e_{\hat a \hat b} \e^{\hat I \hat J},
\quad \{\eta^I_a, \hat\eta^{\hat J}_{\hat b}\} = 0.
\end{equation}

Now, there are again two alternative representations of the on-shell
superfield distinguished by whether the R symmetry  or the little-group
symmetry is manifest. The formula with manifest R
symmetry utilizes the four anticommuting Grassmann coordinates $\eta_-^I$ and
$\hat\eta_{\hat-}^{\hat I}$,
which we simplify to $\eta^I$ and $\hat\eta^{\hat I}$. In terms of these, the expansion is
\begin{equation} \label{App}
\Phi (\eta) = A^{+\hat+} + \eta_I \psi^{\hat+ I} + \hat\eta_{\hat I} \psi^{+ \hat I}
+ \eta_I \hat\eta_{\hat I} \phi^{I\hat I} + \eta^2 A^{-\hat+} + \hat\eta^2 A^{+\hat-} +
\cdots + \eta^2 \hat\eta^2 A^{-\hat-} ,
\end{equation}
where $\eta^2 = \half \e_{IJ} \eta^I \eta^J$ and similarly for $\hat\eta^2$

The alternative representation with manifest little-group symmetry
utilizes the $I=1$ components of $\eta_a^I$, now denoted $\eta_a$,
and the $\hat I =1$ components of $\hat\eta_{\hat a}^{\hat I}$,
now denoted $\hat\eta_{\hat a}$. The on-shell superfield in this representation is
\begin{equation}
\tilde\Phi (\eta) = \phi^{1 \hat1} + \eta_a \psi^{a \hat 1}
+ \hat\eta_{\hat a} \psi^{\hat a 1}
+ \eta_a \hat\eta_{\hat a} A^{a \hat a}
+ \eta^2 \phi^{2 \hat1} + \hat\eta^2 \phi^{1 \hat2} +
\cdots + \eta^2 \hat\eta^2 \phi^{2 \hat2} .
\end{equation}
As before, the two representations are related by a Grassmann Fourier transform. Since the
little group and the R symmetry are both $SU(2) \times SU(2)$ for the D5 theory the
two superfield formulas have the same structure with the role of the R-symmetry and
little-group symmetry interchanged.

\subsection{D3 theory}

Since the D3 theory can be obtained by dimensional reduction of the M5
theory or the D5 theory, let us consider what happens when all of the momenta
are restricted to a 4D Minkowski subspace. The Lorentz group then becomes
$SL(2,\IC)$ and the $\mathbf{4}$ of $Spin(5,1)$ decomposes as $\mathbf{2} + \mathbf{\bar 2}$.
In fact, this is correct for both of the four-dimensional spinor representations
of the 6D Lorentz group, and it is appropriate and
consistent to require that $\l^A_a$ and $\hat\l_{A\hat a}$
become identical when restricted to 4D. In standard notation,
the spinor index $A \to (\a ,\dot\a)$. In terms of $\l_a^A$ the restriction
to 4D is achieved by setting $\l_-^\a=0$ and $\l_+^{\dot\a} =0$. This then gives
$p^{\a\b} = p^{\dot\a \dot\b} =0$ leaving the familiar 4D formula for
an on-shell massless particle in helicity variables:
\begin{equation}
p^{\a \dot\a} = \l_+^\a \l_-^{\dot\a}.
\end{equation}
Now $p^2$ is proportional to the determinant of $p^{\a \dot\a}$,
which vanishes because this matrix has rank one.

Let us now focus on reduction of the M5 theory. The case of the D5 theory is very similar.
The restrictions on the momenta imply that the supercharges in Eq.~(\ref{qAI}) reduce to
\begin{equation} \label{4dsupercharges}
q^{\a I} = \l^\a \eta^I \quad {\rm and} \quad q^{\dot\a}_I
= \tilde\l^{\dot\a}\frac{\pa}{\pa \eta^I} \quad I=1,2,3,4,
\end{equation}
where we have set $\l^\a_+ = \l^\a$ and $\l^{\dot\a}_- = \tilde\l^{\dot\a}$, which is the
standard notation. Also,
an unnecessary constant factor has been removed in the formula for $q^{\dot\a}_I$.
Then $q^{\a I}$ and $q_{\dot\a I}$ form complex-conjugate representations.

The R symmetry
can now be extended to $SU(4)$, with generators given by the traceless expression
\begin{equation}
R^I{}_J =   \eta^I  \frac{\pa}{\pa \eta^J}
-\frac{1}{4} \d^I_J \eta^K  \frac{\pa}{\pa \eta^K}.
\end{equation}
The $SU(4)$ symmetry is manifest in the on-shell supermultiplet expression derived from Eq.~(\ref{Bpp})
\begin{equation}
\Phi (\eta) = A^{--} + \eta^I \psi^-_I + \eta^I \eta^J \phi_{IJ}
+ \frac{1}{6}\e_{IJKL}\eta^I \eta^J \eta^K \psi^{L+}
+ \eta^1\eta^2\eta^3\eta^4 A^{++}.
\end{equation}

The middle term now describes six scalars, one of which descends from $B^{+-}$.
The amplitudes of the D3 theory have an additional $U(1)$
symmetry that can be interpreted as conservation of helicity. Its generator is
\begin{equation}
H =  \frac{1}{4} [ \eta^I , \frac{\pa}{\pa \eta^I} ]
= \half\eta^I  \frac{\pa}{\pa \eta^I} - 1 .
\end{equation}
This is the operator that reads off the helicity of a particle,
and therefore its conservation, $H A_n = (\sum_i H_i) A_n =0$,
implies that the total helicity of the particles participating
in a nonvanishing $n$-particle scattering amplitude must be zero.
Conservation of this charge implies that the amplitude is homogeneous
of degree $2n$ in these $\eta$ coordinates. Moreover, $SU(4)$
R symmetry requires that the total number
of $\eta$'s must be a multiple of four. Together these statements
imply that $n$ must be even for the
D3 theory. In fact, we claim that $n$ must also be even for the M5 and D5 theories
even though this reasoning is not applicable in those cases.

The $U(1)$ symmetry generated by $H$ does not commute with the supercharges.
Therefore, by definition,
it is an additional R symmetry, extending the R-symmetry group to
$SU(4)\times U(1)$. Let us now explain how
the appearance of this symmetry could have been anticipated.
Since the D3 theory can be obtained
by dimensional reduction of the D9-brane theory, the $SU(4)$ subgroup can be
regarded as arising from rotations in the six dimensions transverse to the D3. So
where does the additional $U(1)$ R symmetry come from?
Having posed the question, the answer becomes
clear. The D3 theory can also be obtained by dimensional reduction of the M5 theory,
so the $U(1)$ can be interpreted as rotations in the two extra dimensions of this
construction.

In the case of ${\cal N} =4$ super Yang--Mills (SYM) theory the
$SU(4)$ R symmetry can also be understood by dimensional reduction starting
from SYM in ten dimensions. In fact, like the D3 theory,
that is how this theory was originally obtained. However, ${\cal N} =4$ SYM
cannot be obtained by dimensional reduction of a {\em perturbative} theory in
6D with $(2,0)$ supersymmetry. There are nonperturbative $(2,0)$ theories in 6D
that reduce to ${\cal N} =4$ SYM when placed on a torus. In such a reduction, the 4D
coupling constant is determined by the ratio of the radii of two cycles of the torus,
and different choices are related by dualities. This is not the kind of
dimensional reduction that would give rise to an extra $U(1)$ symmetry.
Even when Kaluza--Klein excitations are omitted, such a reduction does not
retain the transverse rotational symmetry that is
needed to give an additional $U(1)$ R symmetry.
Therefore, in the case of ${\cal N} =4$ SYM, helicity is not conserved and
$n$ does not need to be even.
%It is still true, though, $SU(4)$ R symmetry requires that the number of Grassmann coordinates
%must be a multiple of four.

As in the previous examples, there is an alternative form of the supercharges
and the on-shell superfield that exhibits manifest little-group symmetry. As a
consequence only an $SU(2) \times SU(2)$ subgroup of the $SU(4)$ R symmetry
remains manifest. As before, this representation is
related to the previous one by Fourier transforming two of the four Grassmann
coordinates. In this new basis the 16 supercharges take the form
\begin{equation} \label{l4dsuperchargesa}
q^{\a I} = \l^\a \eta_-^I \quad {\rm and} \quad
q^{\dot\a}_I  = \tilde\l^{\dot\a}\frac{\pa}{\pa \eta_-^I}
\quad I=1,2,
\end{equation}
\begin{equation} \label{l4dsuperchargesb}
\hat q^{\a}_{\hat I} = \l^\a \frac{\pa}{\pa \eta_+^{\hat I}} \quad {\rm and} \quad
\hat q^{\dot\a {\hat I}}  = \tilde\l^{\dot\a}\eta_+^{\hat I}
\quad {\hat I}=1,2.
\end{equation}
The indices $I$ and ${\hat I}$  label doublets of distinct $SU(2)$ subgroups
of the R symmetry group. The indices $\pm$ keep track of $U(1)$ little-group representations,
which corresponds to helicity. In this formulation the on-shell superfield becomes
\[
\tilde\Phi(\eta) =  \phi \, + \, \eta^I_-  \psi^-_I \,
+ \,  \eta^{\hat I}_+  \psi^+_{\hat I}  \,
+ \, \eta^{\hat I}_+ \eta^J_- \phi_{{\hat I}J}\,
+ \, (\eta_+)^2  A^{+} +  (\eta_-)^2 A^{-} \]
\begin{equation} \label{TPhiD3}
+ \, (\eta_+)^2 \eta_-^I \psi_I^{+}\,
+ \, (\eta_-)^2  \eta_+^{\hat I}\psi_{\hat I}^{-} \,
+ \, (\eta_+)^2  (\eta_-)^2  \bar{\phi} \, ,
\end{equation}
where $(\eta_+)^2 = \half \e_{\hat I \hat J} \eta_+^{\hat I} \eta_+^{\hat J}$ and
$(\eta_-)^2 = \half \e_{IJ} \eta_-^I \eta_-^J$.

\section{Four-particle amplitudes} \label{sec:4pts}

\subsection{M5 theory} \label{sec:4ptsM5}

Before discussing the general case, let us consider four-particle amplitudes, starting
with the M5 theory. The plan is to first propose a formula for the result that
corresponds to supermultiplets written in the form given in Eq.~(\ref{tildephi}).
This representation has a manifest
$SU(2)$ little-group symmetry for each external particle. Up to
normalization, the four-particle amplitude with four derivatives
for an abelian tensor supermultiplet with 6D $(2,0)$ supersymmetry is uniquely given by
\begin{equation} \label{A4M5}
A_4 = \d^6 \left(\sum_{i=1}^4 p_i^{AB}\right)\,
\d^8 \left(\sum_{i=1}^4 q_i^{AI} \right).
\end{equation}
As discussed in Sect.~\ref{section:introM5theory}, $p_i^{AB} = \l_{i+}^A \l_{i-}^B - \l_{i-}^A \l_{i+}^B$ and
$q_i^{AI} = \l_{i+}^A \eta_{i-}^I - \l_{i-}^A \eta_{i+}^I$, where $A,B=1,2,3,4$ and $I=1,2$.
The fermionic delta functions are defined for instance in \cite{Dennen:2009vk}.
It will be useful later to write the momentum-conservation condition in matrix notation as
\begin{equation}
\l_+ \l_-^T = \l_- \l_+^T .
\end{equation}
In other words, the matrix $(\l_+ \l_-^T)^{AB} $ is symmetric.
This is valid for any number of particles $n$.
In the special case of $n=4$, $\l_+$  and $\l_-$ are square matrices.
If $n=4$ and $\l_-$ is invertible,
which is generically the case, this implies that $(\l_-^{-1} \l_+)_{ij}$
is symmetric. This fact will be useful later.

The formula for $A_4$ manifestly satisfies several requirements:
total symmetry in the four particles, Lorentz
invariance, conservation of momentum and half of the supercharges,
and little-group symmetry.
Also, the second factor scales as $\l^8$ or $p^4$, as expected.
Conservation of the other half of the supersymmetries is easy to verify.
What one needs to show is that
\begin{equation}
\left(\sum_{i=1}^4 \tilde q_{iI}^A \right)A_4 =0.
\end{equation}
This fact is an immediate consequence of
$\{ \tilde q_{iI}^A , q_j^{BJ}\} = p_i^{AB} \d_{ij} \d^I_J$
and conservation of momentum.

In order to appreciate Eq.~(\ref{A4M5}), let us examine what it implies
for the scattering of four $B$ particles.
They are R-symmetry singlets whose on-shell degrees of freedom are
described by a symmetric tensor $B_{ab} = B_{ba}$
of the $SU(2)$ little group. Eq.~(\ref{A4M5}) implies that their
four-particle amplitudes are given by
\begin{eqnarray} \label{eq:4B}
\langle B_{a_1 a_2} B_{b_1 b_2} B_{c_1 c_2} B_{d_1 d_2} \rangle
=
\langle 1_{a_1} 2_{b_1} 3_{c_1} 4_{d_1}\rangle
\langle 1_{a_2} 2_{b_2} 3_{c_2} 4_{d_2}\rangle  + \mathcal{P}_4 \, ,
\end{eqnarray}
where $\mathcal{P}_4$ denotes the symmetrization over little group indices. Here and throughout, we make use of a Lorentz-invariant bracket:
\begin{eqnarray} \label{4bracket}
\langle 1_{a} 2_{b} 3_{c} 4_{d}\rangle :=
\e_{ABCD} \l^A_{1a} \l^B_{2b} \l^C_{3c} \l^D_{4d} .
\end{eqnarray}
It is interesting to note that any four $B$ particles have a nonzero amplitude.
For example,
\begin{eqnarray}
\langle B_{++} B_{++} B_{++} B_{++} \rangle
\propto
\langle  1_{+} 2_{+} 3_{+} 4_{+}\rangle^2 = (\det \l_+)^2 \, ,
\end{eqnarray}
Similarly, the amplitude for four $B_{--}$ particles is given by $(\det \l_-)^2$.
On reduction to the D3 theory, $B_{++}$ becomes a positive-helicity photon,
and this amplitude vanishes. Indeed, Eq.~(\ref{eq:4B})
gives all the four-photon amplitudes correctly,
with the only nonzero ones involving two positive-helicity and
two negative-helicity photons. It also describes amplitudes involving
additional scalars that arises from reduction of $B_{+-} = B_{-+}$.

Let us now turn to the more difficult issue: verifying $USp(4)$ R symmetry
of an arbitrary four-particle amplitude. We have learned earlier that
this symmetry should be manifest in the representation of the supermultiplet given in Eq.~(\ref{Bpp}). To get to this representation, we rename
$\eta_{i-}^I$ as $\eta_{i}^I$ and $\eta_{i+}^I$ as
$\tilde\eta_{i}^I$. Then we Fourier transform the latter coordinates
to conjugate Grassmann coordinates denoted $\z_{iI}$. Thus, we consider
\begin{equation}
A_4 = \int d^8\tilde\eta^I_i e^{\sum_{iI}\tilde\eta^I_i \z_{iI}}
\d^8 \left(\sum_{i=1}^4 q_i^{AI} \right) .
\end{equation}
Substituting an integral representation of the delta functions and carrying
out the $\tilde\eta$ integrations gives
\begin{equation}
A_4 = \int d^8 \th_{AI} \d^8(\z_{iI}
+ \sum_A \th_{AI} \l^A_{i-} ) e^{\sum_{AIi} \th_{AI} \l^A_{i+} \eta^I_i}.
\end{equation}
If we now assume that the $4\times 4$ matrix $\l^A_{i-}$ is nonsingular, which is
generically the case, then
\begin{equation}
\d^8(\z_{iI} + \sum_A \th_{AI} \l^A_{i-} )
 = (\det \l_-)^2 \d^8((\z \l_-^{-1})_{IA} +  \th_{AI} )
\end{equation}
and thus
\begin{equation}
A_4 = (\det \l_-)^2  \exp(-\tr (\z \l_-^{-1}\l_+\eta))
\end{equation}
More explicitly, the exponent is
\begin{equation}
-\tr (\z \l_-^{-1}\l_+\eta) = \tr(\l_-^{-1}\l_+\eta \z))
= \sum_{ij} (\l_-^{-1}\l_+ )_{ij} (\eta\z)_{ji}
\end{equation}
As was explained earlier, momentum conservation implies that
$(\l_-^{-1}\l_+ )_{ij}$ is a symmetric matrix. Therefore
only the symmetric part of $(\eta\z)_{ji}$ contributes,
which can therefore be replaced by half of
\begin{equation}
E_{ij} = \sum_{I=1}^2 \left(\eta^I_i \z_{Ij} +  \eta^I_j \z_{Ii} \right ).
\end{equation}
We now claim that $E$ (and hence $A_4$) can be rewritten in a form
that has manifest $USp(4)$ symmetry
\begin{equation}
E_{ij} = \sum_{I,J=1}^4 \O_{IJ}\eta^I_i \eta^J_j ,
\end{equation}
where the only nonzero elements above the diagonal of the symplectic
metric are $\O_{13} = \O_{24} =1$.
Note that we have renamed $\z_{Ii} = \eta^{I+2}_i $. Then $\eta^I_i$
belongs to the fundamental representation of the
$USp(4)$ R-symmetry group. The same idea discussed here applies to
more general $n$-particle amplitudes, as we show in Appendix~\ref{appendix:Rsymmetry}.

Note that the amplitude for four $B_{--}$ particles is given by the first
term in the expansion of the exponential, whereas
the amplitude for four $B_{++}$
particles is given by the last (eighth) term in the series expansion of
the exponential. All other four-particle
amplitudes are contained in the intermediate powers. Clearly, this
representation (with manifest R symmetry)
is more complicated than the previous one with manifest little group
$SU(2)$ symmetries for each of the
scattered particles, that amplitudes are no longer homogenous polynomials 
in terms of fermionic variables $\eta$'s.

\subsection{D5 theory} \label{section:introD5thoery}

The four-particle amplitude for this theory is quite similar to the one for the M5 theory.
In the representation with manifest little-group symmetry the four Grassmann coordinates that are
used in the superfield $\tilde\Phi(\eta)$ are $\eta_a$ and $\hat\eta_{\hat a}$. They transform
as $(2,1)$ and $(1,2)$ with respect to the $SU(2) \times SU(2)$ little group. In terms of these
we can define eight anticommuting supercharges
\begin{equation}
q^A =\e^{ab} \l^A_a \eta_b = \l^A_a \eta^a \quad {\rm and} \quad \hat q_A
=\e^{\hat a \hat b} \hat\l_{A\hat a} \eta_{\hat b}
= \hat\l_{A\hat a} \eta^{\hat a}.
\end{equation}
Then the desired amplitude is
\begin{equation} \label{A4D5}
A_4 = \d^6 \left(\sum_{i=1}^4 p_i^{\m} \right)\,
\d^4 \left(\sum_{i=1}^4 q_i^{A} \right)
\d^4 \left(\sum_{i=1}^4 \hat q_{iA} \right).
\end{equation}

In particular, we can read off the amplitude for scattering four vector particles
\begin{eqnarray}
\langle A_{a \hat a} A_{b \hat b} A_{c \hat c} A_{d \hat d} \rangle
= \langle  1_{a} 2_{b} 3_{c} 4_{d}\rangle
\langle  1_{\hat a} 2_{\hat b} 3_{\hat c} 4_{\hat d}\rangle
\end{eqnarray}
where
\begin{eqnarray}
 \langle  1_{a} 2_{b} 3_{c} 4_{d}\rangle :=  \e_{ABCD} \l^A_{1a} \l^B_{2b}
 \l^C_{3c} \l^D_{4d}, \quad
 \langle  1_{\hat a} 2_{\hat b} 3_{\hat c} 4_{\hat d}\rangle :=  \e^{ABCD}
 \hat\l_{1A\hat a} \hat\l_{2B\hat b}
 \hat\l_{3C\hat c} \hat\l_{4D\hat d} .
\end{eqnarray}
For example,
\begin{equation}
\langle A_{+ \hat +} A_{+ \hat +} A_{+ \hat +} A_{+ \hat +} \rangle
\propto \det\l_+ \det \hat\l_{\hat +}.
\end{equation}

As in the case of the M5 theory, the R symmetry of the D5 amplitudes can be verified
by carrying out a Grassmann Fourier transform to the representation in which
that symmetry becomes manifest.

\subsection{D3 theory}

Since the D3 theory can be obtained by dimensional reduction
of the M5 theory, its four-particle amplitude can be deduced
from the preceding results. Specifically, Eq.~(\ref{A4M5}) reduces to
\begin{equation} \label{A4D3}
A_4 = \d^4 \left(\sum_{i=1}^4 p_i^{\a\dot\a} \right)\,
\d^4 \left(\sum_{i=1}^4 q_i^{\a I} \right) \d^4 \left(\sum_{i=1}^4
\hat q_i^{\dot\a \hat I} \right),
\end{equation}
where $p_i^{\a\dot\a} = \l_i^\a \tilde\l_i^{\dot\a}$,
$q_i^{\a I} = \l_i^\a \eta_{i-}^I $, and
${\hat q}_i^{\dot\a \hat I} = \tilde\l_i^{\dot\a} \eta_{i+}^{\hat I}$.
As before, this is easily seen to have all of the
required properties aside from R symmetry. Alternatively, the same result
can be obtained by dimensional reduction of the D5 theory,
whose four-particle amplitude is given in Eq.~(\ref{A4D5}). In this case,
dimensional reduction of $q^A$ gives $q^{\a 1}$ and $q^{\dot\a 1}$,
while dimensional reduction of
$\hat q_A$ gives $q^{\a 2}$ and $q^{\dot\a 2}$.

R symmetry can be investigated, as before, by Fourier transforming the
$\eta_{i+}^{\hat I}$ coordinates. (Recall that $I=1,2$ and $\hat I = 1,2$
label doublets of the two $SU(2)$ factors of an $SU(2) \times SU(2)$ subgroup
of the $SU(4)$ R symmetry group.) However,
the analysis requires some modification, since the matrix $\l_-$,
which was previously assumed to be nonsingular,
is now singular. In fact, two of its four columns are identically zero.

Since $\eta_+$ only occurs in the last delta-function factor,
let us consider its Fourier transform
\begin{equation}
I_4 = \int d^8 \eta^{\hat I}_{i+} e^{\sum_{iI} \eta^{\hat I}_{i+} \z_{i\hat I}}
\d^4 \left(\sum_{i=1}^4 \hat q_i^{\dot\a \hat I} \right)
= \int d^4 \th_{\dot\a \hat I} \d^8(\z_{i \hat I}
+ \sum_{\dot\a} \tilde\l_i^{\dot\a}\th_{\dot\a \hat I} ).
\end{equation}
Momentum conservation can be written as the matrix equation $(\l^T \tilde\l)^{\a\dot\a} =0$.
Therefore the eight delta functions imply the four relations
$\sum_i  \l_i^\a \z_{i \hat I} = 0$. From this it follows that
\begin{equation}
I_4 = J\, \d^4 \left(\sum_{i=1}^4\l_i^\a  \z_{i \hat I} \right),
\end{equation}
where $J$ is a Jacobian factor.It is
straightforward to see that the Jacobian is
\begin{equation} \label{Jform}
J = \left( \frac{[12]}{\langle 34 \rangle }\right)^2.
\end{equation}
Here we are using the standard notation of 4D spinor helicity formalism, $\langle ij\rangle = \e_{\a\b} \l_i^\a \l_j^\b$ and
$[ij] = \e_{\dot\a\dot\b} \tilde\l_i^{\dot\a} \tilde\l_j^{\dot\b}$.  It is important
that $J$ should have total symmetry in the four particles. The proof that
$[12]/ \langle 34 \rangle$ has total antisymmetry, and hence that $J$ has total symmetry,
is straightforward using momentum conservation.

To complete the analysis, we define $\z_{\hat 1} = \eta^3$ and
$\z_{\hat 2} = \eta^4$, as before. Then, assembling the results above,
the Fourier-transformed scattering amplitude becomes
\begin{equation} \label{A4D3b}
A_4 = \left( \frac{[12]}{\langle 34 \rangle }\right)^2  \,
\d^4 \left(\sum_{i=1}^4 p_i^{\a\dot\a} \right)\,
\d^8 \left(\sum_{i=1}^4 q_i^{\a I} \right),
\end{equation}
where the index $I$ on $q_i^{\a I} = \l_i^\a \eta_i^I$ now takes four values.
This version of four-particle superamplitude has appeared before, for instance in \cite{Chen:2015hpa}.
It now has manifest $SU(4)$ R symmetry, because
the Grassmann delta functions contain two factors of
$\e_{IJKL} \eta_i^I \eta_j^J \eta_k^K \eta_l^L$,
which is $SU(4)$ invariant. The amplitude has an additional $U(1)$ R symmetry,
because it contains $2n=8$ factors of $\eta$, as explained earlier.

\section{$\mathbf{n}$-particle amplitudes of the D3 theory} \label{sec:nptD3}

This section briefly reviews the $n$-particle amplitudes for the tree-level
S-matrix of the D3 theory. A nice formula with manifest $SU(4)$ R symmetry
appeared recently in \cite{He:2016vfi}\cite{Cachazo:2016njl}. However, for the purpose of
generalizing to the M5 theory, it is more convenient to break the $SU(4)$ R-symmetry
$SU(4) \rightarrow SU(2)_L \times SU(2)_R$ and make the little-group symmetry manifest.
%The $U(1)$ R symmetry, which has not been discussed previously, will be considered later.
A formula of the required type has appeared previously
for 4D $\mathcal{N}=4$ SYM and $\mathcal{N}=8$ supergravity \cite{Cachazo:2013iaa}.
It contains complex coordinates $\s_i$ (on the Riemann sphere) associated to
the $n$ particles. The formula is required to be invariant under simultaneous
$SL(2, \IC)$ transformations of these coordinates. This implies that only
$n-3$ of them are integrated, while the other three can be set to arbitrarily
chosen distinct values.

The on-shell $n$-particle amplitude formula takes the form
\bea \label{nptamps}
\mathcal{A}_n= \int \frac{d^n \s \, d \mathcal{M} }{ {\rm Vol} (G) }
\D_B(p,\r) \D_F(q,\r,\chi) \, {\cal I} \, ,
\eea
where $\D_B$ is a product of bosonic delta functions
\bea
\D_B(p,\r) = \prod^n_{i=1} \d^4 \left(p^{\a \dot\a }_i
-  { \r^{\a } (\s_i) \tilde{\r}^{ \dot\a} (\s_i) \over P_i(\s) } \right) \, ,
\eea
and $\D_F$ is a product of fermionic (or Grassmann) delta functions
\bea
\D_F(q,\r,\chi) = \prod^n_{i=1} \d^{4}\left(q^{\a I}_i
-   { \r^{ \a } (\s_i) \chi_-^{I} (\s_i)  \over P_i(\s) }  \right)
\d^{4}\left( \hat{q}^{\dot\a \hat I}_i
-   { \tilde{\r}^{ \dot{\a} } (\s_i) \chi_+^{\hat I} (\s_i)  \over P_i(\s) } \right) \, .
\eea
Here $\r^{\a }(\s)$ and $\chi^{\hat I}_+ (\s)$ are degree-$d$ polynomials, while
$\tilde{\r}^{ \dot\a} (\s)$ and $\chi^{I}_- (\s)$ are degree-$\tilde{d}$
polynomials, with
\bea
d+\tilde{d} = n-2.
\eea
Thus, $\r^{ \a } (\s)$ (bosonic) and $\chi^{I}_- (\s)$ (fermionic) take the form
\bea
\r^{ \a } (\s)   = \sum^{d}_{m=0} \r^{ \a }_m \s^m \, ,
\quad
\chi^{I}_- (\s)   = \sum^{\tilde {d} }_{m=0} \chi^{I}_{m,-} \s^m \, ,
\eea
and
\bea
\tilde{\r}^{ \dot{\a} } (\s) = \sum^{\tilde d}_{m=0} \tilde\r^{ \dot\a }_m \s^m \, ,
\quad
\chi_{+}^{\hat I}(\s) = \sum^{ d}_{m=0} \chi^{\hat I}_{m,+} \s^m \, .
\eea
Also,
\bea
P_i(  \s ) = \prod_{j  \neq i} \s_{ji} \quad i,j = 1,2,\ldots,n \, ,
\eea
where $\s_{ij} = \s_i - \s_j$.
Note that $P_i(\s)$ depends on all $n$ of the $\s$ coordinates, but $\s_i$ has a
distinguished role. The integral is taken over the space of punctures and polynomials,
the measure for which contains the following $2n$ bosonic and $2n$ fermionic integrations:
\bea \label{calM}
d \mathcal{M} = \prod^{d}_{m=0} \prod^{ \tilde{d} }_{\tilde{m} =0} d^2 \r^{\a}_m d^2
\tilde{\r}^{\dot{{\a}} }_{\tilde{m}} d^2 \chi^{I}_{\tilde{m},-}
d^2 {\chi}^{\hat I}_{m,+}  \, .
\eea

The integral has a gauge redundancy from the modular and little-group symmetries,
so we must divide by the volume of
\bea
G = SL(2,\IC)\times GL(1, \IC),
\eea
where the modular group $SL(2,\IC)$ acts on the $\s_i$'s and $GL(1, \IC)$,
the complexified little group, acts on the $\r$'s and $\tilde{\r}$'s.

Eq.~(\ref{nptamps}) describes maximally supersymmetric theories with the on-shell states
organized according to Eq.~(\ref{TPhiD3}). It gives the usual scattering amplitude
supplemented by additional delta functions, namely
\bea \label{calAn}
\mathcal{A}_n = \left(\prod^n_{i=1} \d(p^2_i ) \d^{2}(  \langle  \l_i \, q_{i}^I  \rangle )
\d^{2}( [ \tilde{\l}_i \, \hat{q}_{i}^{\hat I}]  )\right) \, A_n \, ,
\eea
where $A_n$ is the usual scattering amplitude including the four momentum-conservation
delta functions and eight supercharge-conservation delta functions. (When the momentum-conservation delta function is also omitted, the amplitude is
denoted $T_n$). The bracket notation is the same as described following Eq.~(\ref{Jform}).
The extra delta functions in Eq.~(\ref{calAn}) impose the conditions that allow us to
introduce the usual on-shell relations of the schematic form $p = \l \tilde\l$
and $q = \l \eta$. So, in practice, to extract the scattering amplitudes $A_n$
from Eq.~(\ref{nptamps}), one should use these relations and remove the extra delta functions.
\hyperref[appendix:Jacobi]{Appendix~A.1} contains the proof that the $4n$ bosonic delta functions $\D_B$ account
for the $n$ mass-shell conditions, four momentum conservation equations, and the
$n-3$ scattering equations. These are precisely the $2n+1$ delta functions that survive
after carrying out the $(2n-1)$-dimensional $\r$ integration.

The choice of the factor ${\cal I}$ in the integrand depends on the theory.
For example, the color-ordered ${\cal N}=4$ SYM amplitudes,
discussed in \cite{Cachazo:2013iaa},
are given by the Parke--Taylor-like factor
\bea \label{eq:4DsYM}
{\cal I}^{\rm YM} = \frac{1}{\s_{12}\s_{23} \cdots \s_{n-1 \, n}\s_{n1}} \, .
\eea
In the case of YM and SYM theories in 4D, the solutions of the scattering
equations can be separated into $n-3$ sectors characterized by the
total helicity (or ``helicity violation") of the
$n$ particles participating in the reaction. The sectors, labeled by $d = 1, 2, \ldots, n-3$,
have $\tilde d - d = n - 2(d+1)$ units of helicity violation. In particular, the $d=1$ sector,
which has $n-4$ units of helicity violation,
is usually referred to as having ``maximal helicity violation" (MHV).
If $n$ is even, the sector with $d = \tilde d = \frac{n}{2} -1$  is helicity conserving.
As was first conjectured in \cite{Spradlin:2009qr} and later proven
in \cite{Cachazo:2013iaa}, the number of solutions of the
scattering equations that contribute to the $(d, \tilde d)$ sector, denoted $N_{d,\tilde d}$,
is given by an Eulerian number. These numbers satisfy $N_{d,\tilde d} = N_{\tilde d, d}$
and $N_{d,1} =1$. They are fully determined by these relations and the recursion relation \cite{Cachazo:2013iaa}
\bea
N_{d,\tilde d} = \tilde d N_{d -1,\tilde d} + d N_{d,\tilde d -1}.
\eea
Furthermore,
\bea
\sum_{d=1}^{n-3} N_{d,\tilde d} = (n-3)!
\eea
which accounts for all the solutions of the scattering equations.

Due to the recent progress in understanding CHY representations of scattering
amplitudes \cite{Cachazo:2013hca, Cachazo:2013iea, Cachazo:2014xea},
it is known that one can pass from
YM theories to DBI theories by simply replacing ${\cal I}^{\rm YM}$ by
\bea
{\cal I}^{\rm DBI} = {\rm det}' S_n \, ,
\eea
where $S_n$ is an $n\times n$ anti-symmetric matrix with
\bea
(S_n)_{ij}= \frac{s_{ij}}{ \s_{ij}}, \quad i,j = 1,2,\ldots,n,
\eea
where
$s_{ij} = (p_i +p_j)^2 = 2 p_i \cdot p_j$ are the familiar Mandelstam invariants. Also,
\bea
{\rm Pf}' S_n = \frac{(-1)^{i+j} }{ \s_{ij}} {\rm Pf} S^{i,j}_{i,j} \, , \quad\quad
{\rm det}' S_n = ({\rm Pf}' S_n)^2 \, .
\eea
Here $S^{i,j}_{i,j}$ means that the $i$-th and $j$-th rows and columns of the matrix
$S_n$ are removed before computing the Pfaffian or determinant. This is
required because $S_n$ has rank $n-2$ if $n$ is even.
Then ${\rm det}' S_n$ is independent of the choice of $i$ and $j$ and transforms with weight
two under $SL(2, \IC)$ transformations of the $\s$ coordinates.
If $n$ is odd, there is no satisfactory nonzero definition. Therefore all nonzero amplitudes of
all DBI-like theories must have $n$ even. This includes all three brane
theories (D3, D5, M5) that are the main emphasis of this paper.

However, if one examines the actions in the literature for these theories, it is only obvious
that $n$ must be even for the bosonic truncation, in each case, but it is not at all obvious
when fermions are involved. These actions, which were derived using
various string theory considerations, contain vertices involving an odd number of bosons
when fermions are also present. Since we claim that on-shell amplitudes with an
odd number of bosons always vanish, it must be possible to eliminate
all terms in the action that have an odd number of boson fields by field redefinitions.
At the leading nontrivial order, the analysis in Sect.~2.1 of \cite{Bergshoeff:2013pia}
implies that this is the case for the D3 theory. Otherwise,
this issue does not seem to have been explored.

In the case of the D3 theory, the extra $U(1)$ R symmetry, discussed earlier,
implies that only the helicity-conserving sector, with
\bea \label{eq:middlesector}
d = \tilde d  = \frac{n}{2} -1,
\eea
is nonvanishing. The number of solutions of the scattering equations that contribute
to this sector is $N_{1,1} = 1$ for $n=4$, $N_{2,2} = 4$ for $n=6$, $N_{3,3} = 66$
for $n=8$, $N_{4,4} = 2416$ for $n=10$ and so forth. These numbers are a significant
fraction of $(n-3)!$.

As indicated in Eq.~(\ref{nptamps}), one should mod out
the volume of $G=SL(2,\IC)\times GL(1, \IC)$, where
$SL(2,\IC)$ acts on the $\s_i$'s and $GL(1, \IC)$ acts on the
$\r$'s and $\chi$'s.
In practice, we may fix any three $\s_i$'s (for instance $\s_1, \s_2, \s_3$)
and one $\r$ (for instance $\r^{1}_0$) to arbitrary values,
with the compensating Jacobian
\bea
J_{SL(2,\IC)\times GL(1, \IC) }
=\r^{1}_0 \, (\s_1 - \s_2 ) (\s_2 - \s_3 ) (\s_3 - \s_1 ) \,.
\eea
We note that the integral formula is not a ``true integral", in the sense that
the number of bosonic delta-functions is equal to the number of
integration variables (after taking account of the $G$ symmetry). This is not a surprise,
of course, since we know that tree amplitudes are entirely algebraic.

As mentioned earlier, the counting of bosonic delta functions is as follows:
the $4n$ bosonic delta-functions in $\D_B$ give
rise to $n$ delta functions for mass-shell conditions in the
coefficient of $A_n$ in Eq.~(\ref{calAn}) and four more for
momentum conservation, $\d^4(\sum_1^n p_i^\m)$, which are included in $A_n$.
The remaining $3n-4$ delta functions determine the $(n-3)$ $\s$'s and $(2n-1)$ $\r$'s
that survive after modding out by the volume of $G$. The Jacobian
that arises from these evaluations is computed explicitly in the Appendix \ref{appendix:Jacobi}.
Also, there are $8n$ fermionic
delta functions in $\D_F$ and $2n$ fermionic integrations in $d {\cal M}$, leaving
an expression of order $6n$ in fermionic coordinates. $4n$ of these appear in the
coefficient of $A_n$ in Eq.~(\ref{calAn}). Therefore the remaining $2n$ $\eta$'s must be
in $A_n$. In fact, half of them are $\eta_+$'s and half are $\eta_-$'s.
This is the number that we argued earlier are required (in this representation)
by the $U(1)$ factor in the R symmetry group of this theory.

The powers of momenta that appear in ${\cal A}_n$ can also be checked. In theories of
Born--Infeld type, such as we are considering, one expects that $T_n \sim p^n$. In four
dimensions this implies that $A_n \sim p^{n-4}$ and ${\cal A}_n \sim p^{3n-4}$.
The latter, given for the D3 theory in Eq.~(\ref{nptamps}),
contains $p^n$ from the measure, $p^{-4n}$
from $\D_B$, $p^{4n}$ from $\D_F$, and $p^{2n-4}$ from $\det' S_n$. These combine to
give $p^{3n-4}$, as desired.

\hyperref[appendix:Jacobi]{Appendix A.1} describes the Jacobian factor generated by pulling out the ``wave functions"
and the momentum conservation delta function. Using these results for the Jacobian,
we have checked explicitly that Eq.~(\ref{nptamps}), with ${\cal I}= {\det}'S_n$,
reproduces the four-point amplitude of the D3 theory given in Eq.~(\ref{A4D3})
as well as the six-point super amplitudes, which may be found in \cite{Chen:2015hpa}.
The appendix also contains the proof that the amplitudes have $SU(4)$ R symmetry
(in addition to the $U(1)$ already demonstrated).

%Appendix A.1 shows that Eq.~(\ref{nptamps}), with ${\cal I}= {\det}'S_n$,
%reproduces the four-point amplitude of the D3 theory given in Eq.~(\ref{A4D3}).
%It also contains the proof that the amplitudes have $SU(4)$ R symmetry (in addition
%to the $U(1)$ already demonstrated).

\section{$\mathbf{n}$-particle amplitudes of the M5 theory} \label{sec:nptM5}

\subsection{The proposed formula}

This section generalizes the twistor-string-like formula of the D3 theory in
Eq.~(\ref{nptamps}) to the M5 theory with $(2,0)$ supersymmetry
in 6D. The $n$-particle tree-level scattering amplitude for this theory takes the form
\begin{eqnarray} \label{eq:(2,0)-old}
{\cal A}_n = \int \frac{d^n \s \, d \mathcal{M} }{ {\rm Vol} (G) }  \,
\D_B(p, \r) \, \D_F (q, \r, \chi)\,
{\rm det}^{\prime} S_n \,  U(\r, \s) ,
\end{eqnarray}
where
\bea \label{eq:DB20}
\D_B (p,\r) = \prod^n_{i=1} \d^6 \left(p^{AB}_i
-  \frac{\r^{A}_a (\s_i) \r^{Ba} (\s_i)}{ P_i(\s)} \right)
\eea
and
\bea \label{eq:DF20}
\D_F (q,\r, \chi) = \prod^n_{i=1} \d^{8}\left(q^{A I}_i
- \frac{ \r^{A}_a(\s_i) \chi^{Ia} (\s_i)} { P_i(\s) } \right).
\eea
These delta functions are the natural $(2,0)$ generalization of the
corresponding D3 formulas.
The factor $ {\rm det}^{\prime} S_n$ is unchanged from the D3 case, since it is a
sensible function of the invariants $s_{ij}$ for any space-time dimension.
A crucial requirement for the M5 theory amplitudes is that they reproduce
the D3 amplitudes under dimensional reduction. The additional factor
$U(\r,\s)$ will be determined by this requirement and 6D Lorentz
invariance later in this section.

The M5 analog of the D3 formula in Eq.~(\ref{calAn}) is
\bea
\mathcal{A}_n = \left(\prod^n_{i=1} \d(p^2_i )
\d^{4}\left(\hat\l_{iA\hat a} q_i^{AI}\right)\right) \, A_n \, .
\eea
The logic here is as follows. The bosonic delta functions
$\D_B$ imply that the $n$ particles are massless, and therefore they
allow us to introduce spinor-helicity variables $\l^A_a$ and $\hat\l_{A\hat a}$
for each of the momenta that are unique up to little-group transformations, 
as explained in Sect.~\ref{section:introD5thoery}.
%Therefore, the bosonic delta functions $\D_B$ imply that $\l^A_a \propto\r_a^A(\s_i)$.
The fermionic delta functions $\D_F$ imply that $\hat\l_{A\hat a} q^{AI}$
should vanish, which accounts
for the delta functions given above. The vanishing of $\hat\l_{A\hat a} q^{AI}$
also implies that $q^{AI}$ can be expressed as
$q^{AI} = \l^A_a \eta^{Ia}$ due to Eq.~(\ref{lalahat}). On reduction to 4D these fermionic delta functions
account for the fermionic delta functions that appear in Eq.~(\ref{calAn}).

Also by analogy with the D3 theory, $\r^{ A}_a (\s) $ and $\chi^{I}_a (\s)$
are bosonic and fermionic polynomials of degree $d$
\begin{eqnarray}
\r^{ A}_a (\s)   = \sum^{d}_{m=0} \r^{ A}_{m,a} \s^m \, ,
\quad
\chi^{I}_a (\s)   = \sum^{d}_{m=0} \chi^{I}_{m,a} \s^m
\, ,
\end{eqnarray}
and the measure $d \mathcal{M} $ for the M5 case is given by
\begin{eqnarray}
d \mathcal{M} = \prod^{d}_{m=0} d^8 \r^{A}_{m,a} \, d^{4} \chi^{I}_{m,b}    \, ,
\end{eqnarray}
where $d =\frac{n}{2} -1$, just as in the D3 theory. The symmetry that needs
to be gauge fixed is now
\bea
G= SL(2, \IC) \times SL(2, \IC).
\eea
The first $SL(2,\IC)$ factor, which concerns the usual modular symmetry
transformations of the $\s$ coordinates, removes the integration over three $\s_i$'s.
This symmetry will be verified later. The second $SL(2,\IC)$ factor,
which is the complexification of the $SU(2)$ little group of the M5 theory,
removes three $\r$ integrations.

The bosonic delta functions completely fix the integration variables,
as in the 4D case, leaving a sum over the solutions of the scattering equations.
Specifically, the $6n$ bosonic delta functions give rise to $n$ on-shell conditions
$p_i^2=0$ and 6D momentum conservation leaving $(5n-6)$ bosonic delta functions.
Since the $\s_i$'s
and $\r^{A}_{m,a}$'s are constrained by $G = SL(2, C) \times SL(2, C)$, there are
$(n-3)$ $\s_i$'s and  $(4n-3)$  $\r^{A}_{m,a}$'s to be integrated, which is the right number
to be fixed by the remaining $(5n-6)$ delta functions. The proof of momentum conservation
and the scattering equations is essentially the same as described for the D3 theory
in Appendix~\ref{appendix:D3Jacobi}.

The gauge-fixing Jacobian for the first $SL(2,\IC)$ factor is
$(\s_1 - \s_2 ) (\s_2 - \s_3 ) (\s_3 - \s_1 )$ as usual. The one for the
second $SL(2,\IC)$ factor will be discussed later. These symmetries, as well as
other properties, will be verified after we have made
a specific proposal for $U(\r,\s)$. It will be determined by considering dimensional
reduction to 4D, with the final result shown in Eq.~(\ref{eq:(2,0)}) or 
equivalently Eq.~(\ref{eq:(2,0)1}).

In contrast to the 4D case, the polynomials $\r^{ Aa } (\s)$ and
$\chi^I_a(\s)$ are required to have degree $d =\frac{n}{2} -1$ due to the
$SU(2)$ little-group
symmetry. Thus, the solutions of the scattering equations, which are
implied by $\D_B (p,\r) =0$, cannot be subdivided into sectors. There is only one sector,
which we find interestingly already contains all $(n-3)!$ solutions of the scattering
equations. (This assertion has been checked explicitly for $n=4,6, 8$.)
When reduced to 4D massless kinematics and for the D3 theory, only a subset of
the $(n-3)!$ solutions contributes, namely those $N_{d,d}$
helicity-conserving solutions.

We have checked explicitly that Eq.~(\ref{eq:(2,0)}) correctly reproduces the
amplitudes with lower multiplicities, such as the four-particle amplitude that was
discussed previously. As we discussed, to extract the amplitudes, one should take
out the ``wave functions" from $\Delta_B$ and $\Delta_F$ defined in Eq.~(\ref{eq:DB20})
and Eq.~(\ref{eq:DF20})
\bea
\mathcal{A}_4 = \left(\prod^4_{i=1} \d(p^2_i )
\d^{4}\left(\hat\l_{iA\hat a} q_i^{AI}\right)\right) \,   A_4 \, ,
\eea
and one can further extract the momentum and (half of the) supercharge conservation delta
functions, namely,
\bea
A_4 =  \d^6 \left(\sum_{i=1}^4 p_i^{AB}\right)\,
\d^8 \left(\sum_{i=1}^4 q_i^{AI} \right) \times J_{4,B} J_{4,F} \times I_4 \, .
\eea
The factors $J_{4,B}$ and $J_{4,F}$ are Jacobians, generated in this process,
which can be found in Appendix~\ref{appendix:Jacobi}.  Finally, $I_4$ is an integral
over the remaining delta functions,
\bea
I_4 &=& \int \frac{d^4 \s \, d \mathcal{M}_4 }{ {\rm Vol} (G) }
\prod^{2}_{i=1} \d^5 \left(p^{A B }_i -  { \r^{A }_a (\s_i)  {\r}^{ Ba} (\s_i)
\over P_i(\s) } \right) \, \d^4 \left(p^{A B}_{4} -   { \r^{A }_a (\s_4)  {\r}^{ Ba} (\s_4)
\over P_4(\s) } \right) \\ \nonumber
&\times& \prod^{2}_{i=1} \d^{2}\left(q^{1 I}_i - { \r^1_a(\s_i) \chi^{Ia} (\s_i)
\over P_i(\s) }  \right) \d^{2}\left(q^{3 I}_i - { \r^3_a(\s_i) \chi^{Ia} (\s_i)
\over P_i(\s) }  \right) \, {\det}' S_4 \, U(\r, \s)
\eea
with $\{A,B\} \neq \{3,4\}$ for the five-dimensional delta-functions,
and the four-dimensional one has $\{A,B\} \neq \{3,4\}, \{1,3\}$. 
The result is of course independent of the choice of $\{A,B\}$ which 
are singled out to be special here.
Performing the integral,\footnote{This means solving for the $\s$'s
and $\r$'s using the bosonic delta functions, together with gauge
fixing the symmetry $G$, and integrating over the eight fermionic
variables $\chi^I_{m,a}$ using the eight fermionic delta functions.}
and using the formula for $U(\r, \s)$ that will be determined later
(Eqs.~(\ref{eq:U(r,s)1}) and (\ref{eq:U(r,s)2}) or equivalently 
Eq.~(\ref{eq:Unew})),  we find that $I_4$
precisely cancels the Jacobian factors $J_{4,B} J_{4,F}$, leaving
\bea
A_4 =  \d^6 \left(\sum_{i=1}^4 p_i^{AB}\right)\,
\d^8 \left(\sum_{i=1}^4 q_i^{AI} \right) \, ,
\eea
which is the result that was obtained in the previous section.

Higher-point amplitudes in the M5 theory have not appeared in the literature to our knowledge.
However, amplitudes with scalars are constrained by soft theorems (as we will
describe in a later subsection), and some of them are completely
determined by recursion relations \cite{Cheung:2015ota}. For instance,
pure-scalar amplitudes are fixed in terms of the four-point ones. We have tested
numerically that Eq.~(\ref{eq:(2,0)}) indeed reproduces such amplitudes correctly
for $n=6,8$. Those results, combined with supersymmetry and R symmetry, which we have
explicitly checked for six and eight particles in Appendix~\ref{appendix:RsymmetryM5}, imply that Eq.~(\ref{eq:(2,0)})
should be valid for the entire supermultiplet for $n=6, 8$. It seems very likely
that they are correct for all $n$, as we find evidence supporting this in the following sections.
%(Note, however, that we have been cavalier about overall normalizations.) {\bf I'm not sure if this sentence is necessary, since the normalization can be absorbed in the definition of the coupling. and here we do not include the coupling explicitly anyway.}

\subsection{Reduction to four dimensions}

This subsection will determine the constraint on $U(\r,\s)$ in Eq.~(\ref{eq:(2,0)-old})
that arises from requiring that its reduction to 4D cancels the Jacobian
that is generated by the dimensional reduction of the M5 amplitude to 4D.
So the key step is to evaluate the relevant Jacobian. What dimensional reduction
does is to set two components of the six-component momenta equal to zero. In our
conventions this means $p^{12}_i \rightarrow 0$  and $p^{34}_i \rightarrow 0$.
This can be implemented by inserting
\bea
\int dp^{12}_n dp^{34}_n
 \prod^{n-1}_{i=1}dp^{12}_i dp^{34}_i \d(p^{12}_i) \d(p^{34}_i)
\eea
into the formula for the $n$-particle amplitude of the M5 theory
in Eq.~(\ref{eq:(2,0)-old}).
Note that $ \d(p^{12}_i) \d(p^{34}_i)$ is only inserted for $n-1$ particles,
even though the
integration is over all $n$ particles, because
momentum conservation in 6D ensures that $p^{12}_n= p^{34}_n=0$ as well, if
$p^{12}_i= p^{34}_i=0$ for $i=1, 2, \ldots, n{-}1$. Since dimensional reduction
requires setting $ \l^1_{i,-}= \l^2_{i,-} =0$
and $\l^3_{i,+} = \l^4_{i,+} = 0$, therefore we should integrate out the corresponding
$\r^1_{m,-}, \, \r^2_{m,-} $ and $\r^3_{m,+}, \, \r^4_{m,+} $. 
Due to the fact that only the middle sector contributes to the scattering 
amplitudes of the D3 theory, we will focus on that sector only in the following 
computations. Explicitly, we have that the M5 amplitudes
given in Eq.~(\ref{eq:(2,0)-old}) reduce to 4D amplitudes of the form given in
Eq.~(\ref{nptamps}), where the factor ${\cal I}$ is given by
\bea
{\cal I}_{\rm R}&=& \int \frac{\prod^{d }_{m=0}
 d\r^1_{m,-} d\r^2_{m,-} d\r^3_{m,+} d\r^4_{m,+}}{ {\rm Vol} (SL(2,\IC)  )}
dp^{12}_n dp^{34}_n
 \prod^{n-1}_{i=1}dp^{12}_i dp^{34}_i \d(p^{12}_i) \d(p^{34}_i) \cr
&\times&
\prod^n_{i=1} \d \left (p^{12}_i -  { \r^{1}_a (\s_i) \r^{2a}
(\s_i) \over P_i(\s) } \right) \d \left(p^{34}_i -  { \r^{3}_a (\s_i)
\r^{4a} (\s_i) \over P_i(\s) } \right) {\det}' S_n \, U(\r,\s) \, .
\eea
The goal here is to determine the condition on $U(\r,\s)$ that will ensure that
${\cal I}_{\rm R} = {\cal I}^{\rm DBI} = {\det}' S_n$.
Here the $SL(2,\IC)$ group is the one that acts on the little-group indices that will be
reduced to $U(1)$ after the dimensional reduction.
%The function
%$F(\r, \s)$ includes anything else in Eq.~(\ref{eq:(2,0)}) that is not shown explicitly
%in the above computation, since it is not so relevant.

The trivial $n-1$ integrations over $p^{12}_i$ and $p^{34}_i$ give
\bea \label{eq:integralR}
{\cal I}_{\rm R} &=&\int  \frac{\prod^{ d }_{m=0}
d\r^1_{m,-} d\r^2_{m,-} d\r^3_{m,+} d\r^4_{m,+} }{ {\rm Vol} (SL(2,\IC)  )}
dp^{12}_n dp^{34}_n
\prod^{n-1}_{i=1} \d \left( { \r^{1}_a (\s_i) \r^{2a} (\s_i) \over P_i(\s) } \right)
\d \left( { \r^{3}_a (\s_i) \r^{4a} (\s_i) \over P_i(\s) } \right)
 \cr
&\times&
 \d \left(p^{12}_n -  { \r^{1}_a (\s_n) \r^{2a} (\s_n) \over P_n(\s) } \right)
 \d \left(p^{34}_n -  { \r^{3}_a (\s_n) \r^{4a} (\s_n) \over P_n(\s) } \right)
{\det}' S_n \, U(\r,\s)  \, .
\eea
The delta functions force $ \r^{1}_a (\s_i) \r^{2a} (\s_i)$ and
$ \r^{3}_a (\s_i) \r^{4a} (\s_i)$ to vanish
for $i=1, 2, \cdots, n{-}1$. However, $\r^{1}_a (\s) \r^{2a} (\s)$ and
$\r^{3}_a (\s) \r^{4a} (\s)$ are polynomials of degree $2d = n{-}2$
\bea
\r^{1}_a (\s) \r^{2a} (\s) = \sum^{2d}_{m=0} c^{12}_m \, \s^{m}  \, , \quad
\r^{3}_a (\s) \r^{4a} (\s) = \sum^{2d}_{m=0} c^{34}_m \, \s^{m}  \, ,
\eea
where
\bea
c^{12}_m  = \sum_{m' =0}^m \r^{1}_{ m', a} \r^{2,a}_{m -m'} \, , \quad
c^{34}_m  = \sum_{m' =0}^m \r^{3}_{ m', a} \r^{4,a}_{m -m'}\, , \quad m=0,1,\ldots, 2d.
\eea
Because the degree of these polynomials is less than the $n-1$ required roots, we conclude that all of the coefficients $c_m^{12}$ and $c_m^{34}$
should vanish.  Since this also implies that
${ \r^{1}_a (\s_n) \r^{2a} (\s_n) }=0$ and ${ \r^{3}_a (\s_n) \r^{4a} (\s_n) }=0$,
the integrations over $p^{12}_n$ and $p^{34}_n$ in Eq.~(\ref{eq:integralR}) are trivial.

The formula for ${\cal I}_{\rm R}$ now contains $2n-2$ delta functions, but there are $2n$
integrations, so we should use $SL(2,\IC)$ to fix two of them.
This leaves a $U(1)$ unfixed, as expected.
Let us now perform the integrations over $\r^1_{m,-}$ and $\r^2_{m,-}$ as well
as $\r^3_{m,+}$ and $\r^4_{m,+}$ explicitly. A convenient method is to change the
integration variables to the coefficients $c^{12}_m$ defined previously,
\bea
&&\frac{\prod^{d }_{m=0}  d \r^1_{m,-} d \r^2_{m,-} d\r^3_{m,+} d\r^4_{m,+}
}{ {\rm Vol} (SL(2,\IC)  )}\prod^{n-1}_{i=1}\left\{ \left[ P_i(\s)\right]^2
\d \left(\sum_{m=0}^{n-2} c^{12}_m \s^m_i\right)
 \d \left(\sum_{m=0}^{n-2} c^{34}_m \s^m_i\right)\right\} \cr
&=& V^{2} (\s) \,  \frac{\prod^{d }_{m=0}  d \r^1_{m,-} d \r^2_{m,-} d\r^3_{m,+} d\r^4_{m,+}
}{ {\rm Vol} (SL(2,\IC)  )}   \prod_{m=0}^{n-2} \d( c^{12}_m )  \d( c^{34}_m)    \cr
&=&
J_C \, J_{SL(2,\IC)} \, V^2(\s) \, d \r^2_{d,-}d \r^4_{d,+}
\prod^{n-2}_{m=0} d c_m^{12} d c_m^{34} \, \prod_{m=0}^{n-2} \d( c^{12}_m )  \d( c^{34}_m) \, ,
\eea
where
\bea
V(\s) = \prod_{i>j } \s_{ij} = \prod_{i=2}^n \prod_{j=1}^{i-1} \s_{ij}
\eea
is the Vandermonde determinant. It arose from the following combination of factors:
\bea
V(\s) = \frac{1}{V_n(\s)}\prod^{n-1}_{i=1} P_i(\s),
\eea
where
\bea
V_n(\s) = \det \s_i^m  = V(\s)/ P_n(\s),
\eea
and
\bea
V^2(\s) = \prod _{i=1}^n P_i(\s).
\eea
There are no minus sign issues, since $n$ is even.

The factor $J_{SL(2,\IC)}$ is due to gauge-fixing the $SL(2,\IC)$ symmetry of
the complexified little-group symmetry.
We have chosen to gauge fix $\r^2_{d,-}, \r^4_{d,+}$ and $\r^2_{d,+} $, and
thus the Jacobian due to the gauge-fixing of the complexified $SU(2)$ symmetry is given by
\bea
J_{SL(2,\IC)} =\r^4_{d,-}  (\r^2_{d,+} \r^4_{d,-}  - \r^2_{d,- } \r^4_{d,+ }  ) \, .
\eea

The factor $J_C$ is the Jacobian that arises due to
the change of variables from $\r$ coordinates to $c$ coordinates.
It contains a product of two resultants, and it is given by
\bea
J^{-1}_C = \r^2_{d,+} \r^4_{d,-} R(\r^1_+, \r^2_+ ) R(\r^3_-, \r^4_- ) \, .
\eea
The resultant has appeared previously in a twistor-string-like formulation of
scattering amplitudes in various theories \cite{Cachazo:2013zc}\cite{Cachazo:2013iaa}, and include the D3 theory \cite{Cachazo:2016njl}.
Its crucial property is that it vanishes if and only if the two polynomials
$\r^A_a(\s)$ and $\r^B_a(\s)$ have a root in common.

A resultant of the form $R(\r^A_a, \r^B_a)$, where $\r^A_a$ and $\r^B_a$ are
both polynomials of degree $d$, is given by the determinant of a Sylvester matrix
$M_a^{(2d)}(A,B)$,
\bea
R(\r^A_a, \r^B_a) = \det M_a^{(2d)}(A,B).
\eea
In particular, $R(\r^1_+, \r^2_+ ) = \det M^{(2d)}_{+}(1,2) $  is the resultant
of the pair of degree $d =\frac{n }{ 2} -1$ polynomials
\bea
\r^1_+(\s) = \sum^{ d}_{m=0} \r^1_{m,+} \s^m \, , \quad
\r^2_+(\s) = \sum^{ d}_{m=0} \r^2_{m,+} \s^m \, .
\eea
Explicitly, the Sylvester matrix $M^{(2d)}_{a}(A,B)$ is given by
\bea \label{eq:Sylvester}
M^{(2d)}_{a}(A, B)  = \begin{pmatrix}
\r_{0,a}^A  &   \r_{1,a}^A&  \r_{2,a}^A&    \cdots &  \cdots &  \r_{d,a}^A & 0  & \cdots &0  \\
0&  \r_{0,a}^A  & \r_{1,a}^A  &   \cdots &  \cdots &   \r_{d-1,a}^A &  \r_{d,a}^A  & \cdots &0 \\
\vdots &  \vdots   &   \cdots &  \cdots &   \vdots  &  \vdots  & \vdots  & \cdots & \vdots
\\
0 & 0& \cdots &   \r_{0,a}^A  &   \r_{1,a}^A   & \cdots &   \cdots  &   \r_{d-1 ,a}^A&
 \r_{d ,a}^A  \\
\r_{0,a}^B  &   \r_{1,a}^B&  \r_{2,a}^B&    \cdots &  \cdots &  \r_{d,a}^B & 0  & \cdots &0  \\
0&  \r_{0,a}^B  & \r_{1,a}^B  &   \cdots &  \cdots &   \r_{d-1,a}^B &  \r_{d,a}^B  & \cdots &0 \\
\vdots &  \vdots   &   \cdots &  \cdots &   \vdots  &  \vdots  & \vdots  & \cdots & \vdots
\\
0 & 0& \cdots &   \r_{0,a}^B  &   \r_{1,a}^B   & \cdots &   \cdots  &
\r_{d-1 ,a}^B&   \r_{d ,a}^B
\end{pmatrix} \, .
\eea
For instance, for $n=6$ or $d=2$, the Sylvester matrices are $4 \times 4$,
\bea
M^{(4)}_{a}(A,B)  = \begin{pmatrix}
\r_{0,a}^A  &   \r_{1,a}^A&   \r_{2,a}^A & 0  \\
0&  \r_{0,a}^A  &   \r_{1,a}^A   &   \r_{2,a}^A \\
 \r_{0,a}^B  &   \r_{1,a}^B  &   \r_{2,a}^B & 0\\
0&  \r_{0,a}^B  &   \r_{1,a}^B   &   \r_{2,a}^B
\end{pmatrix} .
\eea

%Under the constraints $c^{12}_m = c^{34}_m =0$, for all $m$, we find that $R(\r)$ in
%Eq.~(\ref{eq:integralR}) reduces to the product of $R(\r^1_+, \r^2_+ )$ and
%$R(\r^3_-, \r^4_- )$,
%\bea
%R(\r) \rightarrow \left( \frac{ \r^2_{d,+} \r^4_{d,-}
%- \r^2_{d,- } \r^4_{d,+}}
%{ \r^2_{d,+} \r^4_{d,-}   } \right)^{n-2} R(\r^1_+, \r^2_+ )
%   R(\r^3_-, \r^4_- ) \, .
%\eea

To exhibit the residual $U(1)$ little-group symmetry in 4D, we may set
$\r^2_{d,- } =\r^4_{d,+}  =0$ using partly the complexified 6D little-group
symmetry $SL(2,\IC)$. Now we see that all the factors are exactly canceled,
except for $\r^4_{d,-}$, which is precisely the Jacobian for the gauge-fixing
of the left-over $U(1)$ symmetry of the 4D theory.
Furthermore, the fermionic delta functions of $(2,0)$ supersymmetry
also reduce to the 4D ones, without any complications.
Thus, the proposed formula for the M5 amplitude
in Eq.~(\ref{eq:(2,0)-old}) reduces to the D3 amplitude in Eq.~(\ref{calAn})
under dimensional reduction provided that the factor $U(\r,\s)$ reduces according to
\bea
U(\r,\s) \to V^{-2}(\s) R(\r^1_+, \r^2_+ ) R(\r^3_-, \r^4_- )
\eea
in 4D.

\subsection{The extra factor $U(\r,\s)$}

To complete the construction of the M5 amplitudes, we need to determine the
extra factor (relative to the D3 formula) $U(\r,\s)$. We have just learned
what it should give when reduced to 4D. This goes a long way towards
determining it. We claim that the $\s$ and $\r$ dependence factorizes already
in 6D, so that
\bea  \label{eq:U(r,s)1}
U(\r,\s) = V^{-2}(\s) R(\r) \, .
\eea
Note that $V^{-2}$ has total symmetry in the $n$ $\s_i$'s. As will be
verified later, $V^{-2}$ transforms under $SL(2,\IC)$ in the way required to
compensate for the additional bosonic coordinates in the M5 theory.
The factor $R(\r)$ should scale like $p^{2d}$ or $\r^{4d}$ and on reduction
to 4D it should give the product of resultants
$R(\r^1_+, \r^2_+ ) R(\r^3_-, \r^4_- )$. This expression does
not have 6D Lorentz invariance or little-group symmetry, so
it must be embellished by additional pieces that vanish upon dimensional
reduction.

The crucial observation is that the product of resultants
$R(\r^1_+, \r^2_+ ) R(\r^3_-, \r^4_- )$ can be expressed in terms of
${\rm Pf}'S_n$ and the Vandermonde determinant $V(\s)$ \cite{Cachazo:2016njl},
\bea
R(\r^1_+, \r^2_+ ) R(\r^3_-, \r^4_- ) = {\rm Pf}'S_n \, V(\s) \, .
\eea
The above relation is valid for $\r$ and $\s$ under the constraints of the helicity-conserving sector, Eq.~(\ref{eq:middlesector}), which is the case here.
As functions of $s_{ij}$ and $\s_i$, now both ${\rm Pf}'S_n$ and $V(\s)$ can be lifted to 6D straightforwardly without violating Lorentz invariance.

This leads to our proposal for all tree-level scattering amplitudes of the M5 theory,
\begin{align} \label{eq:(2,0)}
{\cal A}_n = \int \frac{d^n \s \, d \mathcal{M} }{ {\rm Vol} (G) }  \,
\D_B(p, \r) \, \D_F (q, \r, \chi)\,
{ \left( {\rm Pf}'S_n \right)^3 \over V(\s)} \, ,
\end{align}
which is the main result of the paper. This formula reduces to the D3 amplitude in
Eq.~(\ref{calAn}) correctly, and it also has many other correct
properties that we will discuss shortly. Importantly, Eq.~(\ref{eq:(2,0)}) produces
known amplitudes as we mentioned.

Alternatively, one can use the definition of the resultant in terms of Sylvester matrix in Eq.~(\ref{eq:Sylvester}). With that, a different possible uplift to 6D is realized by a natural generalization of the resultant and Sylvester matrix. They are given by,
\bea  \label{eq:U(r,s)2}
R(\r) = \det M^{(4d)},
\eea
where $M^{(4d)}$ is the following $ 4d \times 4d$ matrix, a generalization of Sylvester matrix,
\bea \label{eq:U(r,s)}
M^{(4d)} = \begin{pmatrix}
M^{(2d)}_{+}(1,2)     &   M^{(2d)}_{-}(1,2)    \\
M^{(2d)}_{+} (3,4)&   M^{(2d)}_{-} (3,4)
\end{pmatrix} .
\eea
The subscripts $+$ and $-$ are little-group indices, whereas $SU(4)$ Lorentz indices,
$A=1,2,3, 4$, are shown in parentheses.
The four submatrices in $M^{(4d)}$ are $2d \times 2d$ matrices, which take
the form of Sylvester matrices. Upon dimension reduction, the off-diagonal matrices of $M^{(4d)}$ vanish, and thus $R(\r)$ also has the required reduction to 4D.  So in terms of $R(\r)$, the scattering amplitudes of M5 theory then take an alternative form,
\begin{align} \label{eq:(2,0)1}
 {\cal A}_n = \int \frac{d^n \s \, d \mathcal{M} }{ {\rm Vol} (G) }  \,
\D_B(p, \r) \, \D_F (q, \r, \chi)\, {\det}' S_{n} \,
{ R(\r) \over V^2(\s) } \, .
\end{align}
In fact, like the case of 4D where the resultant is related to ${\rm Pf}'S_n$ and the Vandermonde determinant $V(\s)$, we find that, under the support of delta function constraint $\D_B$, $R(\r)$ defined in Eq.~(\ref{eq:U(r,s)2}) is related to ${\rm Pf}'S_n$ and $V(\s)$ in the same way,  namely,
\bea \label{eq:Unew}
R(\r) = {  {\rm Pf}'S_n \, V(\s) }\, .
\eea
Plugging this result into Eq.~(\ref{eq:(2,0)1}) reproduces Eq.~(\ref{eq:(2,0)}). Therefore
these two different approaches actually lead to the same result.

Although the quantity $R(\r)$ can be re-expressed in terms of ${\rm Pf}'S_n$ and $V(\s)$
on the support of delta-function constraints, it may still be of interest on its own right.
Let us make a few comments on it here before closing this subsection. It is straightforward
to show that $R(\r)$ is invariant under little-group and Lorentz-group transformations,
which together act on $R(\r)$ as $SL(2,\IC) \times SL(4, \IC)$. A natural
generalization would be invariant under $SL(k,\IC) \times SL(2k, \IC)$, and
it would relate $2k^2$ polynomials of degree $d$, which transform as bifundamentals.
The generalization to $k>2$ may be relevant for scattering amplitudes of the D-brane
theories in dimension greater than six. We will leave this for the future study. The usual resultant, which corresponds to $k=1$, vanishes whenever the two polynomials
have a common zero. It would be interesting to know the generalization of this
statement when $k>1$. In any case, these remarks suggest introducing the alternative notation
$R^{(k)}_d (\r) = \det M_d^{(k)}$, where the matrix $ M_d^{(k)}$ has $2kd$ rows
and columns. However, we will not utilize that notation in this manuscript.

\subsection{$SL(2,\IC)$ modular symmetry}

Let us examine whether Eq.~(\ref{eq:(2,0)}) has the correct $SL(2,\IC)$ modular
symmetry under the transformations of the form
\bea
\s'_i = \frac{a\, \s_i + b }{ c\, \s_i + d } \quad {\rm with }  \quad ad-bc=1 \, .
\eea
Let us begin with the rescaling symmetry, $\s_i  \rightarrow a\, \s_i$, where
$a$ is a nonzero complex number (the square of the preceding $a$ with $b=c=0$).
To maintain the same delta functions, $\Delta_B(p, \r)$ and $\Delta_F(q, \r, \chi)$
in Eqs.~(\ref{eq:DB20}) and (\ref{eq:DF20}), we rescale
\bea
\r^{Aa}_{m}   \rightarrow a^{{n -1 \over 2} -m} \r^{Aa}_{m} \, , \quad
\chi^{I a}_{m}   \rightarrow a^{{n -1 \over 2} -m} \chi^{I a}_{m} \, .
\eea
With this rescaling
\bea
V^{-1}(\s) \rightarrow a^{- {n^2 \over 2} + { n \over 2}} \, V^{-1}(\s) ,
\eea
\bea
{\rm Pf}' S_n \rightarrow a^{-{n \over 2}} \, {\rm Pf}' S_n ,
\eea
\bea
\prod^n_{i=1} d \s_i  \rightarrow
a^n \prod^n_{i=1} d \s_i ,
\eea
\bea
\prod^{d}_{m=0} d^8 \r^{Aa}_m d^4 \chi^{Ia}_m \rightarrow
a^{{n^2 \over 2} }  \prod^{d}_{m=0} d^8 \r^{Aa}_m d^4 \chi^{Ia}_m \, .
\eea
Thus all the factors of $a$ cancel out, and scale invariance is verified.

Next let us consider inversion, $\s_i \rightarrow  - 1/\s_i$.\footnote{The minus sign
is unnecessary, because we could set $a=-1$ in the preceding scaling symmetry,
but it reduces the need to keep track of minus signs.} First we note that
\bea
P_i(\s) \rightarrow (\prod^n_{j=1} \s^{-1}_j ) \s_i^{2-n} P_i(\s).
\eea
Therefore, we rescale $\r^{Aa}_m$
and $\chi^{Ia}_m$ to keep the delta functions unchanged by
\bea
\r^{Aa}_m \rightarrow  (-1)^m (\prod^n_{j=1} \s^{-1/2}_j ) \, \r^{Aa}_{d -m } \, ,\quad
\chi^{Ia}_m \rightarrow (-1)^m(\prod^n_{j=1} \s^{-1/2}_j ) \, \chi^{Ia}_{ d -m } \, .
\eea
Under such rescalings, we have,
\bea
V(\s) \rightarrow  (\prod^n_{j=1} \s^{1-n}_j ) V(\s)
\eea
and
\bea
{\rm Pf'} S_n \rightarrow (\prod^n_{j=1} \s_j ) \, {\rm Pf}' S_n \, ,
\eea
while the measure behaves as
\bea
\prod^n_{i=1} d \s_i \prod^{d}_{m=0} d^8 \r^{Aa}_m d^4 \chi^{Ia}_m \rightarrow
(\prod^n_{j=1} \s^{-n-2}_j ) \prod^n_{i=1} d \s_i \prod^{d}_{m=0} d^8 \r^{Aa}_m d^4 \chi^{Ia}_m \, .
\eea
Combine all the contributions, the invariance under inversion becomes clear.

Finally, let us consider translation, $\s_i \rightarrow \s_i + b$.
This leaves $V(\s)$, $P_i(\s)$, and ${\rm Pf}' S_n$ invariant.
So we let $\r \rightarrow \r'$ and $\chi \rightarrow \chi'$ such that
\bea
\sum^{d}_{m=0} \r^{Aa}_m (\s_i + b)^m = \sum^{d}_{m=0} \r'^{Aa}_m  \s_i ^m \, , \quad
\sum^{d}_{m=0} \chi^{Ia}_m (\s_i + b)^m = \sum^{d}_{m=0} \chi'^{Ia}_m  \s_i ^m \, .
\eea
It is easy to see that the integration measures are also invariant under this transformation,
since the Jacobian is the determinant of a triangular matrix with 1's on the diagonal.

\subsection{Factorization}

The formula for the amplitude ${\cal A}_n$ in Eq.~(\ref{eq:(2,0)}) is an integral over
sets of polynomials $\r^A_a(\s)$ and $\chi^I_a(\s)$ of
degree $d = {n \over 2}-1$.  To study the multi-particle factorization behavior
of the amplitudes, one may take a limit on the moduli space such that the higher-degree
polynomials degenerate into products of lower-degree
ones \cite{Gukov:2004ei, Vergu:2006np, Cachazo:2012pz}.
Specifically, there is a ``left'' factor containing polynomials of degree
$d_L = \frac{n_L}{2} -1$ and a ``right'' factor containing polynomials
of degree $d_R = \frac{n_R}{2} -1$, where $d_L + d_R = d$ or $n_L + n_R = n+2$.
To achieve this goal, we introduce a parameter $s$ that approaches zero in the desired limit
and perform the following rescaling of the $\r_m$'s\footnote{We thank Ellis Yuan for a
discussion about the factorization limit.}
\bea
&& \r_m \rightarrow  t_L \, s^{d_L-m} \r_{L, d_L- m}\,
 , \quad {\rm for} \quad m=0, 1, \ldots, d_L \cr
&& \r_m \rightarrow  t_R \, s^{m - d_L} \r_{R, m-d_L}\,
, \quad {\rm for} \quad m=d_L, d_L+1, \ldots, d,
\eea
with
\bea
t_L^2 = { (-1)^{n-1} s^{-2 d_R-1} } { \prod_{i \in R} \s_i \over  \prod_{i \in L} \s_i }
\eea
and
\bea
t_R^2 = s^{-2d_R-1},
\eea
where $L$ or $R$ denotes the set of particles on the left- or right-hand side
of a factorization channel.

We will show that the left-hand side of the factorization channel has
polynomials of degree $d_L$ and the right-hand side has polynomials of degree $d_R$ .
Accordingly, we rename the $\r$'s as either $\r_L$ or $\r_R$.
Note $\r_{d_L}$ appears on both sides, but we separate it into two coordinates by
setting $\r_{d_L} = \r_{L, 0}$ and $\r_{d_L} = \r_{R, 0}$, and introducing
$\int d \r_{L, 0} \d ( \r_{L, 0} -  \r_{R, 0})$. Now for the $\sigma_i$'s,
we make the replacements
\bea
&& \sigma_i \rightarrow  {s \over \sigma_i } \, , \quad {\rm for} \quad i \in L \cr
&& \sigma_i \rightarrow  {\sigma_i \over s} \, , \quad {\rm for} \quad i \in R
\eea
In the limit $s \rightarrow 0$, a degree-$d$ polynomial degenerates into a product of
degree $d_L$ or $d_R$ polynomials, depending on whether the particle is on the left-
or the right-hand side, namely
\bea
 \r^{A}_a (\s_i) = \sum^{d}_{m=0} \r^{Aa}_m \sigma^m_i  \rightarrow
  \r^{A}_{L,a} (\s_i) = \sum^{d_L}_{m=0} \r^{Aa}_{L,m} \sigma^m_i
  \, \quad {\rm for} \quad i \in L \, ,   \cr
 \r^{A}_a (\s_i)  = \sum^{d }_{m=0} \r^{Aa}_m \sigma^m_i  \rightarrow
  \r^{A}_{R,a} (\s_i)= \sum^{d_R}_{m=0} \r^{Aa}_{R,m} \sigma^m_i  \,
  \quad {\rm for} \quad i \in R \, .
\eea
It is also straightforward to see that the delta functions reduce to the
corresponding lower-point delta functions, namely,
\bea
p_i^{AB} - { \r^{A}_a (\s_i)\r^{B a} (\s_i) \over P_i(\s) } = 0
\rightarrow p_i^{AB} - { \r^{A}_{L,a} (\s_i)\r^{B a}_L (\s_i)
\over P_{L,i}(\s) } =0 \, ,~~  {\rm or} ~~~ p_i^{AB}
- { \r^{A}_{R,a} (\s_i)\r^{B a}_R (\s_i) \over P_{R, i}(\s) } =0  \nonumber \\
\eea
depending on whether $\s_i$ is on the left or the right. If $ i \in L$,
$P_{L,i}(\s) = (0 - \s_i) \prod^{n_L}_{j\neq i} (\s_{ji})$, where ``0" is the
value of the $\s$ coordinate associated to the internal line in the factorization,
and similarly for $i \in R$.

It is important that the integrand and the integration measure factorize correctly,
and this is straightforward to see for the measure.
On the other hand, the building blocks of the integrand, the Vandermonde determinant $V(\s)$
and ${\rm Pf}' S_n$, have already appeared in literature in the construction of
scattering amplitudes in other theories; they are also known to factorize correctly.
Alternatively for the proposal Eq.~(\ref{eq:(2,0)1}), we find the new mathematical object we constructed, $R_{n}(\r)$, also factorizes properly in
the $s\rightarrow 0$ limit,
\bea
R_n(\r_0, \r_1, \ldots , \r_d)  \rightarrow  t_L^{4d_L} t_R^{4 d_R} \, s^{2(d_L^2+d_R^2)}
R_{n_L}(\r_{L,0}, \r_{L,1} , \ldots , \r_{L,d_L}) \\ \nonumber
\times R_{n_R}( \r_{R, 0}, \r_{R,1} ,
\ldots , \r_{R, d_R}) .
\eea
Here the subscript of $\r_m$ denotes the index $m$ of $\r^{A, a}_m$,
and we have suppressed Lorentz and little-group indices $A$ and $a$

Finally, because $P^2_L \sim s^2$ in the limit $s \rightarrow 0$
at a factorization pole $1/P^2_L$, the amplitude should go as
$ds^2/s^2$~\cite{Cachazo:2012pz}.  By collecting all of the $s$ factors
arising from the integration measure and the various factors in the
integrand,  we have verified that this
is indeed the case. Thus, the general formula Eq.~(\ref{eq:(2,0)})
has the required factorization properties for a tree-level scattering amplitude.

\subsection{Soft theorems} \label{section:soft}

As we  discussed previously, the five scalars of the M5 theory are
Goldstone bosons arising from spontaneous breaking of 11D Poincar\'e symmetry.
More specifically, the relevant broken symmetries are translations in the
five spatial directions that are orthogonal to the M5-brane.
Let us now study how the scattering amplitudes of the M5 theory behave in soft limits,
\ie in the limit where the momentum $p^{AB}$ of a Goldstone boson vanishes.
As shown in \cite{Guerrieri:2017ujb}, amplitudes involving such scalars have
enhanced soft behavior \cite{Cheung:2014dqa}, specifically
\bea
{A}(p_1, \cdots, p_{n-1},  {\tau p_n}) \sim \mathcal{O} (\tau^2)  \, ,
\eea
where particle $n$ is a scalar, with momentum $\t p_n$,
and the soft limit is realized by
$\tau \rightarrow 0$.  Of course, some of the other momenta
should also depend on $\t$, so as to maintain momentum conservation and
masslessness. %Then the limit is unambiguous as long as only $\t p_n$ vanishes.

We claim that the amplitudes obtained from general formula in
Eq.~(\ref{eq:(2,0)}) indeed have this enhanced soft behavior. In particular,
if we rescale $\l^{A, a}_{n} = \tau^{1/2} \, \l^{A, a}_{n}$, so that
the momentum $p_n$ is replaced by $\tau \, p_n$, we find that
the various pieces that contribute to the amplitude scale as follows
\bea
({\rm Pf}' S)^3 \sim \tau^3, \quad J_B
\sim \tau^{-1} \, , \quad J_F \sim \tau^{0}  \, ,
\eea
and the rest, including the Vandermonde determinant $V(\s)$, scales
as $ \tau^{0}$ in the soft limit. As discussed in Appendix~\ref{appendix:M5Jacobi}, $J_B$
and $J_F$ are Jacobians that arise from extracting various ``wave functions"
and momentum-conservation delta functions, and
from performing integrations over $\s$'s, $\r$'s, and $\chi$'s.
$J_F$ also depends on the fact that we are considering a scalar
component of the supermultiplet. Altogether, we obtain the
expected $\mathcal{O} (\tau^2) $ behavior of the amplitudes in the M5 theory.

Just for the comparison, in the case of the D3 theory,
in the soft limit each piece in Eq.~(\ref{nptamps}) behaves as
\bea
 {\rm det}' S \sim \tau^2, \quad J_B  \sim \tau^{0} \, , \quad J_F \sim \tau^{0} \, .
\eea
In total, the amplitudes again scale correctly, namely as $\mathcal{O} (\tau^2)$.

We can also study how the amplitudes behave in the double-soft limit, where we let
two momenta approach zero simultaneously, say, $p_{n+1} \rightarrow \tau p_{n+1}$
and $p_{n+2} \rightarrow \tau p_{n+2}$ with $\tau \rightarrow 0$. For simplicity,
here we only consider the leading soft theorems. The result of the double-soft limit
depends on the species of particles involved as shown here
\bea \label{eq:doublesoft}
A_{n+2} (\phi, \bar{\phi}) &=& \sum^n_{i=1} {(s_{{n+1} \,i} - s_{{n+2}\,i})^2
\over (s_{{n+1} \, i} + s_{{n+2} \,i})} A_n  + \ldots \, , \\ \nonumber
A_{n+2} (\psi_a, \tilde{\psi}_b) &=& \sum^n_{i=1}  \langle  (n{+}1)_a\,  (n{+}2)_b\, i_+\, i_-  \rangle
 {(s_{{n+1} \,i} - s_{{n+2}\,i}) \over (s_{{n+1} \, i} + s_{{n+2} \,i})} A_n  + \ldots \, ,
\\ \nonumber
A_{n+2} (B_{a_1 b_1}, B_{a_2 b_2}) &=& \sum^n_{i=1}
{ \langle  (n{+}1)_{a_1}\,  (n{+}2)_{a_2}\, i_+\, i_-  \rangle
 \langle  (n{+}1)_{b_1}\,  (n{+}2)_{b_2}\, i_+\, i_-  \rangle
 \over (s_{{n+1} \, i} + s_{{n+2} \,i})} A_n + \ldots \, .
\eea
The soft particles $\phi, \bar{\phi}$ (and $\psi, \tilde{\psi}$) are conjugate to each other to
form an R-symmetry singlet.  The ellipsis denotes higher-order terms in the soft limit,
and the lower-point amplitude $A_n$ is the amplitude with the two soft particles removed.
In the case of soft theorems for $B$ fields, on the right-hand side one should symmetrize
the little-group indices $a_1, b_1$ and $a_2 , b_2$.

The double-soft theorems for the scalars and fermions agree with the known
result \cite{Guerrieri:2017ujb} derived from the Ward identity for scalars that
are Goldstone bosons of spontaneously broken higher-dimensional Poincar\'e symmetry,
while the fermions are Goldstinos of broken supersymmetries. The double-soft theorems
for $B$ fields are new; it would be of interest to study the corresponding symmetries.
If we choose both of the soft $B$ fields to be $B_{+-}$ and reduce to 4D, we obtain
the double-soft result for scalars as in the first line of Eq.~(\ref{eq:doublesoft}).
If, instead, we take the two soft $B$ fields to be $B_{--}$ and $B_{++}$, and reduce to 4D,
we obtain the double-soft theorem for photons in Born--Infeld theory, namely
\bea
A_{n+2} (\gamma_+, \gamma_-) &=& \sum^n_{i=1}   {[  n{+}1 \, i ]^2
\langle  n{+}2 \,  i  \rangle^2 \over (s_{{n+1} \, i} + s_{{n+2} \,i})} A_n + \ldots \, ,
\eea
which agrees with what was found in \cite{He:2016vfi}. Similarly, the double-soft theorem
for fermions reproduces that of Volkov-Akulov theory upon reduction to 4D \cite{Chen:2014xoa}. To obtain
these results we have applied the following identities for the dimensional reduction
$6D \rightarrow 4D$, according to our convention,
\bea
&& \langle k_{+} l_{+}  i_{-} j_{-}\rangle  \rightarrow -  \langle k\,l \rangle  [ i\,j ]\, , \quad
[k_{+} l_{+}   i_{-} j_{-}] \rightarrow - \langle k\,l \rangle  [ i\,j]  \, , \nonumber \\
&&  \langle  i_{-} j_{-} k_{-} l_{\pm}\rangle  \rightarrow 0\,,
\quad \langle  i_{+} j_{+} k_{+} l_{\pm}\rangle \rightarrow 0 \, , \\ \nonumber
&& [  i_{-} j_{-} k_{-} l_{\pm} ] \rightarrow 0\,,
\quad [  i_{+} j_{+} k_{+} l_{\pm} ] \rightarrow 0 \, .
\eea

\subsection{Six- and eight-particle amplitudes of the M5 theory}

As an application of the $n$-particle amplitude in
Eq.~(\ref{eq:(2,0)}), this section presents
analytic results for some specific amplitudes of the M5 theory, namely six- and
eight-particle amplitudes of self-dual $B$ fields. To our knowledge, these
amplitudes have not been presented in the literature before. The use of
spinor-helicity variables circumvents the usual difficulties associated
to the lack of a manifestly covariant
formulation of the M5-brane action. Still, it is not easy to directly
compute any higher-point amplitudes analytically, especially due to the fact
that the scattering-equation constraints are high-degree polynomial equations
whose solutions are rather complicated.
The approach that we have used to obtain analytic results
is to write down an ansatz with unknown coefficients for the
amplitude of interest, and then to fix the coefficients by comparing
the ansatz with the result obtained from the general formula in
Eq.~(\ref{eq:(2,0)}).

Let us begin with the six-particle amplitude of $B_{++}$. Recall that the
$B$ particles form a triplet of the $SU(2)$ helicity group. $B_{++}$ corresponds
to the $J_3 =1$ component of this triplet. (The other two components are
$B_{+-} = B_{-+}$ and $B_{--}$.) The ansatz clearly should have correct
factorization properties. Specifically, the amplitude should contain poles
at which the residue factorizes as a product of two four-point amplitudes,
\[
A(B_{++},B_{++},B_{++},B_{++},B_{++},B_{++} )
\]
\bea \label{eq:factorization6}
\rightarrow \sum_{a,b} {A_L(B_{++},B_{++},B_{++},B_{ab} )
A_R(\bar{B}_{ab},B_{++},B_{++},B_{++}) \over P^2_L } \, .
\eea
The summation over $a,b$ denotes the fact that the internal $B_{ab}$
can be $B_{++}, B_{--}$ and $B_{+-}$, whereas $\bar{B}_{ab}$ is the conjugate.
Here we have used the fact that $A(B_{++},B_{++},B_{++},B_{ab} )$ are the
only non-vanishing four-point amplitudes involving three $B_{++}$'s
allowed by R symmetry. Recall the
known result of $A(B_{++},B_{++},B_{++},B_{ab} )$, given in Sect.~\ref{section:introM5theory},
\bea \label{eq:3B++}
A(B_{++},B_{++},B_{++},B_{ab} ) = \langle 1_+ \, 2_+ \, 3_+\,  4_{a} \rangle
\langle 1_+ \, 2_+ \, 3_+\,  4_{b} \rangle \, .
\eea
where we have used the bracket notation defined in Eq.~(\ref{4bracket}).
Using the results of Eq.~(\ref{eq:factorization6}) and Eq.~(\ref{eq:3B++}),
it is straightforward to write an ansatz that has the correct factorization properties,
\bea \label{eq:fact6B++}
A(B_{++},B_{++},B_{++},B_{++},B_{++},B_{++} )  =
{1 \over s_{123} }  \left( \sum^3_{i=1}  \langle 1_+ \, 2_+ \, 3_+\,  i_{a} \rangle
\langle i^{a} \, 4_+ \, 5_+ \, 6_+  \rangle  \right)^2 + \mathcal{P}_6
\eea
here $ \mathcal{P}_6$ means summing over all ten factorization channels
(nine in addition to the one that is shown).

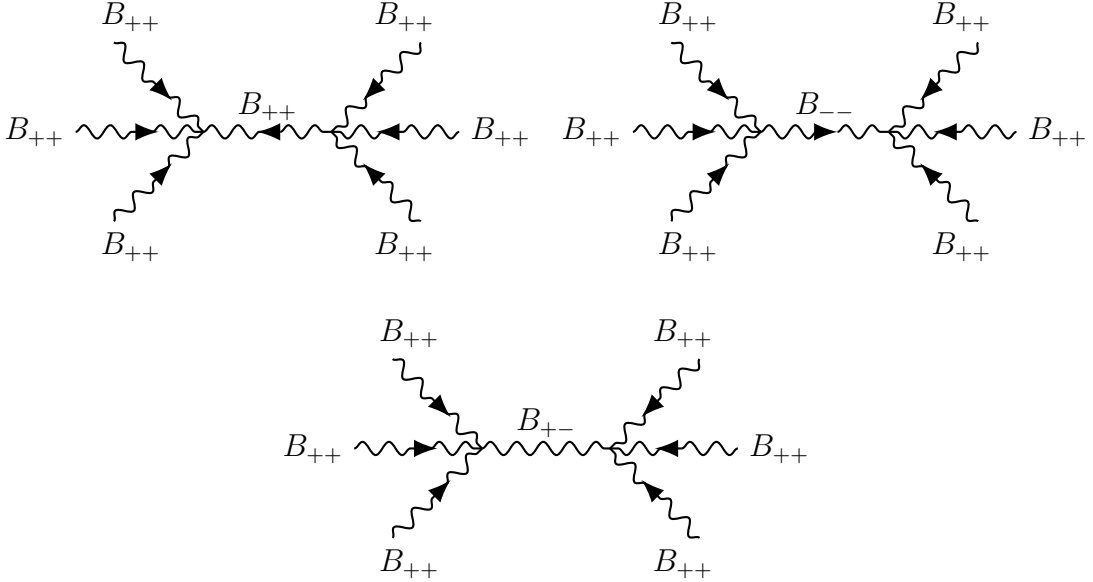
\begin{figure}
\begin{center}
\begin{tikzpicture}[thick, scale = .85]
\draw (-1,0) node[anchor = south] {$B_{++}$};
  \path [draw,decorate, decoration=snake]
    (-4,0) -- (-3,0);
 \path [draw,decorate, decoration=snake]
 (-2.8,0) -- (-2,0);
  \path [draw, opacity = 0, postaction={decorate},
        decoration={markings,mark=at position .65
        with {\arrow[opacity = 1, scale=1.7,>=latex]{>}}}]
    (-4,0) -- (-2,0);  \draw (-4,0) node[anchor = east] {$B_{++}$};
  \path [draw,decorate, decoration=snake]
    (-3.4,1.4) -- (-2.7,.7);
 \path [draw,decorate, decoration=snake]
 (-2.55,.55) -- (-2,0);
  \path [draw,opacity = 0, postaction={decorate},
        decoration={markings,mark=at position .65
        with {\arrow[opacity = 1,scale=1.7,>=latex]{>}}}]
    (-3.4,1.4) -- (-2,0); \draw (-3.14,1.4) node[anchor = south] {$B_{++}$};
  \path [draw,decorate, decoration=snake]
    (-3.4,-1.4) -- (-2.7,-.7);
 \path [draw,decorate, decoration=snake]
 (-2.55,-.55) -- (-2,0);
  \path [draw, opacity = 0, postaction={decorate},
        decoration={markings,mark=at position .65
        with {\arrow[opacity = 1,scale=1.7,>=latex]{>}}}]
    (-3.4,-1.4) -- (-2,0);  \draw (-3.14,-1.4) node[anchor = north] {$B_{++}$};
  \path [draw,decorate, decoration=snake]
    (-2,0) -- (-1,0);
 \path [draw,decorate, decoration=snake]
 (-.8,0) -- (0,0);
  \path [draw,opacity = 0, postaction={decorate},
        decoration={markings,mark=at position .60
        with {\arrow[opacity = 1,scale=1.7,>=latex]{<}}}]
    (-2,0) -- (0,0);
  \path [draw,decorate, decoration=snake]
    (2,0) -- (1,0);
 \path [draw,decorate, decoration=snake]
 (2.8-2,0) -- (2-2,0);
  \path [draw, opacity = 0,postaction={decorate},
        decoration={markings,mark=at position .65
        with {\arrow[opacity = 1,scale=1.7,>=latex]{>}}}]
    (4-2,0) -- (2-2,0);   \draw (4-2,0) node[anchor = west] {$B_{++}$};
  \path [draw,decorate, decoration=snake]
    (3.4-2,1.4) -- (2.7-2,.7);
 \path [draw,decorate, decoration=snake]
 (2.55-2,.55) -- (2-2,0);
  \path [draw, opacity = 0,postaction={decorate},
        decoration={markings,mark=at position .65
        with {\arrow[opacity = 1,scale=1.7,>=latex]{>}}}]
    (3.4-2,1.4) -- (2-2,0);  \draw (3.14-2,1.4) node[anchor = south] {$B_{++}$};
  \path [draw,decorate, decoration=snake]
    (3.4-2,-1.4) -- (2.7-2,-.7);
 \path [draw,decorate, decoration=snake]
 (2.55-2,-.55) -- (2-2,0);
  \path [draw,opacity = 0, postaction={decorate},
        decoration={markings,mark=at position .65
        with {\arrow[opacity = 1,scale=1.7,>=latex]{>}}}]
    (3.4-2,-1.4) -- (2-2,0); \draw (3.14-2,-1.4) node[anchor = north] {$B_{++}$};
\end{tikzpicture}
\begin{tikzpicture}[thick, scale = .85]
  \draw (-1,0) node[anchor = south] {$B_{--}$};
  \path [draw,decorate, decoration=snake]
    (-4,0) -- (-3,0);
 \path [draw,decorate, decoration=snake]
 (-2.8,0) -- (-2,0);
  \path [draw, opacity = 0,postaction={decorate},
        decoration={markings,mark=at position .65
        with {\arrow[opacity = 1,scale=1.7,>=latex]{>}}}]
    (-4,0) -- (-2,0);  \draw (-4,0) node[anchor = east] {$B_{++}$};
  \path [draw,decorate, decoration=snake]
    (-3.4,1.4) -- (-2.7,.7);
 \path [draw,decorate, decoration=snake]
 (-2.55,.55) -- (-2,0);
  \path [draw, opacity = 0,postaction={decorate},
        decoration={markings,mark=at position .65
        with {\arrow[opacity = 1,scale=1.7,>=latex]{>}}}]
    (-3.4,1.4) -- (-2,0); \draw (-3.14,1.4) node[anchor = south] {$B_{++}$};
  \path [draw,decorate, decoration=snake]
    (-3.4,-1.4) -- (-2.7,-.7);
 \path [draw,decorate, decoration=snake]
 (-2.55,-.55) -- (-2,0);
  \path [draw, opacity = 0,postaction={decorate},
        decoration={markings,mark=at position .65
        with {\arrow[opacity = 1,scale=1.7,>=latex]{>}}}]
    (-3.4,-1.4) -- (-2,0);  \draw (-3.14,-1.4) node[anchor = north] {$B_{++}$};
  \path [draw,decorate, decoration=snake]
    (-2,0) -- (-1,0);
 \path [draw,decorate, decoration=snake]
 (-.8,0) -- (0,0);
  \path [draw,opacity = 0, postaction={decorate},
        decoration={markings,mark=at position .60
        with {\arrow[opacity = 1,scale=1.7,>=latex]{>}}}]
    (-2,0) -- (0,0);
  \path [draw,decorate, decoration=snake]
    (2,0) -- (1,0);
 \path [draw,decorate, decoration=snake]
 (2.8-2,0) -- (2-2,0);
  \path [draw,opacity = 0, postaction={decorate},
        decoration={markings,mark=at position .65
        with {\arrow[opacity = 1,scale=1.7,>=latex]{>}}}]
    (4-2,0) -- (2-2,0);   \draw (4-2,0) node[anchor = west] {$B_{++}$};
  \path [draw,decorate, decoration=snake]
    (3.4-2,1.4) -- (2.7-2,.7);
 \path [draw,decorate, decoration=snake]
 (2.55-2,.55) -- (2-2,0);
  \path [draw, opacity = 0,postaction={decorate},
        decoration={markings,mark=at position .65
        with {\arrow[opacity = 1,scale=1.7,>=latex]{>}}}]
    (3.4-2,1.4) -- (2-2,0);  \draw (3.14-2,1.4) node[anchor = south] {$B_{++}$};
  \path [draw,decorate, decoration=snake]
    (3.4-2,-1.4) -- (2.7-2,-.7);
 \path [draw,decorate, decoration=snake]
 (2.55-2,-.55) -- (2-2,0);
  \path [draw, opacity = 0,postaction={decorate},
        decoration={markings,mark=at position .65
        with {\arrow[opacity = 1,scale=1.7,>=latex]{>}}}]
    (3.4-2,-1.4) -- (2-2,0); \draw (3.14-2,-1.4) node[anchor = north] {$B_{++}$};
\end{tikzpicture}
\end{center}
\begin{center}
\begin{tikzpicture}[thick, scale = .85]
  \draw (-1,0) node[anchor = south] {$B_{+-}$};
  \path [draw,decorate, decoration=snake]
    (-4,0) -- (-3,0);
 \path [draw,decorate, decoration=snake]
 (-2.8,0) -- (-2,0);
  \path [draw, opacity = 0,postaction={decorate},
        decoration={markings,mark=at position .65
        with {\arrow[opacity = 1,scale=1.7,>=latex]{>}}}]
    (-4,0) -- (-2,0);  \draw (-4,0) node[anchor = east] {$B_{++}$};
  \path [draw,decorate, decoration=snake]
    (-3.4,1.4) -- (-2.7,.7);
 \path [draw,decorate, decoration=snake]
 (-2.55,.55) -- (-2,0);
  \path [draw,opacity = 0, postaction={decorate},
        decoration={markings,mark=at position .65
        with {\arrow[opacity = 1,scale=1.7,>=latex]{>}}}]
    (-3.4,1.4) -- (-2,0); \draw (-3.14,1.4) node[anchor = south] {$B_{++}$};
  \path [draw,decorate, decoration=snake]
    (-3.4,-1.4) -- (-2.7,-.7);
 \path [draw,decorate, decoration=snake]
 (-2.55,-.55) -- (-2,0);
  \path [draw, opacity = 0,postaction={decorate},
        decoration={markings,mark=at position .65
        with {\arrow[opacity = 1,scale=1.7,>=latex]{>}}}]
    (-3.4,-1.4) -- (-2,0);  \draw (-3.14,-1.4) node[anchor = north] {$B_{++}$};
  \path [draw,decorate, decoration=snake]
    (-2,0) -- (0,0);
  \path [draw,decorate, decoration=snake]
    (2,0) -- (1,0);
 \path [draw,decorate, decoration=snake]
 (2.8-2,0) -- (2-2,0);
  \path [draw, opacity = 0,postaction={decorate},
        decoration={markings,mark=at position .65
        with {\arrow[opacity = 1,scale=1.7,>=latex]{>}}}]
    (4-2,0) -- (2-2,0);   \draw (4-2,0) node[anchor = west] {$B_{++}$};
  \path [draw,decorate, decoration=snake]
    (3.4-2,1.4) -- (2.7-2,.7);
 \path [draw,decorate, decoration=snake]
 (2.55-2,.55) -- (2-2,0);
  \path [draw,opacity = 0, postaction={decorate},
        decoration={markings,mark=at position .65
        with {\arrow[opacity = 1,scale=1.7,>=latex]{>}}}]
    (3.4-2,1.4) -- (2-2,0);  \draw (3.14-2,1.4) node[anchor = south] {$B_{++}$};
  \path [draw,decorate, decoration=snake]
    (3.4-2,-1.4) -- (2.7-2,-.7);
 \path [draw,decorate, decoration=snake]
 (2.55-2,-.55) -- (2-2,0);
  \path [draw, opacity = 0,postaction={decorate},
        decoration={markings,mark=at position .65
        with {\arrow[opacity = 1,scale=1.7,>=latex]{>}}}]
    (3.4-2,-1.4) -- (2-2,0); \draw (3.14-2,-1.4) node[anchor = north] {$B_{++}$};
\end{tikzpicture}
\end{center}
\caption{Exchange diagrams contributing to the 6 $B_{++}$ amplitude.
The internal line may be any of the three states, and we sum over all
the factorization channels as well. It is important to
note that these diagrams do not come directly from Feynman rules as
there is no covariant action available for the M5 theory;
instead, they represent the factorization of the amplitude at the poles
where $s_{ijk} \rightarrow 0$. \label{6BFig1}}
\end{figure}

The ansatz in Eq.~(\ref{eq:fact6B++}) is the simplest guess that has the
correct factorization and little-group properties, and it ends up being correct.
It is instructive to see how one arrives at this conclusion using Feynman diagrams
without recourse to an action. At the poles we can represent the six-point
amplitude in Eq.~(\ref{eq:fact6B++}) as a
sum of exchange diagrams that are the product of four-point amplitudes and an
internal propagator. These diagrams are shown in Fig.~(\ref{6BFig1}).

In evaluating these diagrams, one must sum over all exchange channels
as well as all fields allowed to propagate on the internal lines.
As we have explained, only
$B_{++}, B_{--}$, or $B_{+-} = B_{-+}$ can be exchanged. The pure
positive and negative helicity states are conjugates of each other,
and as with chiral fermions we use an arrow to distinguish them from
the neutral helicity.

The sum of such diagrams must be invariant under the little group of
the internal particle, and this ends up being the case due to a subtlety
in the spinor-helicity formalism. This ``glitch" in the spinor-helicity
formalism as discussed for $6$d SYM in \cite{Bern:2010qa} is that the
spinors cannot distinguish particles and antiparticles, which causes
issues for diagrams with fermions. A new feature of $6$d chiral self-dual
tensors is that the tensor field itself has this issue with the $B_{++}$
and $B_{--}$ polarizations. The resolution, as outlined in \cite{Bern:2010qa}, is
to add extra factors of $i$ to the spinor-helicity variables when we flip
the sign of the momentum for either of these fields:
\begin{equation}
\l^A_a(-p) = i \l^A_a(p)
\end{equation}
so that the momentum is properly
\begin{equation}
\l^A_a(-p) \l^{Ba}(-p) = -p^{AB} \, .
\end{equation}
This introduces additional minus signs for a four-particle amplitude of the form
\begin{equation}
A(B_{++}(+p_1),B_{++}(+p_2),B_{++}(+p_3), B_{\pm \pm}(-p))
=  \langle 1_{+} \, 2_{+} \, 3_{+}\, i  \l_{\pm} \rangle
\langle 1_{+} \, 2_{+} \, 3_{+} \,  i \l_{\pm}\rangle \, .
\end{equation}
Applying this recipe to the exchange diagrams of Fig.~(\ref{6BFig1}),
one is led directly to Eq.~(\ref{eq:fact6B++}), which does not depend on the
little-group structure of the internal line, as it should be.

\begin{figure}[h]
\begin{center}
\begin{tikzpicture}[thick, scale = 1.2]
  \path [draw,decorate, decoration=snake]
    (-4,0) -- (-3,0);
 \path [draw,decorate, decoration=snake]
 (-2.8,0) -- (-2,0);
  \path [draw, opacity = 0, postaction={decorate},
        decoration={markings,mark=at position .62
        with {\arrow[opacity = 1, scale=1.8,>=latex]{>}}}]
    (-4,0) -- (-2,0);  \draw (-4,0) node[anchor = east] {$B_{\pm \pm}$};
  \path [draw,decorate, decoration=snake]
    (-3.4,1.4) -- (-2.7,.7);
 \path [draw,decorate, decoration=snake]
 (-2.55,.55) -- (-2,0);
  \path [draw,opacity = 0, postaction={decorate},
        decoration={markings,mark=at position .62
        with {\arrow[opacity = 1,scale=1.8,>=latex]{>}}}]
    (-3.4,1.4) -- (-2,0); \draw (-3.14,1.4) node[anchor = south] {$B_{\pm \pm}$};
  \path [draw,decorate, decoration=snake]
    (-3.4,-1.4) -- (-2.7,-.7);
 \path [draw,decorate, decoration=snake]
 (-2.55,-.55) -- (-2,0);
  \path [draw, opacity = 0, postaction={decorate},
        decoration={markings,mark=at position .62
        with {\arrow[opacity = 1,scale=1.8,>=latex]{>}}}]
    (-3.4,-1.4) -- (-2,0);  \draw (-3.14,-1.4) node[anchor = north] {$B_{\pm \pm}$};

  \path [draw,decorate, decoration=snake]
    (2-2,0) -- (1-2,0);
 \path [draw,decorate, decoration=snake]
 (2.8-2-2,0) -- (2-2-2,0);
  \path [draw, opacity = 0,postaction={decorate},
        decoration={markings,mark=at position .62
        with {\arrow[opacity = 1,scale=1.8,>=latex]{>}}}]
    (4-2-2,0) -- (2-2-2,0);   \draw (4-2-2,0) node[anchor = west] {$B_{\pm \pm}$};
  \path [draw,decorate, decoration=snake]
    (3.4-2-2,1.4) -- (2.7-2-2,.7);
 \path [draw,decorate, decoration=snake]
 (2.55-2-2,.55) -- (2-2-2,0);
  \path [draw, opacity = 0,postaction={decorate},
        decoration={markings,mark=at position .62
        with {\arrow[opacity = 1,scale=1.8,>=latex]{>}}}]
    (3.4-2-2,1.4) -- (2-2-2,0);  \draw (3.14-2-2,1.4) node[anchor = south] {$B_{\pm \pm}$};
  \path [draw,decorate, decoration=snake]
    (3.4-2-2,-1.4) -- (2.7-2-2,-.7);
 \path [draw,decorate, decoration=snake]
 (2.55-2-2,-.55) -- (2-2-2,0);
  \path [draw,opacity = 0, postaction={decorate},
        decoration={markings,mark=at position .62
        with {\arrow[opacity = 1,scale=1.8,>=latex]{>}}}]
    (3.4-2-2,-1.4) -- (2-2-2,0); \draw (3.14-2-2,-1.4) node[anchor = north] {$B_{\pm \pm}$};
\end{tikzpicture}
\caption{Diagrammatic expression of the local term for a six-particle amplitude.
In the example where all external particles are $B_{ab}$, this
local term vanishes, and the exchange diagrams are the only contribution to the
 total amplitude. \label{6BLocFig1}}
\end{center}
\end{figure}
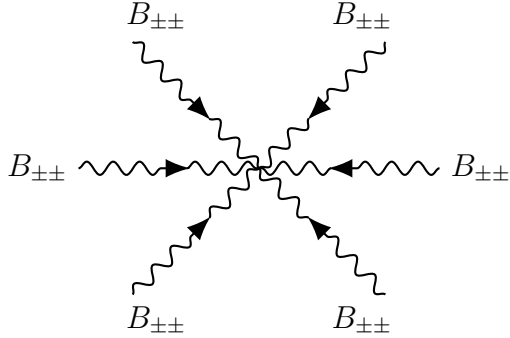

Of course, Eq.~(\ref{eq:fact6B++}) might not be the final
result, since it could differ from the correct answer by terms that have
no poles (thought of as a $6$-particle contact interaction, depicted in
Fig.~\ref{6BLocFig1}).
The only local term allowed by power counting and little-group constraints is
\bea
\langle 1_+\,2_+\,3_+\,4_+ \rangle \langle 1_+\,2_+\,5_+\,6_+ \rangle
\langle 3_+\,4_+\,5_+\,6_+ \rangle + \mathcal{P}_6 \, .
\eea
It turns out that this local term vanishes identically after summing
over the permutations. Thus, we claim that Eq.~(\ref{eq:fact6B++}) is the complete
result for the amplitude of six $B_{++}$'s. Indeed, we find perfect agreement
by comparing Eq.~(\ref{eq:fact6B++}) numerically with the
general integral formula Eq.~(\ref{eq:(2,0)}).

One can perform a similar analysis
for more general amplitudes of self-dual $B$ fields. In all cases we find
that the result takes a form similar to Eq.~(\ref{eq:fact6B++}),
\bea \label{eq:6Bab}
&& A(B_{a_1 b_1},B_{a_2 b_2},B_{a_3 b_3},B_{a_4 b_4},B_{a_5 b_5},B_{a_6 b_6} ) \\
\nonumber
&=& \!  {1 \over s_{123} }  \left( \sum^3_{i=1}  \langle 1_{a_1} \, 2_{a_2} \, 3_{a_3}\,
i_{a} \rangle \langle i^{a} \, 4_{a_4} \, 5_{a_5} \, 6_{a_6}  \rangle  \right)
 \left( \sum^3_{j=1}  \langle 1_{b_1} \, 2_{b_2} \, 3_{b_3}\,
j_{b} \rangle \langle j^{b} \, 4_{b_4} \, 5_{b_5} \, 6_{b_6}  \rangle  \right) + \mathcal{P}_6 \,.
\eea
The symbol $\mathcal{P}_6 $ represents the symmetrization of the little-group
indices $a_i, b_i$ for all $i=1, 2, \ldots, 6$, and the summation over all other
factorization channels.

It is instructive to consider the reduction of these results to the D3 theory.
$B_{++}$ and $B_{--}$ reduce to positive- and
negative-helicity photons $\gamma_+$ and $\gamma_-$ in 4D, while $B_{+-}$ reduces
to a scalar. If we restrict to external $B_{++}$ and $B_{--}$ only, then
$A(B_{++},B_{++},B_{++},B_{--},B_{--},B_{--} )$ is the only amplitude
that is non-vanishing after dimensional reduction to 4D. This is consistent
with the claim that the amplitudes of the D3 theory are helicity conserving.
The helicity-conserving amplitude obtained in this way is
\bea \label{eq:6gamma}
A(\gamma_{+}, \gamma_{+}, \gamma_{+}, \gamma_{-}, \gamma_{-}, \gamma_{-} )  =
{1 \over s_{124} }  [1\,2]^2 \langle 5\,6\rangle^2  \langle 4| 1+2 | 3]^2 + \mathcal{P}_6 \, ,
\eea
where $\langle 4 | 1+2 | 3] := \langle 4 \, 1 \rangle [1 \, 3] +
\langle 4 \, 2 \rangle [2 \, 3]$, and the permutations
$\mathcal{P}_6$  sum over $\gamma_+$'s and $\gamma_-$'s, respectively. The amplitude in Eq.~(\ref{eq:6gamma}) obtained by dimensional reduction agrees
with the amplitude for six photons in the D3 theory computed for instance
in \cite{Boels:2008fc}. We also find that Eq.~(\ref{eq:6Bab}) for the case of six $B_{+-}$'s
reduces correctly to the amplitude for six identical scalars,
\bea \label{eq:6scalar}
\left( {(s_{12}^2 + s_{13}^2 + s_{23}^2 ) (s_{45}^2 + s_{46}^2 + s_{56}^2 )
 \over s_{123}} + \ldots \right)- {1 \over 2 } \left(  s_{123}^3 + \ldots \right) \, ,
\eea
where the ellipsis in the parentheses denote summation over all factorization channels,
as well as all other independent $s_{ijk}$'s for six-particle kinematics. It is straightforward to verify that this
is the unique amplitude for identical scalars determined by the soft theorem.

One can also consider amplitudes of other particles. For instance, we find that the six-particle amplitude of $\phi_{IJ}$ in the spectrum
Eq.~(\ref{tildephi}) agrees with the result in Eq.~(\ref{eq:6scalar}). Also, the
amplitude for six fermions can be expressed as
\bea \label{eq:6fermions}
A(\psi^I_+, \psi^I_+, \psi^I_+, \tilde{ \psi}^I_-,  \tilde{ \psi}^I_-, \tilde{ \psi}^I_- ) =
 A^{(6)}_f  - {1 \over 12} A^{(6)}_c \, ,
\eea
where the factorization term $A^{(6)}_f$ and the local term $A^{(6)}_c$ are given by
\bea
A^{(6)}_f &=& {1 \over s_{124}} \left( \sum_{i=1,2,4}
\langle 1_+ 4_+ 4_- i_a \rangle \langle i^a 5_- 6_+ 6_- \rangle  \right)
\left( \sum_{j=1,2,4}  \langle 2_+ 4_- j_+ j_- \rangle
\langle 3_+ 6_- 5_+ 5_- \rangle \right) + \mathcal{P}_6 \nonumber \\
A^{(6)}_c &=& \langle 1_+ 2_+ 3_+ 4_- \rangle
\langle 5_- 6_- 4_+ 4_- \rangle \langle 5_+ 5_- 6_+ 6_- \rangle   + \mathcal{P}_6 \, ,
\eea
where $ \mathcal{P}_6 $ denotes summing over anti-symmetrizations among all $\psi$ and
$\tilde{ \psi}$ particles separately. Reduced to 4D, the six-fermion amplitude
gives that of Volkov-Akulov theory computed in \cite{Luo:2015tat}.

Let us now consider the amplitudes with eight $B$ particles. For simplicity,
we only consider the amplitude with eight $B_{++}$'s and the amplitude
with seven $B_{++}$'s and one $B_{--}$. As we will see, they take a very
similar form. The strategy is the same as in the case of six-particle amplitudes.
We write down an ansatz that includes factorization parts and local terms,
and then compare the ansatz against the general formula to determine the unknown
coefficients. As before, one can arrive at the ansatz for exchange diagrams
by summing diagrams that are products of amplitudes with fewer particles.
Unlike the case of six $B$ particles, we find
that in general there are contributions from local terms. Explicitly, we find
\bea
 A(B_{++},B_{++},B_{++},B_{++},B_{++},B_{++},B_{++},B_{aa}) = A^{(8)}_f  - 2 A^{(8)}_c \, ,
\eea
where the little-group index $a$ can be $+$ or $-$ depending on whether
$B_{aa}$ is $B_{++}$ or $B_{--}$, and $A^{(8)}_f, A^{(8)}_c$ are the factorization part
 and the local term, respectively. $A^{(8)}_f$ and $A^{(8)}_c$ are given by
\bea
A^{(8)}_f &=& {1 \over s_{123} \, s_{678} }
\left( \sum^3_{i=1} \sum^8_{j=6}  \langle 1_+ \, 2_+ \, 3_+ \,
i_b \rangle \langle i^b \, 4_+\, 5_+ \, j_c \rangle \langle j^c \, 6_+\,
7_+\, 8_a \rangle  \right)^2  \\ \nonumber
&+&
{1 \over s_{123} \, s_{567} }  \left(  \sum^3_{i=1} \sum^7_{j=5} \langle 1_+ \, 2_+ \, 3_+ \,
i_b \rangle \langle i^b \, 4_+\, 8_a \, j_c \rangle \langle j^c \,
5_+ \, 6_+\, 7_+  \rangle  \right)^2
+ \mathcal{P}_8 \, , \\ \nonumber
A^{(8)}_c &=& \left(  \langle 1_+ \, 2_+ \, 3_+ \, 4_+ \rangle
\langle 5_+ \, 6_+\, 7_+ \, 8_a \rangle  \right)^2 + \mathcal{P}_8 \,,
\eea
where $\mathcal{P}_8$ denotes the summation over independent permutations.

As mentioned previously, the amplitudes involving scalars in the M5 theory
should satisfy soft theorems. Some such amplitudes are completely fixed by
the soft theorems. Therefore they can also be computed in a completely different
way via on-shell recursion relations \cite{Cheung:2015ota}. We have
verified that the results agree perfectly with what is obtained from the
proposed formula, Eq.~(\ref{eq:(2,0)}), for such amplitudes containing up to
eight particles.

\section{$\mathbf{n}$-particle amplitudes of the D5 theory} \label{sec:nptD5}

This section describes the tree-level S matrix for the theory of a
single probe D5-brane with 6D $\mathcal{N}=(1,1)$ supersymmetry.
The general formula we propose for the D5 theory takes a form similar
to that of the M5 theory, which we discussed in the previous section.
In particular, the formula contains the same factors of ${\det}' S_n$
and $U(\rho, \s)$,
\begin{eqnarray}
 {\cal A}_n = \int \frac{d^n \s \, d \mathcal{M} }{ {\rm Vol} (G) }  \,
\D_B(p, \r) \, \D_F (q, \r, \chi) \, \hat{\D}_F (\hat{q}, \hat{\r}, \hat{\chi} )\,
{\rm det}^{\prime} S_n \,  {R(\r) \over V^2(\s)} \, ,
\end{eqnarray}
or equivalently
\begin{eqnarray} \label{eq:(1,1)}
 {\cal A}_n = \int \frac{d^n \s \, d \mathcal{M} }{ {\rm Vol} (G) }  \,
\D_B(p, \r) \, \D_F (q, \r, \chi) \, \hat{\D}_F (\hat{q}, \hat{\r}, \hat{\chi} )\,
{({\rm Pf}^{\prime} S_n)^3 \over V(\s) } \, .
\end{eqnarray}
The bosonic delta functions are the same as those in the M5 theory
\bea \label{eq:DeltaB}
\D_B (p,\r) = \prod^n_{i=1} \d^6 \left(p^{AB}_i
- \frac{\r^{A}_a (\s_i) \r^{Ba} (\s_i)}{ P_i(\s)} \right),
\eea
but now there are two kinds of fermionic delta functions due to
$(1,1)$ supersymmetry,
\bea
\D_F (q,\r, \chi) &=& \prod^n_{i=1} \d^{4}\left(q^{A}_i
- \frac{ \r^{A}_a(\s_i) \chi^{a} (\s_i)} { P_i(\s) } \right) \, , \\ \nonumber
\hat{\D}_F (\hat{q}, \hat{\r}, \hat{\chi} )
&=& \prod^n_{i=1} \d^{4}\left(\hat{q}_{iA}
- \frac{ \hat{\r}_{A \hat a}(\s_i) \hat{\chi}^{\hat a} (\s_i)} { P_i(\s) } \right) \, .
\eea
The measure is given by
\bea
d \mathcal{M} = \prod^{d}_{m=0} d^8 \r^{Aa}_m  d^{2} \chi^{b}_m
d^{2} \hat{\chi}^{\hat{b}}_m  \, .
\eea
As before, $d = {n \over 2} - 1$. Note that this integration measure
does not include $d^8\hat{\r}_{mA\hat{a}}$,
even though $\hat{\r}_{mA\hat{a}}$ do appear explicitly in the formula.
The prescription is that the $\hat{\r}_{mA\hat{a}}$ are fixed by the
constraint of the conjugate of $\D_B$ in Eq.~(\ref{eq:DeltaB}), namely,
\bea
{\hat p}_{i AB} - { \hat{\r}_{A \hat{a}}(\sigma_i)  \hat{\r}^{ \hat{a}}_B(\sigma_i)
\over P_i(\s) } =0 \, , \quad
{\rm for} \quad i=1, 2, \ldots, n \, .
\eea
This constraint does not appear explicitly in the general formula
Eq.(\ref{eq:(1,1)}), but we impose it implicitly. To fully fix $\hat{\r}_{mA\hat{a}}$,
we also use the second $SU(2)$ factor of the little-group symmetry
to fix three of the $\hat{\r}_{mA\hat{a}}$ coordinates.
Since Eq.~(\ref{eq:(1,1)}) takes a form that is very similar to Eq.~(\ref{eq:(2,0)})
for the M5 theory, with a simple change to half of the fermionic delta
functions due to the change of chirality for half of the supersymmetry, it is
straightforward to show that Eq.~(\ref{eq:(1,1)}) also has all of the
required properties, such as correct factorizations, soft theorems, and
reduction to the D3 theory. Thus, we will not repeat the analysis and
discussion here.

For computing scattering amplitudes from Eq.~(\ref{eq:(1,1)}), as in the case of the
D3 theory and M5 theory, we again should pull out the bosonic and fermionic
``wave functions" first. For the D5 theory, they are given by
\bea
\mathcal{A}_n = \left(\prod^n_{i=1} \d(p^2_i ) \,
\d^{2}\left(\hat\l_{iA\hat a}\, q_i^{A}\right)
\d^{2}\left(\l^{B}_{i b}\, \hat q_{iB}\right) \right) \, A_n \, .
\eea
The same as in the case of M5 brane, here the constraints $\hat\l_{iA\hat a}\, q_i^{A}=0$ and $\l^{B}_{i b}\, \hat q_{iB}=0$ allow us to express the supercharges in the amplitude $A_n$ as $q_i^{A}= \l^A_{a} \eta^{a}$ and $\hat q_i^{A}= \hat \l^A_{\hat a} \hat \eta^{\hat a}$.
We have checked explicitly that $A_n$, as defined here, produces the
correct fully supersymmetric four-particle amplitudes, as well as many examples of six- and
eight-particle amplitudes in the D5 theory. Here we list the analytical results for some of these amplitudes.

The amplitude for six photons with the same helicity, given by $A_{1\hat{1}}$, is
\bea
&& A(A_{1\hat{1}}, A_{1\hat{1}}, A_{1\hat{1}}, A_{1\hat{1}},
A_{1\hat{1}}, A_{1\hat{1}}) \\  \nonumber
&=&
{1 \over s_{123} }  \left( \sum^3_{i=1} \langle 1_{1} \, 2_{1} \, 3_{1}\,
i_{a} \rangle \langle i^{a} \, 4_{1} \, 5_{1} \, 6_{1}  \rangle  \right)
 \left( \sum^3_{j=1} [1_{\hat{1}} \, 2_{\hat{1}} \, 3_{\hat{1}}\,
\hat{j}_{\hat{a} } ] [ \hat{j}^{\hat{a} } \, 4_{\hat{1}} \, 5_{\hat{1}}\,
6_{\hat{1}} ] \right) + \mathcal{P}_6\,.
\eea
There are similar expressions for other choices of helicities of $A_{a \hat{a}}$.
We have verified that these results agree with the amplitudes obtained directly
from the Born--Infeld action. One can also consider the amplitude of eight
$A_{1\hat{1}}$'s, which takes the form
\bea
 A(A_{1\hat{1}}, A_{1\hat{1}}, A_{1\hat{1}}, A_{1\hat{1}}, A_{1\hat{1}},
 A_{1\hat{1}}, A_{1\hat{1}}, A_{1\hat{1}}) = A_f  - 2 A_c \, .
\eea
The factorization term $A_f$ and the local term $A_c$ are given by
\bea
A_f &=& {1 \over s_{123} \, s_{678} }
\left( \sum^3_{i=1} \sum^8_{j=6}  \langle 1_1 \, 2_1 \, 3_1 \,
i_a \rangle \langle i^a \, 4_1\, 5_1 \, j_b \rangle \langle j^b \, 6_1\,
7_1\, 8_1 \rangle  \right)
\\ \nonumber
&& \times  \left( \sum^3_{i=1} \sum^8_{j=6} [ 1_{\hat{1}} \, 2_{\hat{1}} \, 3_{\hat{1}} \,
\hat{i}_{\hat{a}} ] [\hat{i}^{\hat{a}} \, 4_{\hat{1}} \, 5_{\hat{1}} \,
\hat{j}_{\hat{b}} ] [ \hat{j}^{\hat{b}} \, 6_{\hat{1}}\,
7_{\hat{1}} \, 8_{\hat{1}} ]  \right)  +  \mathcal{P}_8 \, ,  \nonumber  \\
A_c &=& \left(  \langle 1_1 \, 2_1 \, 3_1 \, 4_1 \rangle
\langle 5_1 \, 6_1\, 7_1 \, 8_1 \rangle  \right)\left(  [ 1_{\hat{1}}  \,
2_{\hat{1}}  \, 3_{\hat{1}}  \, 4_{\hat{1}} ] [ 5_{\hat{1}}  \, 6_{\hat{1}} \,
7_{\hat{1}}  \, 8_{\hat{1}}  ] \right) + \mathcal{P}_8 \, .
\eea
These results for photon amplitudes in the D5 theory take a form that is
very similar to the amplitudes of $B_{ab}$  particles in the M5 theory.
They are related to each other by replacing the anti-chiral $\hat{\l}_{\hat{a}}$
by the chiral one $\l_a$.

The similarity between D5 and M5 amplitudes in the above explicit examples,
and more generally the formulas Eq.~(\ref{eq:(1,1)}) and Eq.~(\ref{eq:(2,0)}),
may be surprising, especially given the fact that the classical action
for the M5 theory is more subtle to write down than the one for the D5 theory.
However, one should note that the entire difference between the four-particle
amplitudes, which are completely fixed by the symmetries and power counting
in the D5 theory and the M5 theory, is just a simple modification of the
fermionic delta functions. Since both theories reduce to the same 4D amplitudes,
the similarity is really not so surprising.
The complication of writing the classical M5 action caused by the self-duality
of $B$ field is avoided by considering only the on-shell degrees of freedom
for the S matrix using the spinor-helicity formalism.

\section{Conclusion} \label{sec:conclusion}

This paper has proposed general formulas for $n$-particle on-shell tree-level
scattering amplitudes for three theories: the D3 and D5 theories of type IIB
superstring theory and, especially, the M5 theory of 11D M-theory.
The scattering amplitudes of the M5 theory-- even its bosonic truncation --
have been studied little in the previous literature. In each of these theories
$n$ is required to be even, and the amplitudes
take similar forms, expressed as integrals over rational constraints, built from degree $d = \frac{n}{2}-1$ polynomials.
The integrand contains a new mathematical ingredient, a generalization of resultant
(denoted $R(\r)$ in the text), which is equal to the product of ${\rm Pf}'S_n$ and
the Vandermonde determinant $V(\s)$ on the support of the rational constraints.

The three theories are related to one another in various ways. For instance,
dimensional reduction of each of the 6D $n$-particle amplitudes, which pertain
to the D5 and M5 theories, reduces to the same 4D $n$-particle amplitude,
which pertains to the D3 theory. The function $U(\r,\s)$ in the 6D integrands
cancels the Jacobian factors arising from the dimensional reduction.
As we explained, one consequence
is that the R symmetry of the D3 theory is $SU(4) \times U(1)$. The $U(1)$
factor implies that the D3 amplitudes are helicity conserving.
Interestingly, the formulas for the M5 and D5 amplitudes only differ by
a simple modification of the fermionic delta functions that accounts
for the chirality difference between $(2,0)$ supersymmetry and $(1,1)$
supersymmetry.

We have also checked various general properties such as $SL(2,\IC)$ modular
symmetry, R symmetries, factorization properties, and soft limits.
We have further tested the formulas by explicitly computing amplitudes
that are fixed by the soft theorems, up to $8$ particles. Using the general formulas,
compact analytic expressions for six- and eight-particle amplitudes
of self-dual $B$ particles of the M5 theory for certain choices of the
little-group indices were obtained.
% These amplitudes have not been studied previously in the literature.

Our formulas for scattering amplitudes are similar to those for the
twistor-string formulation of 4D $\mathcal{N}=4$ super Yang--Mills amplitudes
in Witten's twistor-string paper \cite{Witten:2003nn}. Those amplitudes,
and their generalizations, see e.g. \cite{Skinner:2013xp}\cite{Mason:2013sva}, are understood in terms of
two-dimensional world-sheet twistor-string theories. It would
be interesting to explore whether there exists a similar twistor-string theory
for the M5 theory. Such an underlying theory ought to generate the M5
amplitudes in Eq.~(\ref{eq:(2,0)}) directly.  The fact that a
twistor-string-like formulation of the tree-level S-matrix of the M5 theory
does exist already points to some deep structures of the theory.

Finally, we note that the rational constraints in 6D consist of a single
sector of solutions to the scattering equations, which utilizes all $(n-3)!$
solutions of the arbitrary-dimensional scattering equations. We do not have
a general proof of these assertions, but they
have been checked explicitly for the cases $n=4,6,8$.  It would be nice to
prove (or disprove) them and to understand
better this general feature of the 6D rational constraints.
Upon dimensional reduction to the D3 theory, many of these solutions
vanish leaving only contributions from those that correspond to the middle (helicity conserving)
sector in 4D. It would also be interesting to study the rational constraints in dimensions
greater than six, such as 10D or 11D, and to apply them to the D9-brane
theory, as well as the various gauge and gravity theories in those dimensions.

\section*{Acknowledgments}

We would like to thank Clifford Cheung, Song He, Yu-tin Huang,
and Ellis Yuan for helpful discussions.
This work was supported in part by the Walter Burke Institute for Theoretical Physics at
Caltech and by U.S. DOE Grant DE-SC0011632. The work of CW is supported in part by a DOE Early
Career Award under Grant No. DE-SC0010255 and by the NSF under Grant No. NSF PHY-1125915.
Part of this work was performed by JHS at the Aspen Center for Physics,
which is supported by National Science Foundation grant PHY-1607611.

\newpage

\appendix

\section{Further technical details} \label{appendix:Jacobi}

\subsection{D3 theory} \label{appendix:D3Jacobi}

The goal here is to show that the $n$-particle amplitude ${\cal A}_n$ in Eq.~(\ref{nptamps})
contains the delta functions exhibited in the formula
\bea \label{calAn1}
\mathcal{A}_n = \left(\prod^n_{i=1} \d(p^2_i ) \d^{2}(  \langle  \l_i \, q_{i}^I  \rangle )
\d^{2}( [ \tilde{\l}_i \, \hat{q}_{i}^{\hat I}]  )\right) \, A_n \, ,
\eea
as well as additional momentum-conservation and supercharge-conservation delta functions, which
are included in $A_n$. We also wish to
compute the Jacobian $J_B$ that arises from extracting the
momentum-conservation and mass-shell delta functions
%$\d^4(\sum^n_{i=1} p_i) \prod^n_{i=1} \d(p_i^2)$
from the bosonic delta functions,
\bea
\D_B = \prod^n_{i=1} \d^4 \left(p^{\a \dot{\a} }_i -
{\r^{\a} (\s_i) \tilde{\r}^{\dot\a} (\s_i) \over P_i(\s) } \right),
\eea
appearing in the formula for the D3 $n$-particle amplitude ${\cal A}_n$.

It is clear that these delta functions imply masslessness, since they constrain $p_i^{\a\dot\a}$
to take a factorized (rank one) form. It is less obvious that they imply momentum
conservation. The delta functions imply that
\bea
\sum_{i=1}^n p_i^{\a \dot\a} = \sum_{i=1}^n \frac{1}{P_i(\s)}
\sum_{m,m' =0}^d \r_m^\a \tilde\r_{m'}^{\dot\a}
\s_i^{m+m'}.
\eea
This will vanish provided that
\bea \label{vanish}
\sum_{i=1}^n \frac{\s_i^m}{P_i(\s)} =0 \quad {\rm for} \quad m=0,1,2,\ldots, n-2,
\eea
since $2d=n-2$.
To prove that this is the case, let us introduce the Vandermonde determinant
\bea
V(\s) = \prod_{i>j} \s_{ij}.
\eea
Recalling the definition $P_i(\s) = \prod_{j\neq i} \s_{ij}$, we note that
\bea
V_i(\s) = \frac{V(\s)}{P_i(\s)} = (-1)^i \prod_{j>k;\, j,k\neq i} \s_{jk}.
\eea
Then, momentum conservation is a consequence of the following theorem:
\bea
W_m(\s) = \sum_{i=1}^n \s_i^m V_i(\s) =0 \quad {\rm for} \quad m=0,1,\ldots , n-2.
\eea
This is proved by noting that $W_m$ is a symmetric polynomial of the $n$
$\s$ variables whose degree
does not exceed $n-2$ in any of them. Therefore, it vanishes if there
are $n-1$ zeros in each of the coordinates. This is achieved if $W_m$ vanishes
when any pair of variables are equal. For example, when $\s_1 =\s_2$ only
$V_1$ and $V_2$ are nonvanishing. But then $W_m(\s) = \s_1^m (V_1 + V_2)$.
This vanishes because $V_1 +V_2 = 0$ when $\s_1 =\s_2$. This completes the proof
of momentum conservation.

We have seen that $n+4$ of the $4n$ delta functions in $\D_B$ account for the
mass-shell conditions and momentum conservation. The integrations over the $\r$
and $\tilde \r$ coordinates use up $2n-1$ more of the delta functions, leaving
$n-3$ to account for. The important fact is that the remaining delta functions
lead to the scattering equations
\bea
E_i = \sum_{j \neq i} \frac{p_i \cdot p_j}{\s_{ij}} =0, \quad i= 1,2,\ldots ,n
\eea
and the $n-3$ integrations over the $\s$ coordinates imply that one should
sum over the solutions of these equations. Only $n-3$ of
the scattering equations are linearly independent, since the mass-shell and
momentum-conservation conditions imply that
\bea
\sum _{i=1}^n E_i = \sum _{i=1}^n \s_i E_i = \sum _{i=1}^n \s_i^2 E_i =0.
\eea
Thus, there is just the right number of delta functions to account for the
scattering equations.
As discussed earlier, the scattering equations have $(n-3)!$ solutions,
but only $N_{dd}$ of them give nonzero contributions to the amplitudes.
These are the ones that are helicity conserving, as required by the $U(1)$
R symmetry.

Let us now verify that the delta functions in $\D_B$ actually do
imply the scattering equations. Substituting for $p_i \cdot p_j$ gives
\bea
E_i = \sum_{j\neq i} \sum_{mnm'n'=0}^d \frac{\langle \r_m \r_n \rangle
[\tilde\r_{m'} \tilde\r_{n'}] \s_i^{m+m'} \s_j^{n+n'}}{\s_{ij} P_i(\s) P_j(\s)}
\eea
However, $ \langle \r_m \r_n \rangle = - \langle \r_n \r_m \rangle$ and
$ \langle \tilde\r_{m'} \tilde\r_{n'} \rangle
= - \langle \tilde\r_{n'} \tilde\r_{m'} \rangle$.
Therefore we can replace $\s_i^m \s_j^n$ by
\bea
\half (\s_i^m \s_j^n - \s_j^m \s_i^n) =  \s_{ij} Q_{mn}(\s_i,\s_j)
\eea
where $Q_{mn}$ is a polynomial. It then follows that
\bea
E_i = \frac{1}{P_i(\s)} \sum_{j=1}^n \frac{\s_{ij} Q(\s_i, \s_j)}{P_j(\s)}
\eea
where
\bea
Q(\s_i,\s_j) = \sum_{mnm'n'}
 \langle \r_m \r_n \rangle [\tilde\r_{m'} \tilde\r_{n'}]
Q_{mn}(\s_i,\s_j) Q_{m'n'}(\s_i,\s_j).
\eea
Since $\s_{ij} Q(\s_i,\s_j)$ is a polynomial function of $\s_j$ of degree $n-3$, the
scattering equations $E_i =0$ follow as a consequence of Eq.~(\ref{vanish}).

The structure of the $4n$ delta functions in $\D_B$ ensures masslessness,
momentum conservation, and the scattering equations, which is a total of $2n+1$
conditions. They can be expressed as delta functions and used to rewrite
$\D_B$ as these $2n+1$ delta functions times $2n-1$ additional
delta functions and a Jacobian factor, which will be described later.
Given this, it is natural to examine next
what can be learned from the structure of the $8n$ fermionic delta functions
\bea
\D_F(q,\r,\chi) = \prod^n_{i=1} \d^{4}\left(q^{\a I}_i
-   { \r^{ \a } (\s_i) \chi^{I} (\s_i)  \over P_i(\s) }  \right)
\d^{4}\left( \hat{q}^{\dot\a \hat I}_i
-   { \tilde{\r}^{ \dot{\a} } (\s_i) {\hat\chi}^{\hat I} (\s_i)  \over P_i(\s) } \right).
\eea
First of all, the delta functions in $\D_F$ imply the conservation of eight supercharges:
\bea
\sum_{i=1}^n q^{\a I}_i = \sum_{i=1}^n \hat{q}^{\dot\a \hat I}_i =0.
\eea
This is proved by exactly the same reasoning that was used to establish momentum
conservation earlier in this appendix.
Note that these eight supercharges are mutually anticommuting,
as are the other eight, but there are nonzero anticommutators between the two sets.
The conservation of the second set of eight supercharges needs to be established
separately.

Next we wish to account for the factors $\prod_i \d^{2}(\langle \l_i \, q_{i}^I \rangle)
\d^{2}([ \tilde{\l}_i \, \hat{q}_{i}^{\hat I}] )$ in Eq.~(\ref{calAn1}). The first set
should derive from the first set of delta functions in $\D_F$ and the second set from
the second factor (by identical reasoning). It is important that the bosonic analysis
has already been completed, so that masslessness, \ie the presence of the
factors $\prod_i \d(p_i^2)$, can be invoked to justify writing
$p_i^{\a\dot\a} = \l_i^\a \tilde\l_i^{\dot\a}$. Therefore the
fermionic delta functions imply that $\langle \l_i \, q_{i}^I \rangle =
[ \tilde{\l}_i \, \hat{q}_{i}^{\hat I}] =0$. These relations are implemented by
the $4n$ fermionic delta functions exhibited in Eq.~(\ref{calAn1}).
They provide the justification for using the relations
\bea
q_i^{\a I} = \l_i^\a\, \eta_i^I \quad {\rm and} \quad  \hat{q}^{\dot\a \hat I}_i
= \tilde\l_i^{\dot\a}\, \hat\eta_{i}^{\hat I}
\eea
in the amplitude $A_n$.

Having established masslessness and momentum conservation, we can now write
\bea
\D_B  = J_B \,
\d^4(\sum^n_{i=1} p_i) \prod^n_{i=1} \d(p_i^2)
\prod^{n-2}_{i=1} \d^3 \left(p^{\a \dot{ \a} }_i -
{\r^{\a} (\s_i) \tilde{\r}^{\dot\a} (\s_i) \over P_i(\s) } \right) \,
\d^2 \left(p^{\a\dot{ \a} }_{n} -  { \r^{\a} (\s_n) \tilde{\r}^{\dot\a}
(\s_n) \over P_n(\s) } \right),
\eea
where the three-dimensional delta functions can be chosen, for instance, to be
$\{\a \, \dot{\a} \}= \{ 1 \dot{1} \}, \{ 2 \dot{1}  \}, \{ 2 \dot{2}  \}$,
and the two-dimensional delta function of particle $n$ can be chosen to be
$ \{\a \, \dot{\a} \}= \{ 1 \dot{1} \}, \{ 2 \dot{1}  \}$. For these choices,
the Jacobian $J_B$ is
\bea
J_B =  \tilde{\l}^{\dot{1}}_{n-1} \tilde{\l}^{\dot{1}}_{n}
\langle n{-}1 \, n\rangle \,  \prod^{n-2}_{i=1} p^{2 \dot{1}}_i \, .
\eea
By the same kind of reasoning, the first set of fermionic delta functions in $\D_F$
can be recast in the form
\bea
J_F \, \d^{4} ( \sum^{n}_{i=1}  q^{\a I}_{i} )
\prod^n_{i=1} \d^{2}( \langle \l_i \, q^I_{i} \rangle ) \prod^{n-2}_{i=1}
\d^{2}\left( \l^1_i \eta^{I}_{i} - { \r^{ 1 } (\s_i) \chi^{I} (\s_i) \over P_i(\s) } \right) \, ,
\eea
with $J_F$ given by
\bea
J_F ={1 \over \langle n{-}1 \, n \rangle^2} \prod^{n-2}_{i=1}
 \left( {1 \over \l^1_i} \right)^2 \, ,
\eea
and similarly for the second set of fermionic delta functions.

\subsection{M5 theory} \label{appendix:M5Jacobi}

Let us now consider the 6D formula for the M5-theory amplitudes.
Beginning with the bosonic delta functions, we can extract the mass-shell and
momentum-conservation delta functions as follows
\bea
&& \prod^n_{i=1}  \d^6 \left(p^{AB }_i
-  { \r^{A }_a (\s_i)  {\r}^{ Ba} (\s_i) \over P_n(\s_i) } \right)  =
\d^6(\sum^n_{i=1} p_i) \prod^n_{i=1} \d(p_i^2) \\ \nonumber
&\times &
J_B \, \prod^{n-2}_{i=1} \d^5 \left(p^{A B }_i -  { \r^{A }_a (\s_i)  {\r}^{ Ba} (\s_i)
\over P_i(\s) } \right) \, \d^4 \left(p^{A B}_{n}
-   { \r^{A }_a (\s_n)  {\r}^{ Ba} (\s_n) \over P_n(\s) } \right) \, .
\eea
If we choose the five-dimensional delta function with $\{A,B\} \neq \{3,4\}$
and the four-dimensional one with
$\{A,B\} \neq \{3,4\}, \{1,3\}$, $J_B$ is given by
\bea
J_B = \prod^{n}_{i=1} p_i^{12} \, \left( { p^{24}_{n-1}
\over p_{n-1}^{12} } - { p^{24}_{n} \over p_{n}^{12} } \right).
\eea
Next, we proceed similarly for the fermionic delta functions.
Extracting the fermionic ``wave functions" and supercharge conservation from the
fermionic delta functions gives
\bea
&& \prod^n_{i=1} \d^{8} \left(q^{A I}_i -  { \r^A_a(\s_i) \chi^{Ia} (\s_i)  \over P_i(\s) } \right)
=
\d^{8}( \sum^n_{i=1} q^{A I}_i  ) \prod^n_{i=1} \d^{4}( \hat{\l }_{i A\hat{a}} q^{A I}_i  )
 \\ \nonumber
& \times &
\, J_F \,
 \prod^{n-2}_{i=1} \d^{2}\left(q^{1 I}_i -   { \r^1_a(\s_i) \chi^{Ia} (\s_i)
  \over P_i(\s) }  \right) \d^{2} \left(q^{3 I}_i -   { \r^3_a(\s_i) \chi^{Ia} (\s_i)
  \over P_i(\s) }  \right) \, ,
\eea
with the Jacobian
\bea
J_F = { 1 \over [ \hat{\l}_{n-1\, \hat{a}}  \hat{\l}^{\hat{a}}_{n} \,
 \hat{\l}_{n-1\, \hat{b}}  \hat{\l}^{\hat{b}}_{n}    ]^2 }
  \prod^{n-2}_{i=1}  \left ({ 1 \over [ \hat{\l}^2_i  \,
  \hat{\l}^4_i  ] } \right)^2 \, ,
\eea
where $[ \hat{\l}^2_i  \, \hat{\l}^4_i  ] = \e^{\hat{a} \hat{b}}
\hat{\l}^{2}_{i \hat{a}}  \hat{\l}^{4}_{i \hat{b}} $.

\section{R symmetry} \label{appendix:Rsymmetry}

\subsection{D3 theory}

Let us now verify the $SU(4)$ R symmetry of the D3 theory. (The $U(1)$ factor of the R symmetry
was established in the main text.) As presented in Sect.~\ref{sec:nptD3}, the formula for the amplitudes
only makes an $SU(2) \times SU(2)$ subgroup manifest. However, as we saw in the case of the
four-particle amplitude, the full $SU(4)$ symmetry can be made manifest by performing
an appropriate Grassmann Fourier transform. For this purpose, it is useful to first recast
the fermionic delta functions as follows
\bea
\prod^n_{i=1}\d^{4} \left(q^{\a I}_{i} -
{ \r^{ \a } (\s_i) \chi^{I} (\s_i)  \over P_i(\s) }  \right)
= J_F \times  \prod^n_{i=1}\left\{ \d^{2}( \langle \l_i \, q^I_{i} \rangle ) \,
  \d^{2}\left( \eta^{I}_{i}
  -   { \r^{ 1 } (\s_i) \chi^{I} (\s_i)  \over \l^1_i  P_i(\s) }  \right)\right\}
\eea
and similarly for the $\hat q$ and $\tilde{\l}$ sector. Here the explicit expression for the fermionic Jacobian $J_F$ is not important for the following discussion.

Now let us consider the Grassmann Fourier transformation
\bea
I_F = \int \left(\prod^{d}_{m=0}  d^2 \chi^I_{m} d^2 \hat\chi^{\hat I}_{m}\right)
 \,
{\rm exp}( \sum^n_{i=1} \hat\eta^{\hat I}_{i} \z_{i \hat I} )
\prod^{n}_{i=1} d^2 \hat\eta^{\hat I}_{i}
\d^{2}( \eta^{I}_{i} -   t_i \, \chi^{I} (\s_i)    )
\d^{2}( \hat\eta^{\hat I}_{i} -   \tilde{t}_i \,  \hat\chi^{\hat I} (\s_i)    ) \, ,
\eea
where we have Fourier transformed $\hat\eta^{\hat I}_{i}$ and defined
\bea
t_i =  { \r^{ 1 } (\s_i)  \over \l^1_i  P_i(\s) }  \quad {\rm and} \quad
\tilde{t}_i = { \tilde{\r}^{ \dot{1} } (\s_i)  \over \tilde{\l}^{ \dot{1} } _i P_i(\s) }  \, .
\eea
Since the bosonic delta functions (not displayed in this Appendix) imply that
\bea
p_i^{1 \dot 1} = \frac{\r^{ 1 } (\s_i) \tilde{\r}^{ \dot{1} } (\s_i)}{P_i(\s)}
= \l_i^1 \tilde\l_i^{\dot 1},
\eea
we have
\bea \label{tteqn}
t_i \tilde{t}_i = 1 / P_i(\s).
\eea
Integration over
$d^2 \hat\eta^{\hat I}_{i}$ gives
\bea
I_F = \int \left(\prod^{d}_{m=0}  d^2 \chi^I_{m} d^2 \hat\chi^{\hat I}_{m}\right) \,
{\rm exp}( \sum^n_{i=1} \tilde{t}_i \,  \hat\chi^{\hat I} (\s_i)  \z_{i \hat I} )
\prod^n_{i=1} \d^{2}( \eta^{I}_{i} - t_i \, \chi^{I} (\s_i)) \, ,
\eea
and further integration over $d^2 \hat\chi^{\hat I}_{m} $ leads to
\bea
I_F=  \prod^{d }_{m=0}\d^{2} (\sum^n_{i=1} \tilde{t}_i \z_{i \hat I} \s_i^m )  \,
\int  \prod^{d}_{m=0}  d^2 \chi^I_{m}
  \prod^n_{i=1} \d^{2}( \eta^{I}_{i} -   t_i \, \chi^{I} (\s_i)    ) \, .
\eea

The final integration over $d^2 \chi_m^I$ involves $n$ integrals of $2n$
delta functions, thereby leaving
$n$ delta functions. Using Eqs.~(\ref{vanish}) and (\ref{tteqn}), it is
\bea
\int  \prod^{d  }_{m=0}  d^2 \chi^I_{m}
  \prod^n_{i=1} \d^{2}( \eta^{I}_{i} -   t_i \, \chi^{I} (\s_i)    )
  = (V_n \prod^n_{i=1} \tilde{t}_i)^{-1}
  \prod^d_{m=0} \d^{2}(  \sum^n_{i=1}  \tilde{t}_i \,  \eta^I_i \s_i^m   ) \, ,
\eea
Renaming $\z_{i\hat 1} = \eta_i^3$ and $\z_{i\hat 2} = \eta_i^4$, as before, we now have a complete
$SU(4)$ multiplet $\eta^I_i$ with $I=1,2,3,4$, and
\bea
I_F \sim  \prod^{d}_{m=0}\d^{4} (\sum^n_{i=1} \tilde{t}_i \eta^I_{i} \s_i^m ) \, ,
\eea
which is now manifestly $SU(4)$ invariant.

\subsection{M5 theory} \label{appendix:RsymmetryM5}

Next we wish to verify the $USp(4)$ R symmetry of the M5 theory.
As in the case of 4D, it is useful to begin by decomposing
the supercharge-conservation delta functions as follows
\bea \label{eq:fermiondelta}
&& \int \prod^{ d }_{m=0} d^2 \chi^I_{m+} d^2 \chi^I_{m-} \prod^n_{i=1} \d^{8} \left(q^{A I}_i
-  { \r^A_a(\s_i) \chi^{Ia} (\s_i)  \over P_i(\s) }  \right)  \\ \nonumber
& =&
J_F \, \prod^n_{i=1} \d^{4}( \tilde{\l }_{i A\dot{a}} q^{A I}_i  )  \, \int \prod^{ d }_{m=0} d^2 \chi^I_{m+} d^2 \chi^I_{m-}  \,
 \prod^{n}_{i=1} \d^{2} \left(q^{1 I}_i - { \r^1_a(\s_i) \chi^{Ia} (\s_i)
 \over P_i(\s) } \right) \d^{2} \left(q^{3 I}_i -   { \r^3_a(\s_i) \chi^{Ia} (\s_i)
 \over P_i(\s) }  \right)
\eea
where $d={n\over 2} -1$ and the Jacobian is given by
\bea
J_F = \prod^{n}_{i=1}  \left({ 1 \over \langle \tilde{\l}^2_i  \, \tilde{\l}^4_i  \rangle } \right)^2 \, .
\eea
Again the choice of singling out Lorentz indices $1,3$ is arbitrary.
Ignore all the Jacobi, which are not relevant to the R symmetry, the integration over $\chi$'s in the second line of Eq.~(\ref{eq:fermiondelta}) reduces to
\bea
\int \prod^{ d }_{m=0} d^2 \chi^I_{m+} d^2 \chi^I_{m-} \prod^{n}_{i=1} \d^{2} \left( \eta^{I}_{ i+} -  { \langle X^I_i \,
\l_{i+} \rangle_{13} \over  p^{13}_i } \right) \d^{2} \left( \eta^{I}_{ i-}
- { \langle X^I_i \, \l_{ i-} \rangle_{13} \over  p^{13}_i  } \right)
\, ,
\eea
where
\bea
\langle X^I_i \,  \l_{i a} \rangle_{13}  = X^{I3}_i \,
\l^1_{i a} - X^{I1}_i \,  \l^3_{i a} \, ,
\eea
with
\bea
X^{I1}_i = { \r^1_a(\s_i) \chi^{Ia} (\s_i) \over P_i(\s) }
\eea
and similarly for $ X^{I 3}_i $. Fourier transforming over $\eta^I_{i-}$ now gives
\bea
\int \prod^{ d }_{m=0} d^2 \chi^I_{m+} d^2 \chi^I_{m-}
\exp \left( \sum^n_{i=1}  { \z_{iI}   \langle X^I_i \,  \l_{i-} \rangle_{13}
\over  p^{13}_i   }    \right)     \prod^{n}_{i=1}
\d^{2} \left( \eta^{I}_{i+} -   {  \langle X^I_i \,
\l_{i+} \rangle_{13} \over  p^{13}_i } \right) \, .
\eea

The remaining $2n$ delta functions are exactly enough to integrate out
the $\chi^I_+$'s and $\chi^I_-$'s. Explicitly, the delta functions lead to,
\bea \label{eq:eta}
\eta^I_{i+} = \sum^{d}_{m=0}
\kappa_{i, m, a} (\lambda_+) \, \chi_{m}^{I\,a} \, ,
\eea
where the matrix $\kappa$ is a square $n \times n$ matrix (with $i$ running from $1$ to $n$, and $m, a$ together from $1$ to $n$), and it is given by
\bea
\kappa_{i, m, a} (\lambda_+) = {  ( \r^1_a(\s_i) \lambda^3_{i\,+}  -  \r^3_a(\s_i) \lambda^1_{i\,+} ) \s_i^m  \over p^{13}_{i} P_i( \s )  }  \, .
\eea
Solve $\chi_{m}^{I\,a}$ in terms of $\eta^I_{i+}$ using Eq.~(\ref{eq:eta}), and plug the result into the
exponent (again ignoring the Jacobian, which is not relevant here), we arrive at
\bea
   \exp \left( \sum^n_{i,j = 1}    \z_{iI}   M_{ij} \eta^I_{j+}    \right)    \, ,
\eea
with the matrix $M_{ij}$ given by
\bea
M_{ij} =\sum^{d}_{m=0} \kappa^a_{i, m} (\lambda_-)  \kappa^{-1}_{j, m, a} (\lambda_+) \,.
\eea
If the matrix $M_{ij}$ is symmetric, then (as we showed for the case of $n=4$ in Sect.~\ref{sec:4ptsM5}), the
expression has manifest R symmetry. We have checked explicitly that is indeed the case for $n=6,8$.  We also note that the matrix $M_{ij}$ has following property of converting
$\lambda^A_{j +}$ into $\lambda^A_{j -}$,
\bea \label{eq:M+-}
\sum_j M_{ij} \lambda^A_{j +} = \lambda^A_{i -} \, .
\eea
Multiplying $\lambda^B_{i +}$ on both sides of the equation and summing over $i$ gives
\bea \label{eq:M+-2}
\sum_{i,j} \lambda^B_{i +} M_{ij} \lambda^A_{j +} = \sum_i \lambda^B_{i +}  \lambda^A_{i -} \, .
\eea
Due to momentum conservation, the right-hand side of this equation is symmetric in exchanging
$A$ and $B$, which is consistent with the fact that  $M_{ij}$ is symmetric.  Curiously, the complete formula for the amplitude
with manifest R symmetry is somewhat more complicated than the original one, which only makes
a subgroup manifest.


\newpage

\begin{thebibliography}{99}

%\cite{Volkov:1973ix}
\bibitem{Volkov:1973ix}
  D.~V.~Volkov and V.~P.~Akulov,
  ``Is the Neutrino a Goldstone Particle?,''
  Phys.\ Lett.\  {\bf 46B}, 109 (1973).
  %doi:10.1016/0370-2693(73)90490-5
  %%CITATION = doi:10.1016/0370-2693(73)90490-5;%%
  %1547 citations counted in INSPIRE as of 15 Aug 2017

%\cite{Kallosh:1997aw}
\bibitem{Kallosh:1997aw}
  R.~Kallosh,
  ``Volkov-Akulov Theory and D-branes,''
  Lect.\ Notes Phys.\  {\bf 509}, 49 (1998)
 % doi:10.1007/BFb0105228
  [hep-th/9705118].
  %CITATION = doi:10.1007/BFb0105228;%%
  %20 citations counted in INSPIRE as of 17 Feb 2016

    %\cite{Rosly:2002jt}
\bibitem{Rosly:2002jt}
  A.~A.~Rosly and K.~G.~Selivanov,
  ``Helicity Conservation in Born-Infeld Theory,''
  hep-th/0204229.
  %%CITATION = HEP-TH/0204229;%%
  %7 citations counted in INSPIRE as of 26 Oct 2017

  %\cite{Gibbons:1995cv}
\bibitem{Gibbons:1995cv}
  G.~W.~Gibbons and D.~A.~Rasheed,
  ``Electric - Magnetic Duality Rotations in Nonlinear Electrodynamics,''
  Nucl.\ Phys.\ B {\bf 454}, 185 (1995)
%  doi:10.1016/0550-3213(95)00409-L
  [hep-th/9506035].
  %%CITATION = doi:10.1016/0550-3213(95)00409-L;%%
  %297 citations counted in INSPIRE as of 26 Oct 2017

%\cite{Aganagic:1996nn}
\bibitem{Aganagic:1996nn}
  M.~Aganagic, C.~Popescu and J.~H.~Schwarz,
  ``Gauge-Invariant and Gauge-Fixed D-brane Actions,''
  Nucl.\ Phys.\ B {\bf 495}, 99 (1997)
  % doi:10.1016/S0550-3213(97)00180-6
[hep-th/9612080].
  %CITATION = doi:10.1016/S0550-3213(97)00180-6;%%
  %316 citations counted in INSPIRE as of 18 Jan 2016

%\cite{Cederwall:1996pv}
\bibitem{Cederwall:1996pv}
  M.~Cederwall, A.~von Gussich, B.~E.~W.~Nilsson and A.~Westerberg,
  ``The Dirichlet Super Three-brane in Ten-dimensional Type IIB Supergravity,''
  Nucl.\ Phys.\ B {\bf 490}, 163 (1997)
 % doi:10.1016/S0550-3213(97)00071-0
  [hep-th/9610148].
  %%CITATION = doi:10.1016/S0550-3213(97)00071-0;%%
  %256 citations counted in INSPIRE as of 17 févr. 2016

%\cite{Aganagic:1996pe}
\bibitem{Aganagic:1996pe}
  M.~Aganagic, C.~Popescu and J.~H.~Schwarz,
  ``D-brane Actions with Local Kappa Symmetry,''
  Phys.\ Lett.\ B {\bf 393}, 311 (1997)
%  doi:10.1016/S0370-2693(96)01643-7
  [hep-th/9610249].
  %%CITATION = doi:10.1016/S0370-2693(96)01643-7;%%
  %279 citations counted in INSPIRE as of 17 Feb 2016

%\cite{Cederwall:1996ri}
\bibitem{Cederwall:1996ri}
  M.~Cederwall, A.~von Gussich, B.~E.~W.~Nilsson, P.~Sundell and A.~Westerberg,
  ``The Dirichlet Super p-branes in Ten-dimensional Type IIA and IIB Supergravity,''
  Nucl.\ Phys.\ B {\bf 490}, 179 (1997)
 % doi:10.1016/S0550-3213(97)00075-8
  [hep-th/9611159].
  %CITATION = doi:10.1016/S0550-3213(97)00075-8;%%
  %361 citations counted in INSPIRE as of 17 févr. 2016

%\cite{Bergshoeff:1996tu}
\bibitem{Bergshoeff:1996tu}
  E.~Bergshoeff and P.~K.~Townsend,
  ``Super D-branes,''
  Nucl.\ Phys.\ B {\bf 490}, 145 (1997)
 % doi:10.1016/S0550-3213(97)00072-2
  [hep-th/9611173].
  %CITATION = doi:10.1016/S0550-3213(97)00072-2;%%
  %448 citations counted in INSPIRE as of 17 févr. 2016

%\cite{Bergshoeff:2013pia}
\bibitem{Bergshoeff:2013pia}
  E.~Bergshoeff, F.~Coomans, R.~Kallosh, C.~S.~Shahbazi and A.~Van Proeyen,
  ``Dirac-Born-Infeld-Volkov-Akulov and Deformation of Supersymmetry,''
  JHEP {\bf 1308}, 100 (2013)
 % doi:10.1007/JHEP08(2013)100
  [arXiv:1303.5662 [hep-th]].
  %%CITATION = doi:10.1007/JHEP08(2013)100;%%
  %23 citations counted in INSPIRE as of 17 févr. 2016

%\cite{He:2016vfi}
\bibitem{He:2016vfi}
  S.~He, Z.~Liu and J.~B.~Wu,
  ``Scattering Equations, Twistor-string Formulas and Double-soft Limits in Four Dimensions,''
  JHEP {\bf 1607}, 060 (2016)
 % doi:10.1007/JHEP07(2016)060
  [arXiv:1604.02834 [hep-th]].
  %%CITATION = doi:10.1007/JHEP07(2016)060;%%
  %24 citations counted in INSPIRE as of 03 Aug 2017

  %\cite{Cachazo:2016njl}
\bibitem{Cachazo:2016njl}
  F.~Cachazo, P.~Cha and S.~Mizera,
  ``Extensions of Theories from Soft Limits,''
  JHEP {\bf 1606}, 170 (2016)
 % doi:10.1007/JHEP06(2016)170
  [arXiv:1604.03893 [hep-th]].
  %%CITATION = doi:10.1007/JHEP06(2016)170;%%
  %42 citations counted in INSPIRE as of 03 Oct 2017

%\cite{Perry:1996mk}
\bibitem{Perry:1996mk}
  M.~Perry and J.~H.~Schwarz,
  ``Interacting Chiral Gauge Gields in Six Dimensions and Born--Infeld Theory,''
  Nucl.\ Phys.\ B {\bf 489}, 47 (1997)
%  doi:10.1016/S0550-3213(97)00040-0
  [hep-th/9611065].
  %%CITATION = doi:10.1016/S0550-3213(97)00040-0;%%
  %163 citations counted in INSPIRE as of 03 Aug 2017

%\cite{Aganagic:1997zq}
\bibitem{Aganagic:1997zq}
  M.~Aganagic, J.~Park, C.~Popescu and J.~H.~Schwarz,
  ``World Volume Action of the M-Theory Five-brane,''
  Nucl.\ Phys.\ B {\bf 496}, 191 (1997)
 % doi:10.1016/S0550-3213(97)00227-7
  [hep-th/9701166].
  %%CITATION = doi:10.1016/S0550-3213(97)00227-7;%%
  %295 citations counted in INSPIRE as of 03 Aug 2017

%\cite{Howe:1996yn}
\bibitem{Howe:1996yn}
  P.~S.~Howe and E.~Sezgin,
  ``D = 11, p = 5,''
  Phys.\ Lett.\ B {\bf 394}, 62 (1997)
 % doi:10.1016/S0370-2693(96)01672-3
  [hep-th/9611008].
  %%CITATION = doi:10.1016/S0370-2693(96)01672-3;%%
  %206 citations counted in INSPIRE as of 28 Sep 2017

%\cite{Pasti:1997gx}
\bibitem{Pasti:1997gx}
  P.~Pasti, D.~P.~Sorokin and M.~Tonin,
  ``Covariant Action for a D = 11 Five-brane with the Chiral Field,''
  Phys.\ Lett.\ B {\bf 398}, 41 (1997)
  %doi:10.1016/S0370-2693(97)00188-3
  [hep-th/9701037].
  %%CITATION = doi:10.1016/S0370-2693(97)00188-3;%%
  %283 citations counted in INSPIRE as of 28 Sep 2017

%\cite{Bandos:1997ui}
\bibitem{Bandos:1997ui}
  I.~A.~Bandos, K.~Lechner, A.~Nurmagambetov, P.~Pasti, D.~P.~Sorokin and M.~Tonin,
  ``Covariant Action for the Superfive-brane of M Theory,''
  Phys.\ Rev.\ Lett.\  {\bf 78}, 4332 (1997)
 % doi:10.1103/PhysRevLett.78.4332
  [hep-th/9701149].
  %%CITATION = doi:10.1103/PhysRevLett.78.4332;%%
  %348 citations counted in INSPIRE as of 28 Sep 2017

%\cite{Howe:1997fb}
\bibitem{Howe:1997fb}
  P.~S.~Howe, E.~Sezgin and P.~C.~West,
  ``Covariant Field Equations of the M Theory Five-brane,''
  Phys.\ Lett.\ B {\bf 399}, 49 (1997)
 % doi:10.1016/S0370-2693(97)00257-8
  [hep-th/9702008].
  %%CITATION = doi:10.1016/S0370-2693(97)00257-8;%%
  %226 citations counted in INSPIRE as of 28 Sep 2017

  %\cite{Huang:2010rn}
\bibitem{Huang:2010rn}
  Y.~t.~Huang and A.~E.~Lipstein,
  ``Amplitudes of 3D and 6D Maximal Superconformal Theories in Supertwistor Space,''
  JHEP {\bf 1010}, 007 (2010)
  %doi:10.1007/JHEP10(2010)007
  [arXiv:1004.4735 [hep-th]].
  %%CITATION = doi:10.1007/JHEP10(2010)007;%%
  %55 citations counted in INSPIRE as of 10 Sep 2017

  %\cite{Czech:2011dk}
\bibitem{Czech:2011dk}
  B.~Czech, Y.~t.~Huang and M.~Rozali,
  ``Chiral Three-point Interactions in 5 and 6 Dimensions,''
  JHEP {\bf 1210}, 143 (2012)
  %doi:10.1007/JHEP10(2012)143
  [arXiv:1110.2791 [hep-th]].
  %%CITATION = doi:10.1007/JHEP10(2012)143;%%
  %25 citations counted in INSPIRE as of 10 Sep 2017

    %\cite{Elvang:2012st}
\bibitem{Elvang:2012st}
  H.~Elvang, D.~Z.~Freedman, L.~Y.~Hung, M.~Kiermaier, R.~C.~Myers and S.~Theisen,
  ``On Renormalization Group Flows and the a-theorem in 6d,''
  JHEP {\bf 1210}, 011 (2012)
  %doi:10.1007/JHEP10(2012)011
  [arXiv:1205.3994 [hep-th]].
  %%CITATION = doi:10.1007/JHEP10(2012)011;%%
  %75 citations counted in INSPIRE as of 01 Sep 2017

  %\cite{Chen:2015hpa}
\bibitem{Chen:2015hpa}
  W.~M.~Chen, Y.~t.~Huang and C.~Wen,
  ``Exact Coefficients for Higher Dimensional Operators with Sixteen Supersymmetries,''
  JHEP {\bf 1509}, 098 (2015)
  %doi:10.1007/JHEP09(2015)098
  [arXiv:1505.07093 [hep-th]].
  %%CITATION = doi:10.1007/JHEP09(2015)098;%%
  %9 citations counted in INSPIRE as of 01 Sep 2017

  %\cite{Bianchi:2016viy}
\bibitem{Bianchi:2016viy}
  M.~Bianchi, A.~L.~Guerrieri, Y.~t.~Huang, C.~J.~Lee and C.~Wen,
  ``Exploring Soft Constraints on Effective Actions,''
  JHEP {\bf 1610}, 036 (2016)
  %doi:10.1007/JHEP10(2016)036
  [arXiv:1605.08697 [hep-th]].
  %%CITATION = doi:10.1007/JHEP10(2016)036;%%
  %10 citations counted in INSPIRE as of 01 Sep 2017

%\cite{Strominger:1995ac}
\bibitem{Strominger:1995ac}
  A.~Strominger,
  ``Open p-branes,''
  Phys.\ Lett.\ B {\bf 383}, 44 (1996)
  %doi:10.1016/0370-2693(96)00712-5
  [hep-th/9512059].
  %%CITATION = doi:10.1016/0370-2693(96)00712-5;%%
  %572 citations counted in INSPIRE as of 27 Sep 2017

 %\cite{Witten:2003nn}
\bibitem{Witten:2003nn}
  E.~Witten,
  ``Perturbative Gauge Theory as a String Theory in Twistor Space,''
  Commun.\ Math.\ Phys.\  {\bf 252}, 189 (2004)
 % doi:10.1007/s00220-004-1187-3
  [hep-th/0312171].
  %%CITATION = doi:10.1007/s00220-004-1187-3;%%
  %959 citations counted in INSPIRE as of 10 Sep 2017

  %\cite{Roiban:2004vt}
\bibitem{Roiban:2004vt}
  R.~Roiban, M.~Spradlin and A.~Volovich,
  ``A Googly Amplitude from the B Model in Twistor Space,''
  JHEP {\bf 0404}, 012 (2004)
%  doi:10.1088/1126-6708/2004/04/012
  [hep-th/0402016].
  %%CITATION = doi:10.1088/1126-6708/2004/04/012;%%
  %117 citations counted in INSPIRE as of 20 Nov 2017

  %\cite{Roiban:2004ka}
\bibitem{Roiban:2004ka}
  R.~Roiban and A.~Volovich,
  ``All Conjugate-Maximal-Helicity-Violating Amplitudes from Topological Open String Theory in Twistor Space,''
  Phys.\ Rev.\ Lett.\  {\bf 93}, 131602 (2004)
%  doi:10.1103/PhysRevLett.93.131602
  [hep-th/0402121].
  %%CITATION = doi:10.1103/PhysRevLett.93.131602;%%
  %84 citations counted in INSPIRE as of 20 Nov 2017

    %\cite{Roiban:2004yf}
\bibitem{Roiban:2004yf}
  R.~Roiban, M.~Spradlin and A.~Volovich,
  ``On the Tree Level S Matrix of Yang-Mills Theory,''
  Phys.\ Rev.\ D {\bf 70}, 026009 (2004)
 % doi:10.1103/PhysRevD.70.026009
  [hep-th/0403190].
  %%CITATION = doi:10.1103/PhysRevD.70.026009;%%
  %192 citations counted in INSPIRE as of 02 Oct 2017

%\cite{Cachazo:2013iaa}
\bibitem{Cachazo:2013iaa}
  F.~Cachazo, S.~He and E.~Y.~Yuan,
  ``Scattering in Three Dimensions from Rational Maps,''
  JHEP {\bf 1310}, 141 (2013)
  %doi:10.1007/JHEP10(2013)141
  [arXiv:1306.2962 [hep-th]].
  %%CITATION = doi:10.1007/JHEP10(2013)141;%%
  %60 citations counted in INSPIRE as of 18 Aug 2017

%\cite{Cachazo:2013gna}
\bibitem{Cachazo:2013gna}
  F.~Cachazo, S.~He and E.~Y.~Yuan,
  ``Scattering Equations and Kawai-Lewellen-Tye Orthogonality,''
  Phys.\ Rev.\ D {\bf 90}, no. 6, 065001 (2014)
 % doi:10.1103/PhysRevD.90.065001
  [arXiv:1306.6575 [hep-th]].
  %%CITATION = doi:10.1103/PhysRevD.90.065001;%%
  %155 citations counted in INSPIRE as of 02 Oct 2017

  %\cite{Cheung:2009dc}
\bibitem{Cheung:2009dc}
  C.~Cheung and D.~O'Connell,
  ``Amplitudes and Spinor-Helicity in Six Dimensions,''
  JHEP {\bf 0907}, 075 (2009)
  %doi:10.1088/1126-6708/2009/07/075
  [arXiv:0902.0981 [hep-th]].
  %%CITATION = doi:10.1088/1126-6708/2009/07/075;%%
  %105 citations counted in INSPIRE as of 11 Sep 2017

  %\cite{Dennen:2009vk}
\bibitem{Dennen:2009vk}
  T.~Dennen, Y.~t.~Huang and W.~Siegel,
  ``Supertwistor Space for 6D Maximal Super Yang--Mills,''
  JHEP {\bf 1004}, 127 (2010)
  %doi:10.1007/JHEP04(2010)127
  [arXiv:0910.2688 [hep-th]].
  %%CITATION = doi:10.1007/JHEP04(2010)127;%%
  %57 citations counted in INSPIRE as of 26 Sep 2017

  %\cite{Bern:2010qa}
\bibitem{Bern:2010qa}
  Z.~Bern, J.~J.~Carrasco, T.~Dennen, Y.~t.~Huang and H.~Ita,
  ``Generalized Unitarity and Six-Dimensional Helicity,''
  Phys.\ Rev.\ D {\bf 83}, 085022 (2011)
  %doi:10.1103/PhysRevD.83.085022
  [arXiv:1010.0494 [hep-th]].
  %%CITATION = doi:10.1103/PhysRevD.83.085022;%%
  %83 citations counted in INSPIRE as of 26 Sep 2017

  %\cite{Brandhuber:2010mm}
\bibitem{Brandhuber:2010mm}
  A.~Brandhuber, D.~Korres, D.~Koschade and G.~Travaglini,
  ``One-loop Amplitudes in Six-Dimensional (1,1) Theories from Generalised Unitarity,''
  JHEP {\bf 1102}, 077 (2011)
 % doi:10.1007/JHEP02(2011)077
  [arXiv:1010.1515 [hep-th]].
  %%CITATION = doi:10.1007/JHEP02(2011)077;%%
  %27 citations counted in INSPIRE as of 26 Sep 2017

%\cite{Spradlin:2009qr}
\bibitem{Spradlin:2009qr}
  M.~Spradlin and A.~Volovich,
``From Twistor String Theory to Recursion Relations,''
  Phys.\ Rev.\ D {\bf 80}, 085022 (2009)
%  doi:10.1103/PhysRevD.80.085022
  [arXiv:0909.0229 [hep-th]].
  %%CITATION = doi:10.1103/PhysRevD.80.085022;%%
  %35 citations counted in INSPIRE as of 20 Nov 2017

%\cite{Cachazo:2013hca}
\bibitem{Cachazo:2013hca}
  F.~Cachazo, S.~He and E.~Y.~Yuan,
  ``Scattering of Massless Particles in Arbitrary Dimensions,''
  Phys.\ Rev.\ Lett.\  {\bf 113}, no. 17, 171601 (2014)
 % doi:10.1103/PhysRevLett.113.171601
  [arXiv:1307.2199 [hep-th]].
  %%CITATION = doi:10.1103/PhysRevLett.113.171601;%%
  %105 citations counted in INSPIRE as of 20 janv. 2016

%\cite{Cachazo:2013iea}
\bibitem{Cachazo:2013iea}
  F.~Cachazo, S.~He and E.~Y.~Yuan,
  ``Scattering of Massless Particles: Scalars, Gluons and Gravitons,''
  JHEP {\bf 1407}, 033 (2014)
 % doi:10.1007/JHEP07(2014)033
  [arXiv:1309.0885 [hep-th]].
  %%CITATION = doi:10.1007/JHEP07(2014)033;%%
  %90 citations counted in INSPIRE as of 20 janv. 2016

%\cite{Cachazo:2014xea}
\bibitem{Cachazo:2014xea}
  F.~Cachazo, S.~He and E.~Y.~Yuan,
  ``Scattering Equations and Matrices: From Einstein to Yang--Mills, DBI and NLSM,''
  JHEP {\bf 1507}, 149 (2015)
 % doi:10.1007/JHEP07(2015)149
  [arXiv:1412.3479 [hep-th]].
  %%CITATION = doi:10.1007/JHEP07(2015)149;%%
  %36 citations counted in INSPIRE as of 20 janv. 2016

    %\cite{Cheung:2015ota}
\bibitem{Cheung:2015ota}
  C.~Cheung, K.~Kampf, J.~Novotny, C.~H.~Shen and J.~Trnka,
  ``On-Shell Recursion Relations for Effective Field Theories,''
  Phys.\ Rev.\ Lett.\  {\bf 116}, no. 4, 041601 (2016)
  %doi:10.1103/PhysRevLett.116.041601
  [arXiv:1509.03309 [hep-th]].
  %%CITATION = doi:10.1103/PhysRevLett.116.041601;%%
  %27 citations counted in INSPIRE as of 10 Sep 2017

  %\cite{Cachazo:2013zc}
\bibitem{Cachazo:2013zc}
  F.~Cachazo,
  ``Resultants and Gravity Amplitudes,''
  arXiv:1301.3970 [hep-th].
  %%CITATION = ARXIV:1301.3970;%%
  %17 citations counted in INSPIRE as of 11 Sep 2017

%\cite{Gukov:2004ei}
\bibitem{Gukov:2004ei}
  S.~Gukov, L.~Motl and A.~Neitzke,
  ``Equivalence of Twistor Prescriptions for Super Yang--Mills,''
  Adv.\ Theor.\ Math.\ Phys.\  {\bf 11}, no. 2, 199 (2007)
  %doi:10.4310/ATMP.2007.v11.n2.a1
  [hep-th/0404085].
  %%CITATION = doi:10.4310/ATMP.2007.v11.n2.a1;%%
  %68 citations counted in INSPIRE as of 27 Aug 2017

%\cite{Vergu:2006np}
\bibitem{Vergu:2006np}
  C.~Vergu,
  ``On the Factorisation of the Connected Prescription for Yang--Mills Amplitudes,''
  Phys.\ Rev.\ D {\bf 75}, 025028 (2007)
  %doi:10.1103/PhysRevD.75.025028
  [hep-th/0612250].
  %%CITATION = doi:10.1103/PhysRevD.75.025028;%%
  %19 citations counted in INSPIRE as of 27 Aug 2017

%\cite{Cachazo:2012pz}
\bibitem{Cachazo:2012pz}
  F.~Cachazo, L.~Mason and D.~Skinner,
  ``Gravity in Twistor Space and its Grassmannian Formulation,''
  SIGMA {\bf 10}, 051 (2014)
  %doi:10.3842/SIGMA.2014.051
  [arXiv:1207.4712 [hep-th]].
  %%CITATION = doi:10.3842/SIGMA.2014.051;%%
  %47 citations counted in INSPIRE as of 27 Aug 2017

   %\cite{Guerrieri:2017ujb}
\bibitem{Guerrieri:2017ujb}
  A.~L.~Guerrieri, Y.~t.~Huang, Z.~Li and C.~Wen,
  ``On the Exactness of Soft Theorems,''
  arXiv:1705.10078 [hep-th].
  %%CITATION = ARXIV:1705.10078;%%
  %1 citations counted in INSPIRE as of 01 Sep 2017

  %\cite{Cheung:2014dqa}
\bibitem{Cheung:2014dqa}
  C.~Cheung, K.~Kampf, J.~Novotny and J.~Trnka,
  ``Effective Field Theories from Soft Limits of Scattering Amplitudes,''
  Phys.\ Rev.\ Lett.\  {\bf 114}, no. 22, 221602 (2015)
 % doi:10.1103/PhysRevLett.114.221602
  [arXiv:1412.4095 [hep-th]].
  %%CITATION = doi:10.1103/PhysRevLett.114.221602;%%
  %44 citations counted in INSPIRE as of 01 Sep 2017

  %\cite{Chen:2014xoa}
\bibitem{Chen:2014xoa}
  W.~M.~Chen, Y.~t.~Huang and C.~Wen,
  ``New Fermionic Soft Theorems for Supergravity Amplitudes,''
  Phys.\ Rev.\ Lett.\  {\bf 115}, no. 2, 021603 (2015)
 %doi:10.1103/PhysRevLett.115.021603
  [arXiv:1412.1809 [hep-th]].
  %%CITATION = doi:10.1103/PhysRevLett.115.021603;%%
  %29 citations counted in INSPIRE as of 28 Sep 2017

%\cite{Boels:2008fc}
\bibitem{Boels:2008fc}
  R.~Boels, K.~J.~Larsen, N.~A.~Obers and M.~Vonk,
  ``MHV, CSW and BCFW: Field Theory Structures in String Theory Amplitudes,''
  JHEP {\bf 0811}, 015 (2008)
 % doi:10.1088/1126-6708/2008/11/015
  [arXiv:0808.2598 [hep-th]].
  %%CITATION = doi:10.1088/1126-6708/2008/11/015;%%
  %44 citations counted in INSPIRE as of 10 Sep 2017

  %\cite{Luo:2015tat}
\bibitem{Luo:2015tat}
  H.~Luo and C.~Wen,
  ``Recursion Relations from Soft Theorems,''
  JHEP {\bf 1603}, 088 (2016)
 % doi:10.1007/JHEP03(2016)088
  [arXiv:1512.06801 [hep-th]].
  %%CITATION = doi:10.1007/JHEP03(2016)088;%%
  %19 citations counted in INSPIRE as of 30 Sep 2017

  %\cite{Skinner:2013xp}
\bibitem{Skinner:2013xp}
  D.~Skinner,
  ``Twistor Strings for N=8 Supergravity,''
  arXiv:1301.0868 [hep-th].
  %%CITATION = ARXIV:1301.0868;%%
  %40 citations counted in INSPIRE as of 10 Sep 2017

  %\cite{Mason:2013sva}
\bibitem{Mason:2013sva}
  L.~Mason and D.~Skinner,
  ``Ambitwistor strings and the scattering equations,''
  JHEP {\bf 1407}, 048 (2014)
%  doi:10.1007/JHEP07(2014)048
  [arXiv:1311.2564 [hep-th]].
  %%CITATION = doi:10.1007/JHEP07(2014)048;%%
  %136 citations counted in INSPIRE as of 05 Oct 2017

\end{thebibliography}
\end{document}